\documentclass[12pt]{article}%
\usepackage{heck}
\usepackage{graphicx}
\usepackage{makeidx}
\usepackage{multicol}
\usepackage{geometry}
\usepackage{amsfonts}
\usepackage{mathrsfs}
\usepackage{amssymb}
\usepackage{amsmath}%
\providecommand{\U}[1]{\protect\rule{.1in}{.1in}}
\numberwithin{equation}{section}

\def\^{{\wedge}}
\def\*{{\star}}
\def\ad{{\mathop{\rm ad}}}

\def\Tr{{\mathop{\rm Tr}}}
\def\Spin{{\mathop{\rm Spin}}}

\def\rk{{\mathop{\rm rk}}}
\def\Pf{{\mathop{\rm Pf}}}
\def\bar{\overline}
\def\wt{\widetilde}
\def\ha{{\frac{1}{2}}}
\def\llangle{{\langle\langle}}
\def\rrangle{{\rangle\rangle}}

\def\BC{{\mathbb C}}
\def\BE{{\mathbb E}}

\def\BR{{\mathbb R}}
\def\BZ{{\mathbb Z}}

\def\CD{{\cal D}}

\def\CG{{\cal G}}
\def\CH{{\cal H}}
\def\CK{{\cal K}}

\def\CN{{\cal N}}
\def\CO{{\cal O}}

\def\SU{{\mathscr U}}

\begin{document}

\date{February, 2008}

\preprint{arXiv:0802.3391}

\institution{HarvardU}{Jefferson Physical Laboratory, Harvard University, Cambridge,
MA 02138, USA}%

\title{GUTs and Exceptional Branes in \\F-theory - I}%
%

\authors{Chris Beasley\footnote{e-mail: {\tt
beasley@physics.harvard.edu}},
Jonathan J. Heckman\footnote{e-mail: {\tt
jheckman@fas.harvard.edu}} and
Cumrun Vafa\footnote{e-mail: {\tt vafa@physics.harvard.edu}%
}}%

\abstract{Motivated by potential phenomenological applications, we develop the necessary
tools for building GUT models in F-theory.  This approach is quite flexible because the local
geometrical properties of singularities in F-theory compactifications encode the physical content of the theory.
In particular, we show how geometry determines the gauge group, matter content and Yukawa
couplings of a given model.  It turns out that these features are beautifully captured by a four-dimensional topologically twisted
${\cal N }=4$ theory which has been coupled to a surface defect theory on which chiral matter can propagate.
From the vantagepoint of the four-dimensional topological theory, these defects are surface operators.
Specific intersection points of these defects lead to Yukawa couplings.  We also find that the unfolding
of the singularity in the F-theory geometry precisely matches to properties of the  topological theory
with a defect.}%

\maketitle

\tableofcontents

\pagebreak

\section{Introduction}

String theory appears to provide a large number of consistent vacua which can
accomodate the observed features of the Universe. \ Given this perhaps
embarassment of riches, it is natural to ask whether the deeper understanding
of non-perturbative features of string theory obtained in the post-duality era
provides some degree of uniqueness or at least some novel concrete predictions
for upcoming experiments.

There are encouraging signs that string theory naturally includes many of
the qualitative features of the Standard Model such as classical unitary gauge
groups and bifundamental chiral matter. \ Indeed, these are ubiquitous
features of D-brane realizations of gauge theories. \ Reviews of the vast
literature of models which attempt to realize the Standard Model via D-branes
may be found in \cite{CveticShiuReview,BlumenhagenReview,MarchesanoReview,MalyshevVerlindeReview}.

\begin{figure}
[ptb]
\begin{center}
\includegraphics[
height=6.4247in,
width=3.4886in
]%
{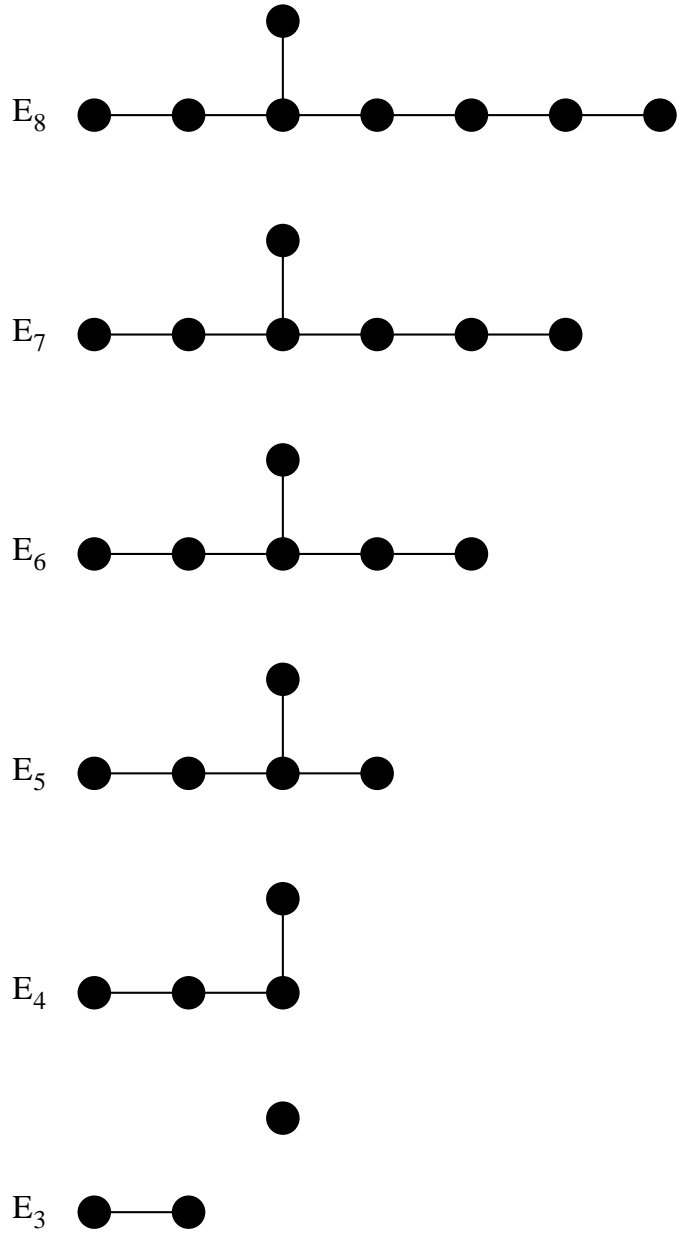}%
\caption{Dynkin diagrams for the $E$-series of Lie Groups. \ Starting from
$E_{8}$, deleting the rightmost node of each successive diagram produces the
next entry. \ The entry $E_{3}=SU(3)\times SU(2)$ is the non-abelian gauge
group of the Standard Model.}%
\label{DYNKIN}%
\end{center}
\end{figure}

Independent of its connection to string theory, a compelling motivation for
low energy supersymmetry is that the particle content of the MSSM\ improves
the unification of the gauge coupling constants observed in the Standard Model.
\ But because the a priori independent volumes of cycles wrapped by D-branes control the values of the
gauge coupling constants, D-brane constructions of Standard Model-like vacua
tend to obscure this fact. \ Both gauge coupling unification as well
as the matter content of the Standard Model hint at the
presence of a unified gauge group structure at high energies.\footnote{This is
not to say that the unification of gauge couplings in GUT-like models cannot
be realized in D-brane constructions. \ See \cite{Lykkenbraneguts} for one
early realization of a D-brane GUT.} \ In fact, this unification
naturally suggests the presence of an exceptional gauge group looming in the
background. \ Indeed, there is a natural sequence of $E$-group embeddings\footnote{Note that
even though in four dimensions only $E_6$ and lower rank exceptional groups can contain chiral matter, in
the context of higher dimensional theories coming from string theory, all of these groups can have chiral matter.
In this sense string theory completes the link between GUT theories and all exceptional groups by bringing
the higher dimensions into play.} depicted in figure \ref{DYNKIN} which
give the Standard Model gauge group and matter structure in an elegant manner:%
\begin{equation}
E_{3}\times U(1)\subset E_{4}\subset E_{5}\subset
E_{6}\subset... \label{nested}%
\end{equation}
where $E_{3}=SU(3)\times SU(2)$ denotes the non-abelian gauge group of the Standard Model,
$E_{4}=SU(5)$ and $E_{5}=SO(10)$. \ Some early field theory realizations
of this paradigm may be found in \cite{EBREAKI,EBREAKII,EBREAKIII}.%

Moreover, packaging the field content of the Standard Model into the
appropriate GUT\ multiplet is not always possible in D-brane realizations.
\ Indeed, although the classical groups appear in this sequence of embeddings,
only one and two index tensor representations of the gauge groups can appear
in D-brane constructions. \ In particular, for $SO(10)$ GUTs, the matter
content of the Standard Model organizes into the ${\bf 16}$ spinor representation. This
representation is conspicuously absent from perturbative D-brane setups.
Further issues pertaining to D-brane realizations of GUT models are in discussed in
\cite{BerensteinGUTS}.

From this perspective, we interpret the unification of the gauge couplings
in the minimal supersymmetric extension of the Standard Model as evidence
that instead of the infinitely complex variety of classical groups and matter
which can appear in string theory, the list of relevant simple gauge groups is
limited to the finite number of exceptional gauge groups and their subgroups.\footnote{Alternatively,
even if one views gauge coupling unification as an accident,
non-trivial qualitative criteria such as the existence of a self-similar
duality cascade structure can emerge from properties natural to D-branes in
the large $N$ limit of the MSSM or some minimal extension thereof \cite{SMCASCADE}.}
This is a vast simplification!

In this paper we take as given that the gauge groups unify into a GUT and
further that the matter content of the Standard Model descends naturally from
the representation content of an exceptional gauge group.\footnote{By this we
do not mean that we will restrict our model building efforts to four-dimensional $E_{6}$
GUTs. Indeed, as we shall argue below, many of the necessary features of even
more conventional GUTs require the presence of an $E_{6,7,8}$-type gauge group
which may be broken by the geometry of the F-theory compactification.} For
this reason, we now focus on how exceptional gauge groups arise in string
theory. Perhaps the most obvious answer is the $E_{8}\times E_{8}$ heterotic
string, as was already exploited more than two decades ago
\cite{GrossHarveyHeterotic}. Nevertheless, as reviewed in
\cite{WittenStrong}, the simplest approach in this direction does not quite
succeed because the perturbative heterotic $E_{8}\times E_{8}$ string
compactified on an isotropic Calabi-Yau threefold makes an (incorrect)
prediction for the relation between $M_{GUT}$ and $M_{pl}$:%
\begin{equation}
\frac{M_{GUT}^{2}}{M_{pl}^{2}}\gtrsim\alpha_{GUT}^{4/3}\text{.}%
\label{pertbound}%
\end{equation}
Witten has shown in \cite{WittenStrong} that the above bound can be
significantly weakened in the strongly coupled regime of the heterotic
$E_{8}\times E_{8}$ string described by M-theory compactified on $S^{1}/%
\mathbb{Z}
_{2}$. In the Ho\v{r}ava-Witten description, each $E_{8}$ gauge group factor
confines to a ten-dimensional \textquotedblleft end of the Universe\textquotedblright%
\ boundary of the eleven-dimensional M-theory compactification \cite{HoravaWitten}. \ As
shown in \cite{WittenStrong}, separating the two boundaries far away from one
another significantly weakens the above bound. \ At a pragmatic level this
allows us to focus on one of the $E_{8}$ boundary walls as leading to the
observed gauge symmetries of the Standard Model.

But many properties of the $E_{8}$ wall are mysterious.  In particular, the
argument for the choice of the $E_{8}$ gauge group is based primarily on
anomaly considerations. \ This fact presents a hindrance towards a
more detailed description of the properties of the matter content and their interactions.

Further insight is possible for exceptional gauge groups which are realized by
the geometry of exceptional singularities in string theory. For example,
real codimension four gauge theories defined by type IIA in six dimensions or M-theory in seven dimensions
can provide a geometric understanding of exceptional gauge groups when the
internal compactified directions are local singularities of the type:
\begin{equation}%
\mathbb{C}
^{2}/\Gamma
\end{equation}
where $\Gamma$ is one of the three exceptional subgroups of $SU(2)$, leading
to $E_{6,7,8}$ gauge symmetries. \ Similarly, compactifications of F-theory on such a
singularity descend to eight-dimensional gauge theories. \ This eight-dimensional theory is interpreted as
the worldvolume of a non-perturbative seven-brane in a type IIB\ compactification.\footnote{
\ In fact, the Ho\v{r}ava-Witten $E_{8}$ wall can also be related to such
singularities upon further compactification \cite{CachazoVafa}.} \ As opposed
to the $E_{8}$ wall of Ho\v{r}ava-Witten theory, in the geometric approach there
are a number of analytic tools available for describing the unfolding of
geometric singularities.

An $\mathcal{N}=1$ supersymmetric GUT theory with exceptional gauge group
derived from a geometric singularity can arise either from M-theory
compactified on a seven-dimensional manifold of $G_{2}$ holonomy
\cite{AcharyaWitten} or from F-theory compactified on an elliptic Calabi-Yau
fourfold \cite{VafaFTHEORY}. \ In the $G_{2}$ case, the absence of holomorphic
structure limits the analytic control over detailed properties of the
geometry. \ By contrast, in the Calabi-Yau fourfold case, the
complex/K\"{a}hler geometry allows a more powerful array of techniques. \ For
this reason, compactifications of F-theory on Calabi-Yau fourfolds preserving $\mathcal{N}=1$
supersymmetry in four dimensions have been extensively studied. \ A foundational example
of this work is \cite{BershadskyFOURD}. \ To the best of our knowledge, however,
no systematic study of the connection between F-theory and GUTs has been
undertaken (see, however, \cite{AC} for some progress in this direction).
Some work in this direction has been done in connection
with models with a heterotic dual. Even in these examples, though, a direct
analysis from the vantagepoint of F-theory has not been given yet.
However, as we explain below, we do not wish to assume that a given model has
a well-defined heterotic dual. Our aim in this work and the followup paper
\cite{BHV} is to take a first step in filling this gap. We note that independent work
on extracting the chiral matter content directly from F-theory and matching
these results to heterotic duals has recently appeared in \cite{DonagiWijnholt}.

At a foundational level, there is another reason to study models constructed from $E$-type singularities
in F-theory in their own right, as opposed to appealing to a potential heterotic
dual.  Recall that F-theory compactifications with a heterotic dual derive from
the basic duality between compactifications of F-theory on an elliptic $K3$ and its
heterotic dual on a $T^{2}$.  Extending this duality fiberwise over a
complex surface $S$, we achieve a duality between heterotic strings compactified
on Calabi-Yau threefolds elliptically fibered over $S$ and F-theory compactified
on a Calabi-Yau fourfold given by an elliptic $K3$-fibration over $S$.  Geometrically,
the condition that an elliptic fibration of $S$ yields a Calabi-Yau threefold
requires that $S$ be of Fano type. Said differently, there exist a large class
of heterotic and F-theory compactifications which may not possess a dual description.
In order to maintain maximal flexibility for future model building applications, we
shall therefore not limit our considerations to models with a well-defined heterotic
dual.\footnote{In particular, in Section \ref{CONCLUDE} we speculate on one possible model building application when $S$ is not a Fano variety.}

As mentioned previously, gauge groups in F-theory arise from codimension one
singularities in the base which are in turn identified with the worldvolume of
some seven-branes. \ In order to maintain a finite gauge coupling constant in
the four-dimensional effective theory on $\mathbb{R}^{3,1}$, we assume that the seven-brane wraps a compact complex
surface $S$ of real dimension four. \ Turning on a supersymmetric gauge field
configuration on $S$ in some subgroup $H_{S}\subset G_{S}$ breaks the gauge group $G_{S}$
to the commutant of $H_{S}$ in $G_{S}$. \ This provides an economical way to break the
exceptional gauge group to gauge symmetries closer to the MSSM. \ Moreover,
when this supersymmetric gauge field configuration has a non-trivial overall
$U(1)$ factor, the resulting spectrum can contain chiral matter originating from
the zero modes of fields propagating in the bulk of $S$.

Along a codimension two subspace of the threefold base defined by a Riemann
surface $\Sigma\subset S$, the rank of the singularity type can increase.
As discussed in \cite{BershadskyKachruSadov,KatzVafa}, this leads to matter
living on $\Sigma$. The Riemann surface $\Sigma$ can be viewed as the
intersection locus of two singularities fibered over two different complex
surfaces $S$ and $S^{\prime}$. From the perspective of the threefold base,
$S^{\prime}$ can be viewed as the locus of another seven-brane which may be
non-compact. The fields localized on $\Sigma$ are a direct extension of the
bi-fundamental fields obtained from the intersection of D-branes to the more general
case of colliding singularities.  A non-trivial background gauge field configuration on
$\Sigma$ can also induce a four-dimensional chiral matter spectrum.

The fields of the MSSM interact via cubic couplings. \ In the present class of models,
Yukawa couplings among chiral matter come about in three different ways.  The first type
correspond to couplings between three bulk fields on $S$.  When $S$ is a Hirzebruch or
del Pezzo surface, these couplings all vanish. The second type of coupling comes from
the interaction between two fields living on $\Sigma$ and one living on $S$ ($\Sigma \Sigma S$).
The last type of Yukawa coupling occurs when the zero mode wave functions for fields localized on three such
$\Sigma$'s intersect at a point ($\Sigma \Sigma \Sigma$).  This type of coupling
turns out to be generic in the case of exceptional singularities.\footnote{In some very special situations studied in
\cite{MorrisonFlorea}, a further enhancement in the singularity may not
correspond to an $ADE$ type singularity. \ We discuss some of the physics
of this case in Appendix \ref{EXOTICA}.}

It turns out that there is a beautiful interplay between the topological field
theories which dictate the matter content coming from $S$ and $\Sigma$.
\ The relevant degrees of freedom living on $S$ are captured by an
$\mathcal{N}=4$ topologically twisted theory of the type studied in
\cite{VafaWitten}. \ Furthermore, the matter fields coming from $\Sigma$ are
well-described by another topological field theory defined on $\Sigma$ which
naturally couples to the bulk theory on $S$ as a defect theory. \ The coupling
of these two topological theories leads to sources for some of the fields in
the bulk theory on $S$ when the fields on $\Sigma$ develop a vev. \ This turns
out to nicely correspond to the unfolding of the singularity in F-theory.

The primary aim of the present paper is to flesh out these ideas and to
set the groundwork for potential applications.  Although we shall present some examples of
semi-realistic models, more phenomenologically viable constructions of GUTs from F-theory will
appear in the followup paper \cite{BHV}.

The rest of this paper is organized as follows. \ In Section \ref{GENOVERVIEW}
we describe the basic setup. \ In Section \ref{SEVENBRANE} we describe the
simplest F-theory geometry with exceptional singularities localized on an
isolated seven-brane. \ Furthermore, we introduce the eight-dimensional partially twisted topological
theory corresponding to the worldvolume of a general seven-brane
and show that the geometric unfolding of the singularity
exactly matches to degrees of freedom in the gauge theory. We
also consider the possibility of turning on supersymmetric background gauge
fields on the seven-brane and analyze the resulting (chiral) matter and existence
(or absence) of Yukawa couplings among them.  In Section
\ref{COUPLINGDEFECT} we introduce the partially twisted theory describing a
more general class of compactifications in which seven-branes intersect along Riemann surfaces in $S$.
In this same section, we also discuss how four-dimensional chiral matter
can arise on such intersections and compute the Yukawa couplings between pairs
of chiral matter fields living on a Riemann surface and bulk gauge fields in $S$.
Even in this more general class of models, we observe a harmonious match between
gauge theory and geometric degrees of freedom. In Section \ref{MULTINT} we study
subloci of real codimension four in $S$ along which the singularity type enhances to even higher rank.
\ We find that for certain geometries, such points signal the presence of additional
Yukawa couplings among chiral matter fields with wave functions localized on Riemann surfaces.
\ In Section \ref{FINALTOY} we present a toy model which incorporates many of the ingredients
developed in previous Sections. Section \ref{CONCLUDE} presents our conclusions. Additional
background and more technical material is included in the Appendices.

\section{General Overview\label{GENOVERVIEW}}

In this section we present an overview of the types of models we shall treat.
\ Our setup is locally given by the worldvolume of a seven-brane of generalized
$ADE$-type in a compactification on a Calabi-Yau fourfold.
\ Letting $S$ denote a K\"{a}hler manifold of complex dimension two which is
wrapped by the seven-brane, the resulting local model will reduce to an
$\mathcal{N}=1$\ supersymmetric theory in four dimensions. \ In order to work
in the limit in which gauge dynamics decouple from gravity, we sometimes
restrict our attention to geometries where $S$ can shrink to zero size inside
a general threefold base. This is also natural from the viewpoint of the
strongly coupled limit of the heterotic $E_{8}\times E_{8}$ string in that the
bound of (\ref{pertbound}) can be evaded. \ In addition, in this
limit there are no additional massless scalar fields corresponding to
the motion of $S$ inside the threefold base of the F-theory
compactification. \ For all of these reasons, we shall sometimes require that
the anti-canonical bundle of the surface $S$ is ample so that such a
contraction is possible. \ This amounts to the condition that $S$ is a del
Pezzo surface.

As will be discussed in greater detail in later Sections,
there are in general two ways in which chiral matter can arise from such a
theory. \ The first corresponds to turning on a gauge field configuration on
$S$ with non-trivial first Chern class. \ Given a field which transforms in a
representation of the structure group for the holomorphic gauge bundle
on $S$, the number of net generations will be given by an index computation.

\begin{figure}
[ptb]
\begin{center}
\includegraphics[
height=1.785in,
width=1.5091in
]%
{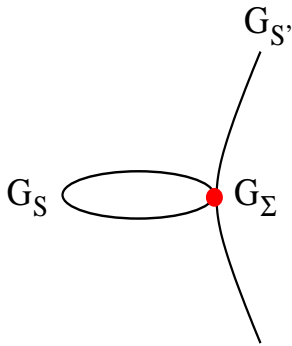}%
\caption{Depiction of intersecting seven-branes wrapping a compact surface $S$
with gauge group $G_{S}$ and a non-compact surface $S^{\prime}$ with
non-dynamical gauge group $G_{S^{\prime}}$. \ In the threefold base of the
F-theory compactification, the intersection locus of $S$ and $S^{\prime}$ is a
Riemann surface where the singularity type enhances further to $G_{\Sigma
}\supset G_{S}\times G_{S^{\prime}}$.}%
\label{INTSEVENBRANES}%
\end{center}
\end{figure}

Another source of matter originates from six-dimensional chiral fields localized along Riemann surfaces in $S$.  In many cases
this can be interpreted as the intersection locus of distinct seven-branes.
\ This chirality can be preserved in four dimensions when a suitable background gauge field
configuration has been turned on along the Riemann surface. \ To see how
this comes about, recall that an F-theory compactification will generically
contain three types of singularities. \ Over the locus given by the
holomorphic surface $S$, the corresponding singularity of $ADE$ type
will give rise to a gauge group which we denote by $G_{S}$. \ This singularity
may collide with another singularity with gauge group $G_{S^{\prime}}$
supported on a non-compact complex surface $S^{\prime}$. \ See figure
\ref{INTSEVENBRANES} for a depiction of the intersecting seven-brane locus in the
case where $S$ is compact and $S^{\prime}$ is non-compact. \ Over the Riemann
surface $\Sigma$ defined by the intersection of $S$ and $S^{\prime}$, the
singularity type enhances to $G_{\Sigma}$ such that:%
\begin{equation}
G_{\Sigma}\supset G_{S}\times G_{S^{\prime}}\text{.}%
\end{equation}
For the case of $SU(N+M)\supset SU(N)\times SU(M)$, this corresponds to
intersecting D7-branes and the matter localized at the intersection transforms
in the $(N,\overline{M})$. \ More generally, the six-dimensional chiral matter localized on
$\Sigma$ transforms under the analogous projection of the adjoint of
$G_{\Sigma}$ to $G_{S}$ as in \cite{KatzVafa} for breaking $G_{\Sigma}$ to a
subgroup with some number of $U(1)$ factors. \ When the bulk gauge field on
$\Sigma$ has non-trivial first Chern class, the resulting spectrum in four dimensions can be
chiral. \ The net number of generations localized on $\Sigma$ is given by an
index computation on the Riemann surface for fields charged under the given
combination of background $U(1)$ gauge fields.%

We now explain the origin of Yukawa couplings in this setup. There are a
priori three ways in which non-trivial interaction terms can arise. \ The
first possibility corresponds to interactions between three bulk
fields of the eight-dimensional theory. \ As we will argue in Section 3, such interaction
terms are identically zero when $S$ is either a Hirzebruch or del Pezzo
surface. \ The second type of interaction term originates from
the coupling between a single bulk field on $S$ with two matter fields localized
along a Riemann surface $\Sigma$ in $S$. \ The final type of interaction
originates from the triple overlap at a single point in $S$ of matter fields
localized along three matter curves. \ We find that in many cases of interest,
such triple overlaps are a generic feature of F-theory compactifications.

One may also wonder if there are instanton corrections to the Yukawa couplings.
In F-theory the natural source for such instantons are wrapped
Euclidean D3-branes \cite{witsup}, which in the present case means
wrapping D3-branes on $S$.  Note, however, that this is nothing but the usual
contribution from gauge theory instantons. In the regime of interest for GUT models
where $1/g_{YM}^2 >>1$, such contributions lead to very small corrections on the order of
${\rm exp}(-a/g_{YM}^2)$ for some $O(1)$ constants $a$, so for the purposes of this paper we
shall ignore these effects.

\section{Partially Twisted Gauge Theory on a Seven-Brane}\label{SEVENBRANE}

As a warmup for the rest of the paper, in this Section we
determine the four-dimensional effective theory for a class of
particularly simple F-theory compactifications which preserve only
${\CN=1}$ supersymmetry in four dimensions.  Specifically, in Section
\ref{LOCALMODEL} we consider local F-theory geometries which describe a
small neighborhood of a seven-brane with worldvolume ${\BR^{3,1} \times S}$,
where $S$ is a compact K\"ahler surface and the worldvolume gauge
group $G_{S}$ is of arbitrary $ADE$-type.

In Section \ref{TWISTONS} we argue that the dynamics
of the low-energy degrees of freedom in F-theory are captured by a
partially twisted version of the maximally supersymmetric Yang-Mills
theory on ${\BR^{3,1} \times S}$.  We also present
evidence that the partially twisted Yang-Mills theory
correctly describes the moduli of the given background in F-theory.

In Section \ref{REDUCTION} we determine some basic properties of the four-dimensional effective theory for this
special class of F-theory models. In particular, we present the BPS equations of motions for the
eight-dimensional fields in the partially twisted Yang-Mills theory and subsequently study supersymmetric vacua in the
presence of a non-trivial background gauge field configuration on $S$.
We find that while many vacua support a chiral matter spectrum in four dimensions,
when $S$ is a Hirzebruch or del Pezzo surface, all Yukawa
couplings identically vanish.

\subsection{A Local Model for $X$}\label{LOCALMODEL}

To set our notation, we consider F-theory
\cite{VafaFTHEORY, MorrisonVafaI, MorrisonVafaII}
on a background of the form  ${\BR^{3,1} \times X}$, where $X$ is a
Calabi-Yau fourfold.  By assumption, the fourfold $X$ fibers
elliptically with a section over a complex threefold $B$,
\begin{equation}\label{FIBX}
\begin{matrix}
&\BE\,\longrightarrow\,X\cr
&\mskip 65mu\Big\downarrow\lower 0.5ex\hbox{$^\pi$}\cr
&\mskip 55mu B\cr
\end{matrix}\,,
\end{equation}
and in general, the elliptic fiber $\BE$ of $X$ degenerates over a
locus $\Delta$ of complex codimension one in $B$,
\begin{equation}
\Delta \,\subset\, B\,.
\end{equation}

Physically, $\Delta$ encodes the location of various seven-branes in
$B$, and the nature of the singularities in $\BE$ over $\Delta$
determines the worldvolume gauge group on each seven-brane.  From the
perspective of the present paper, what is extremely interesting about
F-theory is that with the appropriate singularity in $\BE$,
seven-branes wrapping $\Delta$ can carry a worldvolume gauge
group of arbitrary\footnote{In the presence of monodromies, even
  non-simply-laced gauge groups are possible, though we will not
  exploit this possibility in the present paper.}
$ADE$-type.  While the gauge groups $SU(n+1)$ or $SO(2n)$
can be realized using D7-branes and (in the latter case) orientifold
planes, a seven-brane with worldvolume gauge group $E_{6,7,8}$
is a rather mysterious object that cannot be described
perturbatively in string theory.  We shall refer to such
seven-branes as ``exceptional''.

In addition to seven-branes, a general F-theory background may also contain
spacetime filling D3-branes which sit at points in $B$, as well as
supersymmetric fluxes associated to either the bulk supergravity
fields on $B$ or to the worldvolume gauge fields on $\Delta$.
However, for reasons to become clear momentarily, only the worldvolume
fluxes on $\Delta$ will play a role in the F-theory models we study.

F-theory on $X$ typically represents a strongly-coupled background of
string theory, and to make progress in even our low-energy analysis,
we make two simplifying assumptions.  First, we focus
attention throughout on local, non-compact models which describe only
a small neighborhood of $\Delta$ inside $B$, with neither D3-branes
nor bulk fluxes present.\footnote{The D3-branes on $\Delta$ are included in
our construction as they would correspond to internal point-like instantons for gauge fields on $\Delta$,
which is a special case of what
we will study in this paper.}  With this assumption, gravity decouples in four
dimensions, and modulo exotic possibilities such as that reviewed in
Appendix \ref{EXOTICA}, we expect to obtain an ${\CN=1}$
supersymmetric gauge theory on $\BR^{3,1}$ which captures the
effective dynamics of the light worldvolume degrees-of-freedom living
on $\Delta$.  This gauge theory is the analogue for seven-branes in
F-theory of the well-known quiver gauge theories which describe
D-branes at singularities in perturbative compactifications of type
IIB string theory on Calabi-Yau threefolds.

Even in local models for $B$, the geometry of the seven-brane
configuration represented by $\Delta$ can be quite complicated.  For
instance, $\Delta$ might be reducible and hence appear as a union of
several components associated to loci of colliding singularities in
the elliptic fibration over $B$.  As our second simplifying
assumption, we suppose that $\Delta$ consists only of an irreducible,
smooth, compact, complex surface $S$ embedded in $B$.  So in this
Section, we study the worldvolume theory on a single seven-brane which
wraps ${\BR^{3,1} \times S}$ in F-theory on $X$.  Later in Sections
\ref{COUPLINGDEFECT} and \ref{MULTINT}, we extend our analysis to the
case that $\Delta$ is reducible and multiple seven-branes intersect in $B$.

Before we proceed to a fairly detailed analysis of the worldvolume
gauge theory living on a seven-brane of arbitrary $ADE$-type wrapping
${\BR^{3,1} \times S}$, let us first present a local model for the
Calabi-Yau fourfold $X$ which describes F-theory in the background of
such a brane.  Since we are already working with a local model for
$B$, we will also work with a local model for $X$.  We thus take $X$
to be a local elliptic $K3$-fibration over $S$ of the form
\begin{equation}\label{FIBY}
\begin{matrix}
&Y \,\longrightarrow\,X\cr
&\mskip 65mu\Big\downarrow\lower 0.5ex\hbox{$^\pi$}\cr
&\mskip 55mu S\cr
\end{matrix}\,,
\end{equation}
where we model the local elliptically-fibered $K3$-surface $Y$ on a
hypersurface in $\BC^3$ with an isolated $ADE$ singularity at the origin.

Not suprisingly, the $ADE$ singularities play a prominent role
throughout the paper.  In terms of coordinates $(x,y,z)$ on $\BC^3$,
we recall that the $ADE$ singularities can be presented canonically as
below.
\begin{equation}\label{ADES}
\begin{tabular}{|l|l|}
\hline
$A_n$ & ${y^2 \,=\, x^2 \,+\, z^{n+1}}$ \\
$D_n$ & ${y^2 \,=\, x^2 z \,+\, z^{n-1}}$ \\
$E_6$ & ${y^2 \,=\, x^3 \,+\, z^4}$ \\
$E_7$ & ${y^2 \,=\, x^3 \,+\, x z^3}$ \\
$E_8$ & ${y^2 \,=\, x^3 \,+\, z^5}$ \\
\hline
\end{tabular}
\end{equation}
In the following work we shall sometimes rescale these coordinates by overall numerical coefficients in order to
more easily compare our results with the unfolding of singularities in \cite{KatzMorrison}.

If $X$ is to be Calabi-Yau, then the hypersurface $Y$ must fiber
appropriately over $S$.  To describe the fibering,  we promote the
local coordinates $(x,y,z)$ in \eqref{ADES} to transform as sections
of a rank three bundle $V$ over $S$ which is given as a sum of tensor
powers of the canonical bundle $K_S$,
\begin{equation}\label{RKTHV}
V \,=\, K_S^{a} \oplus K_S^{b} \oplus K_S^{c}\,.
\end{equation}
Here $(a,b,c)$ are three integers associated to $(x,y,z)$ which we
must determine.

The integers $(a,b,c)$ necessarily satisfy two conditions for the
simple local model of $X$ to exist.  First, in order that the defining
equations for $Y$ in \eqref{ADES} make sense, the individual terms in
each equation must transform as sections of the same line bundle over
$S$.  This yields the homogeneity conditions:
\begin{equation}\label{ADESII}
\begin{tabular}{|l|l|}
\hline
$A_n$ & ${2 b \,=\, 2 a \,=\, (n+1) c}$ \\
$D_n$ & ${2 b \,=\, 2 a + c \,=\, (n-1) c}$ \\
$E_6$ & ${2 b \,=\, 3 a \,=\, 4 c}$ \\
$E_7$ & ${2 b \,=\, 3 a \,=\, a + 3 c}$ \\
$E_8$ & ${2 b \,=\, 3 a \,=\, 5 c}$ \\
\hline
\end{tabular}
\end{equation}
Second, the Calabi-Yau condition on $X$ implies that the holomorphic
two-form ${(1/y) \, dx \^ dz}$ on $Y$ transforms over $S$ as a section
of $K_S$.  Thus $(a,b,c)$ also satisfy
\begin{equation}\label{HOLX}
a \,-\, b \,+\, c \,=\, 1\,.
\end{equation}
Together, the equations in \eqref{ADESII} and \eqref{HOLX} admit the
following unique solutions.
\begin{equation}\label{ADESIII}
\begin{tabular}{|l|l|l|l|}
\hline
$A_n$ & ${a\,=\,{{n+1}\over 2}}$ & ${b\,=\,{{n+1}\over 2}}$ &
${c\,=\,1}$ \\
$D_n$ & ${a\,=\,n-2}$ & ${b\,=\,n-1}$ & ${c\,=\,2}$ \\
$E_6$ & ${a\,=\,4}$ & ${b\,=\,6}$ & ${c\,=\,3}$ \\
$E_7$ & ${a\,=\,6}$ & ${b\,=\,9}$ & ${c\,=\,4}$ \\
$E_8$ & ${a\,=\,10}$ & ${b\,=\,15}$ & ${c\,=\,6}$ \\
\hline
\end{tabular}
\end{equation}
For the cases other than $A_n$ with $n$ even, $(a,b,c)$ in \eqref{ADESIII}
are integers, so the local model for $X$ which describes a
seven-brane of $ADE$-type wrapping $S$ clearly exists.  To treat the
case of $A_n$ with $n$ even, for which we see that $a$ and $b$ in
\eqref{ADESIII} are only half-integral, we make an elementary change of
variables to realize the $A_n$ singularity as ${x y = z^{n+1}}$.
Homogeneity of the latter equation implies that ${a + b = (n+1)\,c}$,
and the Calabi-Yau condition on $X$ now fixes ${c=1}$.  Hence in this
form, a local model for $X$ with $A_n$ singularity over $S$ exists for
arbitrary integers $a$ and $b$ which satisfy ${a + b = n+1}$.

Now, one interesting fact about the $E$-type singularities in
\eqref{ADES} is that these singularities appear automatically in the
Weierstrass form:
\begin{equation}\label{WEIER}
y^2 \,=\, x^3 \,+\, f(z) \, x \,+\, g(z)\,.
\end{equation}
Here $x$ and $y$ can be interpreted as affine coordinates
parameterizing the elliptic fiber $\BE$ of $X$, and $f(z)$ and $g(z)$ are
determined as simple monomials in $z$ by \eqref{ADES}.  In particular, since
$x$ and $y$ parameterize $\BE$, the coordinate $z$ must parameterize the
normal direction to $S$ inside the non-compact threefold $B$.  Thus
for $E_{6,7,8}$, the base $B$ is given by the total spaces of the
respective line bundles $K_S^{3}$, $K_S^{4}$, and
$K_S^{6}$ over  $S$.  The particular numerology that occurs here
will later be quite significant for the worldvolume description of
the exceptional seven-brane.

For $A$- and $D$-type singularities, the coordinate $z$ is
similarly distinguished as parameterizing the normal direction to $S$
inside $B$.  This assertion can be checked directly by analyzing
global Weierstrass models for the elliptic singularities.
Alternatively, in the $A_n$ case, one can simply note that according to
\eqref{ADESIII}, $z$ transforms as a section of the canonical bundle
$K_S$.  Hence, if $z$ parameterizes the normal direction to $S$
inside $B$, the threefold $B$ is itself Calabi-Yau, as one expects for
seven-branes which carry a perturbative gauge group of type
$A_n$.

Additionally, in the case of a $D_n$ singularity, $B$ is the total
space of $K_S^{2}$ over $S$.  The total space of
$K_S^{2}$ can be considered as the quotient of the local
Calabi-Yau threefold associated to $K_S$ by a $\BZ_2$ involution of
the fiber.  This involution fixes $S$ as the zero-section of $K_S$ and
acts freely elsewhere, so we naturally obtain a local orientifold
geometry, as expected for seven-branes which carry a gauge
group of type $D_n$.

\subsection{Twisting on $S$}\label{TWISTONS}

To determine the effective worldvolume description of the seven-brane
wrapping ${\BR^{3,1} \times S}$, let us begin with the trivial case
that ${S\,=\,\BC^2}$.  In this case, F-theory reduced to ${\BR^{7,1} =
  \BR^{3,1} \times \BC^2}$ on the hypersurface $Y$ with an isolated
$ADE$ singularity is described at low energies by the maximally
supersymmetric Yang-Mills theory with gauge group $G_{S}$ of
corresponding $ADE$-type.\footnote{Note that because $Y$ is non-compact, gravity
decouples in eight dimensions.}

If we now regard $\BC^2$ as a local patch of $S$, the standard
adiabatic argument suggests that $F$-theory on $X$ is still described
at low energies by eight-dimensional Yang-Mills theory on
${\BR^{3,1}\times S}$.  However, in order to preserve ${\CN=1}$
supersymmetry, this Yang-Mills theory must be topologically twisted on
$S$.  By the end of the Section, we will present some very suggestive
evidence that the partially twisted Yang-Mills theory on
${\BR^{3,1}\times S}$ does describe the light degrees of freedom of
F-theory on $X$.

The idea of studying F-theory on $X$ by means of the topological gauge
theory living on the worldvolume of the seven-brane wrapping
${\BR^{3,1} \times S}$ is not new.  This idea, including certain
elaborations we consider later, was sketched over ten years ago in one
of the foundational papers \cite{BershadskyFOURD} on ${\CN=1}$ F-theory
compactifications.  However, the details behind some of the ideas in
\cite{BershadskyFOURD} seem never to have been fleshed out, and doing
so is one of our goals here.

Actually, to refer to the twisted Yang-Mills theory on ${\BR^{3,1}
  \times S}$ as ``topological'' is a bit of a misnomer.  In order to
twist the eight-dimensional Yang-Mills theory on $S$, we will have to
use the fact that $S$ is not a generic Riemannian four-manifold but rather,
through the embedding of $S$ into $B$ and hence $X$,
carries an induced K\"ahler structure.  As we explain
later, the observables of primary interest which determine the
effective superpotential in four dimensions will be insensitive to the
particular K\"ahler metric on $S$.  On the other hand, these observables
will certainly depend on the complex structure of $S$.

For supersymmetric gauge theories in dimensions three and four, many
qualitatively distinct possiblities for twisting exist.  However,
the possibilities for twisting supersymmetric gauge theories in higher
dimensions are much more restricted.  Given that $S$ is K\"ahler, the
maximally supersymmetric Yang-Mills theory in eight dimensions admits
a unique twist on ${\BR^{3,1}\times S}$ which preserves ${\CN=1}$
supersymmetry in four dimensions.  As a result, once we argue that
F-theory on $X$ is described by a twisted Yang-Mills theory on
${\BR^{3,1} \times S}$, we have no choice about which twist to
consider.

The procedure of twisting the maximally supersymmetric Yang-Mills
theory on ${\BR^{3,1} \times S}$ is entirely standard, but since this
gauge theory and its later elaborations provide our basic tool for
studying F-theory on $X$, we now discuss the twist in some detail.

We start with the maximally supersymmetric Yang-Mills theory in ten
dimensions on $\BR^{9,1}$.  In ten dimensions, the super Yang-Mills
multiplet consists of a gauge field and an adjoint-valued fermion
which transforms under $SO(9,1)$ in the positive-chirality
spinor representation ${\bf 16}_{+}$.  This theory preserves sixteen
supersymmetries, which transform in the representation ${\bf 16}_{+}$.

Under reduction to $\BR^{7,1}$, the Yang-Mills multiplet decomposes
into an eight-dimensional gauge field which we denote by $A$, two
real scalar fields $\Phi_8$ and $\Phi_9$, and two fermions $\Psi_\pm$.
Each of the scalars and fermions transforms in the adjoint
representation of the gauge group.  For later use, we now introduce the
following complex-linear combinations of $\Phi_8$ and $\Phi_9$,
\begin{equation}\label{VPHI}
\varphi \,=\, \Phi_8 \,+\, i \, \Phi_9\,,\qquad\qquad
\bar\varphi \,=\, \Phi_8 \,-\, i \, \Phi_9\,.
\end{equation}

The eight-dimensional Yang-Mills theory preserves ${SO(7,1) \times
U(1)_R}$ as a global symmetry, under which the fermions $\Psi_\pm$
transform as respective summands in the reducible representation
\begin{equation}\label{EDS}
{\bf 16}_+ \,\longmapsto\, \left({\bf S}_+, +\ha\right)
\oplus \left({\bf S}_-, -\ha\right)\,.
\end{equation}
Here ${\bf S}_{\pm}$ denote the positive and negative chirality
spinor representations of $SO(7,1)$.  Also, the complex scalar fields
$\varphi$ and $\bar\varphi$ transform trivially under $SO(7,1)$ and
with charges $\mp 1$ under $U(1)_R$.

To twist the gauge theory on ${\BR^{7,1} = \BR^{3,1} \times \BC^2}$,
we further reduce the global symmetry group from ${SO(7,1) \times
U(1)_R}$ to
\begin{equation}\label{GLBS}
SO(3,1) \times SO(4) \times U(1)_R\,.
\end{equation}
Under \eqref{GLBS}, both the fermions $\Psi_\pm$ and the corresponding
supersymmetry generators $\epsilon_\pm$ transform as
\begin{align}\label{EDSII}
\left({\bf S}_+,+\ha\right)\,&\longmapsto\, \left[({\bf 2},{\bf 1}),
  ({\bf 2},{\bf 1}), +\ha\right] \oplus \left[({\bf 1},{\bf 2}),
  ({\bf 1},{\bf 2}), +\ha\right]\,,\cr
\left({\bf S}_-,-\ha\right)\,&\longmapsto\, \left[({\bf 2},{\bf 1}),
  ({\bf 1},{\bf 2}), -\ha\right] \oplus \left[({\bf 1},{\bf 2}),
  ({\bf 2},{\bf 1}), -\ha\right]\,.
\end{align}
Here we make use of the standard local isomorphism ${SO(4) \cong SU(2)
  \times SU(2)}$, and similarly for $SO(3,1)$, to describe the
representation content in \eqref{EDSII}.  Thus, $({\bf 2},{\bf
1})$ describes the left handed chiral spinor of either $SO(4)$ or
$SO(3,1)$, and $({\bf 1},{\bf 2})$ describes the right handed
anti-chiral spinor.

If $S$ were an arbitrary Riemannian four-manifold, we would now specify
the twist on $S$ by an embedding of the global $U(1)_R$ symmetry into
$SO(4)$, identified with the structure group of the tangent bundle
of $S$.  However, at this point we make use of the fact that
$S$ is K\"ahler so that the structure group of the tangent bundle of
$S$ actually reduces from $SO(4)$ to $U(2)$.  Hence the possible
twists on $S$ are specified by embeddings of $U(1)_R$ into the smaller
group $U(2)$.

Up to isomorphism, a unique topological twist is possible, under which
$U(1)_R$ is embedded into the central $U(1)$ subgroup of $U(2)$.  To
present the twist, we let $R$ be the generator of $U(1)_R$.
Similarly, we let $J$ be the generator of the central $U(1)$ subgroup
in $U(2)$.  We normalize $J$ so that under the reduction from
$SO(4)$ to $U(2)$, the chiral and anti-chiral spinor representations
of $SO(4)$ decompose as
\begin{equation}\label{EDSIII}
({\bf 2},{\bf 1}) \,\rightarrow\, {\bf 2}_0\,,
  \qquad\qquad ({\bf 1},{\bf 2}) \,\rightarrow\, {\bf 1}_{+1} \oplus
  {\bf 1}_{-1}\,,
\end{equation}
where the subscripts in \eqref{EDSIII} denote the $U(1)$ charges under
the central generator $J$.

Comparing \eqref{EDSIII} to \eqref{EDSII}, we see that in order to
obtain four scalar supercharges on $S$ which descend to the standard
${\CN=1}$ supersymmetry generators $Q_\alpha$ and $\bar Q{}_{\dot\alpha}$ on
$\BR^{3,1}$, the new central $U(1)$ generator $J_{top}$ in the twisted
$U(2)$ must be given by one of the following linear combinations of
the original generators $J$ and $R$,
\begin{equation}\label{TWTJ}
J_{top} \,=\, J \,\pm\, 2 R\,.
\end{equation}
As one can check, either sign in \eqref{TWTJ} leads to an isomorphic
twist, so we take  ${J_{top} = J + 2 R}$ without loss of generality.  This choice
leads to somewhat more natural conventions regarding holomorphy in the
twisted Yang-Mills theory on $S$.

In the twisted theory, the fermions $\Psi_\pm$ and supersymmetry
generators $\epsilon_\pm$ now transform under ${SO(3,1) \times U(2)}$ as
\begin{align}\label{EDSIV}
&\big[({\bf 2},{\bf 1}) \otimes {\bf 2}_{+1}\big] \,\oplus\,
\big[({\bf 1},{\bf 2}) \otimes ({\bf 1}_{+2} \oplus {\bf 1}_{0})\big]\,,\cr
&\big[({\bf 1},{\bf 2}) \otimes {\bf 2}_{-1}\big] \,\oplus\,
\big[({\bf 2},{\bf 1}) \otimes ({\bf 1}_{0} \oplus {\bf
1}_{-2})\big]\,.
\end{align}
Here the subscripts in \eqref{EDSIV} denote the charges under the
central $U(1)$ generator $J_{top}$ in $U(2)$.  In particular, from the
representations ${({\bf 1},{\bf 2})\otimes{\bf 1}_0}$ and ${({\bf
    2},{\bf 1})\otimes{\bf 1}_0}$ appearing in \eqref{EDSIV}, we see
that the twisted theory on $S$ possesses four scalar supercharges
$(Q_\alpha\,, \bar Q{}_{\dot\alpha})$, associated to the fact that
F-theory on $X$ preserves ${\CN=1}$ supersymmetry in four dimensions.

By the same token, the complex scalar fields $\varphi$ and
$\bar\varphi$ now transform under ${SO(3,1) \times U(2)}$ as:
\begin{equation}\label{EDPH}
({\bf 1},{\bf 1}) \otimes {\bf 1}_{\mp 2}\,.
\end{equation}

To interpret the twisted fermions in \eqref{EDSIV} geometrically, we fix
conventions under which the central $U(1)$ in $U(2)$ acts on vectors
in the holomorphic tangent bundle $TS$ with charge $+1$ and dually on
covectors in the holomorphic cotangent bundle $\Omega^1_S$ with charge
$-1$.  The twisted fermions then appear as forms of
holomorphic/anti-holomorphic type $(p,0)$ and $(0,p)$ on $S$ for
${p=0,1,2}$.  Corresponding to \eqref{EDSIV}, we denote the twisted
fermions by:
\begin{equation}\label{TWF}
\begin{tabular}{l l}
$\bar\eta{}_{\dot\alpha}$ & \qquad section of $\ad(P)$\,, \\
${\psi_{\alpha} \,=\, \psi_{\alpha\,\bar m}\,d\bar s{}^{\bar m}}$ &
\qquad section of $\bar\Omega{}^1_S \otimes \ad(P)$\,, \\
${\bar\chi{}_{\dot\alpha} \,=\, \bar\chi{}_{\dot\alpha\, \bar m \bar n}
\, d\bar s{}^{\bar m} \^ d\bar s{}^{\bar n}}$ & \qquad section of
$\bar\Omega{}^2_S \otimes \ad(P)$\,,
\end{tabular}
\end{equation}
and
\begin{equation}\label{TWFC}
\begin{tabular}{l l}
$\eta_{\alpha}$ & \qquad section of $\ad(P)$\,, \\
${\bar\psi{}_{\dot\alpha} \,=\, \bar\psi{}_{\dot\alpha\,m}\,ds^m}$ &
\qquad section of $\Omega{}^1_S \otimes \ad(P)$\,, \\
${\chi_{\alpha} \,=\, \chi_{\alpha\, m n}\,ds^m \^ ds^n}$ & \qquad
section of $\Omega{}^2_S \otimes \ad(P)$\,.
\end{tabular}
\end{equation}
Here, $s^m$ and ${\bar s}{}^{\bar m}$ are local holomorphic and
anti-holomorphic coordinates on $S$ which we use to indicate the
transformations of the twisted fermions as differential forms on
$S$.  As standard, we also use $\alpha$ and $\dot\alpha$ for
${\alpha,\dot\alpha = 1,2}$ to indicate the transformations of the
twisted fermions as either chiral or anti-chiral spinors on
$\BR^{3,1}$.  Finally, $\ad(P)$ is the adjoint bundle associated to a
fixed principal $G_{S}$-bundle $P$ over $S$.  Here we anticipate the
possibility of turning on a background instanton on $S$ associated to
the choice of a topologically non-trivial $G_{S}$-bundle $P$.

From \eqref{EDPH}, we also see that the twisted complex scalar
$\varphi$ now transforms on $S$ as a section of ${\Omega^2_S \otimes
  \ad(P)}$,
\begin{equation}
\varphi \,=\, \varphi_{m n}\,ds^m \^ ds^n\,,
\end{equation}
 and the conjugate scalar $\bar\varphi$ transforms as a section of
 ${\bar\Omega{}^2_S \otimes \ad(P)}$,
\begin{equation}
\bar\varphi \,=\, \bar\varphi{}_{\bar m \bar n} \, d\bar s{}^{\bar m}
\^ d\bar s{}^{\bar n}\,.
\end{equation}
Equivalently, $\varphi$ transforms as a section of
${K_S\otimes\ad(P)}$, and $\bar\varphi$ transforms as a section of
${\bar K_S\otimes\ad(P)}$.

\bigskip\noindent{\it A Small Paradox, and Its Resolution}\smallskip

Before delving further into the details of the twisted Yang-Mills
theory on $S$, let us discuss the basic physical
interpretation of the twist.  As we will explain, the transformation
of $\varphi$ as a section of ${K_S \otimes \ad(P)}$ presents a small
paradox, whose resolution proves to illuminate a basic feature of
exceptional seven-branes in F-theory.

In the case that the seven-brane is a D7-brane in a
perturbative compactification of type IIB string theory, the
transformation of the worldvolume scalar $\varphi$ as a section of
${K_S \otimes \ad(P)}$ is very easy to understand.  In that situation,
$B$ is a Calabi-Yau threefold, and by the standard adjunction
formula, the canonical bundle $K_S$ of $S$ is isomorphic to the
holomorphic normal bundle $N_{S / B}$ of $S$ inside $B$.  Hence
the twist of $\varphi$ amounts to the geometric statement that
$\varphi$ describes normal motion of the D7-brane wrapping $S$ inside
$B$ \cite{BSV}.

More generally, if $B$ is not Calabi-Yau, the adjunction formula
provides an isomorphism
\begin{equation}\label{ADJCT}
K_S \,=\, K_B\big|_S \otimes N_{S / B}\,,
\end{equation}
where $K_B\big|_S$ denotes the restriction of the canonical bundle of
$B$ to $S$.  For a seven-brane of $D$- or $E$-type, the threefold
$B$ is not Calabi-Yau and $K_B$ is non-trivial, so we see from
\eqref{ADJCT} that the normal bundle $N_{S / B}$ is generally {\em not}
isomorphic to the canonical bundle $K_S$ in those cases.

On the other hand, the topological twist of Yang-Mills theory on
$S$ uniquely fixes $\varphi$ to transform as a section
of ${K_S \otimes \ad(P)}$ for a seven-brane of arbitrary $ADE$-type.  So
if our geometric intuition for a D7-brane in a perturbative Calabi-Yau
compactification is taken at face value, the fact that $\varphi$
generally transforms as a section of ${K_S \otimes \ad(P)}$ as opposed
to ${N_{S / B} \otimes \ad(P)}$ on the exceptional seven-brane presents a
small paradox.

The resolution of this paradox turns out to be very instructive and
provides immediate evidence that the topologically twisted Yang-Mills
theory on ${\BR^{3,1} \times S}$ serves as a valid worldvolume
description for exceptional seven-branes in F-theory.

To explain why $\varphi$ generally transforms as a section of ${K_S
\otimes \ad(P)}$ as opposed to ${N_{S / B} \otimes \ad(P)}$, we
recall that the moduli for the positions of seven-branes in $F$-theory
are encoded by complex structure moduli of the Calabi-Yau fourfold $X$
itself, since these moduli determine the location in $B$ of the
discriminant locus $\Delta$ on which the seven-branes wrap.  Thus to
understand how $\varphi$ describes the position of seven-branes in
$B$, we must really ask how $\varphi$ can be used to deform the
complex structure of $X$.  In our local model, $X$ is a fibration of
the $ADE$ hypersurface $Y$ over $S$, so we effectively want to consider
how an expectation value for $\varphi$ translates into a
deformation of the canonical equations for the $ADE$-singularities in
\eqref{ADES}.

As a simple example, let us consider the case that $Y$ describes an
$A_n$ singularity over $S$, for which the seven-brane is an ordinary
D7-brane.  The general deformation of the $A_n$ singularity in
\eqref{ADES} can be parameterized by ${n+1}$ complex variables
$(t_1,\ldots,t_{n+1})$ subject to the single constraint ${t_1 + \cdots
  + t_{n+1} = 0}$, in terms of which the deformation is given by
\begin{align}\label{DEFAN}
y^2 \,&=\, x^2 \,+\, \prod_{j=1}^{n+1}\,(z + t_j)\,\cr
&=\, x^2 \,+\, z^{n+1} \,+\, \sum_{k=2}^{n+1} \,
s_k(t_1\,,\ldots\,,t_{n+1}) \, z^{n+1-k}\,.
\end{align}
Here $s_k$ is the elementary symmetric polynomial which is homogeneous
of degree $k$ in the variables $t_1\,,\ldots\,,t_{n+1}$.

Not coincidentally, the parameters $(t_1,\ldots,t_{n+1})$ can also be
interpreted as coordinates on the Cartan subalgebra of the group ${G_{S}
  = SU(n+1)}$, in which form we identify them with the eigenvalues of
the adjoint-valued scalar field $\varphi$.  Equivalently, we identify
the degree $k$ symmetric polynomial $s_k$ with the degree $k$ Casimir
of $\varphi$,
\begin{align}\label{EIGPHI}
s_2(t_1\,,\ldots\,,t_{n+1}) \,&=\, -\ha \Tr(\varphi^2)\,,\cr
&\,\,\,\vdots\cr
s_{n+1}(t_1\,,\ldots\,,t_{n+1}) \,&=\, \det(\varphi)\,.
\end{align}

Now, if the deformation in \eqref{DEFAN} is to make sense when $z$
transforms as a section of $K_S$, the degree $k$ polynomial
$s_k(t_1,\cdots,t_{n+1})$ must transform as a section of $K_S^{k}$,
simply so that the deformed equation in \eqref{DEFAN} remains
homogeneous as an equation on sections of $K_S^{(n+1)}$.  Under the
identification in \eqref{EIGPHI}, this requirement is equivalent to
the condition that $\varphi$ transform as a section of ${K_S \otimes
  \ad(P)}$.  Thus, as we certainly expect, the automatic twisting of
$\varphi$ in the Yang-Mills theory on ${\BR^{3,1} \times S}$ is
consistent with the non-trivial fibration of $Y$ over $S$ which
defines $X$.

For the $D$- and $E$-type singularities, precisely the same geometric
reasoning implies that $\varphi$ still transforms as a section of
${K_S \otimes \ad(P)}$.  Just as for the $A_n$ singularities, we
recall \cite{KatzMorrison} that the deformations of the general $D$- or
$E$-type singularities are parameterized by coordinates $t_j$ for
${j=1,\ldots, r}$ on the Cartan subalgebra of the group $G_{S}$, where ${r
  = \rk(G_{S})}$ is the rank of $G_{S}$.  In terms of $(t_1,\cdots,t_r)$, the
deformations of the $ADE$ singularities in \eqref{ADES} can be presented
as in the table below.
\begin{equation}\label{ADESDF}
\begin{tabular}{|l|l|}
\hline
$A_n$ &\qquad ${y^2 \,=\, x^2 + z^{n+1} \,+\, \sum_{k=2}^{n+1}\, \alpha_k
  \, z^{n+1-k}}$ \\ \hline
$D_n$ &\qquad ${y^2 \,=\, -x^2 \, z \,+\, z^{n-1} \,+\,
  \sum_{k=1}^{n-1} \delta_{2k} \, z^{n-k-1} \,-\, 2 \,
  \gamma_n \, x}$ \\ \hline
$E_6$ &\qquad ${y^2 \,=\, x^3 \,+\, \frac{z^4}{4} \,+\, \varepsilon_2 \, x z^2
  \,+\, \varepsilon_5 \, x z \,+\, \varepsilon_6 \, z^2 \,+\, \varepsilon_8 \,
  x \,+\, \varepsilon_9 \, z \,+\, \varepsilon_{12}}$ \\ \hline
$E_7$ &\qquad ${y^2 \,=\, -x^3 \,+ 16\, x\,z^3 \,+\, \varepsilon_2 \, x^2 z
  \,+\, \varepsilon_6 \, x^2 \,+\, \varepsilon_8 \, x z \,+\,}$ \\
&\qquad\qquad ${+\, \varepsilon_{10} \, z^3 \,+\, \varepsilon_{12} \, x
  \,+\, \varepsilon_{14} z \,+\, \varepsilon_{18}}$ \\ \hline
$E_8$ &\qquad ${y^2 \,=\, x^3 \,-\, z^5 \,+\, \varepsilon_2 \, x z^3
  \,+\, \varepsilon_8 \, x z^2 \,+\, \varepsilon_{12} \, z^3 \,+\,
  \varepsilon_{14} \, x z \,+\,}$ \\
&\qquad\qquad ${+\, \varepsilon_{18} \, z^2 \,+\, \varepsilon_{20}\, x \,+\,
  \varepsilon_{24} \, z \,+\, \varepsilon_{30}}$ \\
\hline
\end{tabular}
\end{equation}
Up to shifts of coordinates by an overall constant, the particular
choices of minus signs and numerical coefficients in the table
above have been chosen in order to conform with the presentation
of the singularities of table 3 in \cite{KatzMorrison} so that for example,
$\alpha_{k} = s_{k}$.\footnote{Here we deviate slightly from \cite{KatzMorrison}
by writing $y^2 = x^2 + z^{n+1}$ as opposed to $xy = z^{n+1}$.
This distinction is unimportant, though, because the deformations of the singularity
are all independent of $x$ and $y$.  The presentation given here proves convenient for providing
a uniform treatment of the Calabi-Yau fourfold condition for all singularities.} For ease of comparison,
we have retained the conventions of \cite{KatzMorrison} for labelling the
deformation parameters of the $D$- and $E$- type singularities.

By analogy with the role of the symmetric polynomials $s_k$ in
\eqref{DEFAN}, each of the $\delta_{2k}$, $\gamma_k$, and $\varepsilon_k$
appearing in \eqref{ADESDF} is one of the fundamental invariant
polynomials of degree $k$ on the Cartan subalgebra of $G_{S}$.
Specifically, for the case of a $D_n$ singularity, the polynomials
$\delta_{2k}$ are invariant polynomials of degree $2k$ on the Cartan
subalgebra of $SO(2n)$, given in terms of the elementary symmetric
polynomials by ${\delta_{2k}\,=\,s_k(t_1^2,\ldots,t_n^2)}$, and
${\gamma_n\,=\, t_1 t_2\cdots t_n}$ is the invariant polynomial of
degree $n$ which represents the Pfaffian of $SO(2n)$.  Equivalently
in terms of Casimirs of $SO(2n)$,
\begin{align}
\delta_2(t_1,\cdots\,t_n) \,&=\, -\ha \Tr(\varphi^2)\,,\cr
&\,\,\,\vdots\cr
\gamma_n(t_1,\cdots,t_n) \,&=\, \Pf(\varphi)\,.
\end{align}
For the $E$-type singularities, each $\varepsilon_k$ is an invariant
polynomial of degree $k$ in the coordinates $t_{i}$ which parameterize
the Cartan subalgebra of $E_{6,7,8}$.
The invariant polynomials for $E_{6,7,8}$ are notoriously
complicated, but thankfully at the moment we only require knowledge of
their total degrees, indicated by the subscripts in \eqref{ADESDF}.

In order for the deformations in \eqref{ADESDF} to make sense
geometrically when the affine coordinates $(x,y,z)$ transform in the
specific powers of $K_S$ given in \eqref{ADESIII}, each invariant
polynomial $\delta_{2k}$, $\gamma_k$, and $\varepsilon_k$ must transform in
the tensor power of $K_S$ whose degree matches the degree of the
polynomial in $(t_1,\ldots,t_r)$.  For instance, in the case
of the $E_8$ singularity, each term appearing in the deformed equation in
\eqref{ADESDF} must transform as a section of $K_S^{30}$,
since $y^2$, $x^3$, and $z^5$ all transform as sections of
$K_S^{30}$ according to \eqref{ADESIII}.  As one can easily
check, this condition implies that the invariant polynomial
$\varepsilon_k$ of degree $k$ on the Cartan subalgebra of $E_8$ must
transform as a section of the corresponding tensor power
${K_S^{k}}$, where $k$ runs over
\begin{equation}
(2, 8, 12, 14, 18, 20, 24, 30)\,.
\end{equation}

Once again, we physically interpret the coordinates
${(t_1,\ldots,t_r)}$ on the Cartan subalgebra of $G_{S}$ as the
eigenvalues of the adjoint-valued scalar field $\varphi$.  Because
each degree $k$ Casimir of $\varphi$ then transforms as a section of
$K_S^{k}$, consistency requires that $\varphi$ itself must
transform as a section of ${K_S\otimes\ad(P)}$, just as we found from
the twisted Yang-Mills theory on $S$.

This simple observation is our first check that the degrees of freedom of the
twisted  gauge theory on ${\BR^{3,1} \times S}$ agree with the
geometric description of F-theory on $X$.  Reversing our observation,
we also obtain a powerful means to determine the low-energy
description of F-theory on more general $X$.  Namely, if
$X$ is given by an arbitrary unfolding of the singularity over $S$ as
in \eqref{ADESDF}, then F-theory on $X$ is still described at
low-energies by the partially twisted Yang-Mills theory on ${\BR^{3,1}
  \times S}$, but that Yang-Mills theory is now in a background for
which $\varphi$ has a non-trivial, necessarily holomorphic expectation
value, determined by the coefficients in the unfolding of the ADE-singularity.

\subsection{The Four-Dimensional Effective Theory}\label{REDUCTION}

In this Section we begin our analysis of four-dimensional supersymmetric vacua
on $\mathbb{R}^{3,1}$ associated to the partially twisted seven-brane
theory.  To this end, we first explain the regime of F-theory compactifications
for which we expect our analysis to remain valid. In many topological quantum
field theories, the primary purpose of the Lagrangian is to enforce the BPS
equations of motion when the theory is studied classically at weak-coupling.
In the present case, however, the role of the eight-dimensional partially twisted
Lagrangian is more central because it determines the Yukawa couplings in the
low-energy effective theory on $\BR^{3,1}$. Additional features of the
partially twisted eight-dimensional theory are presented in Appendices \ref{TWISTEDSUSYS} and
\ref{8DACTION}.

In this subsection we extract the massless spectrum and F-terms of the four-dimensional effective theory.
Because this information only depends on holomorphic data and not the K\"{a}hler metric, general arguments
suggest that we can reliably compute this result at large volume and then extrapolate to the regime of
small volume, if desired. On the other hand, the D-terms of the four-dimensional effective theory will in
general receive quantum corrections.  In the regime of large volume, these effects can in principle be
computed by integrating out the corresponding massive Kaluza-Klein modes which arise from compactification
on $S$.

More precisely, the essential point which allows us to perform our analysis at large volume
is that we may view the supersymmetry
transformation ${\bar\delta \,=\, \bar\delta_{\dot 1} \,+\,
  \bar\delta_{\dot 2}}$ as a BRST-operator. Because
the states in the cohomology\footnote{Because we effectively
consider cohomology with respect to the pair of supercharges
$\bar\delta_{\dot\alpha}$, this topological field theory is of the
``balanced'' type discussed in \cite{DijkgraafMooreBalanced}.}
of $\bar\delta$ transform in massless ${\CN=1}$ chiral multiplets,
for the purposes of analyzing these states and their superpotential
interactions, we are free to take $\omega$ to be arbitrarily large.
This observation is very important, since it underlies our classical
analysis of the effective theory on $\BR^{3,1}$ obtained by
compactification of the eight-dimensional Yang-Mills theory on $S$.
Indeed, this fact allows us to reliably compute both the massless
spectrum and superpotential of the resulting theory.

From the perspective of the four-dimensional effective description in $\mathbb{R}^{3,1}$,
the eight-dimensional fields of the partially twisted theory are packaged as two $\mathcal{N}=1$
chiral multiplets $(A_{\overline{m}},\psi_{\alpha \overline{m}})$ and $(\varphi_{mn},\chi_{\alpha mn})$, and a
single vector multiplet $(A_{\mu},\eta_{\alpha})$.
In rigid $\mathcal{N}=1$ superspace, the chiral multiplets then organize into
chiral superfields ${\bf \Phi}_{mn}$ and ${\bf A}_{\overline{m}}$ with lowest bosonic components
$\varphi_{mn}$ and $A_{\overline{m}}$, respectively.  Letting $F_{S}$ denote the restriction of
the eight-dimensional field strength $F$ to $S$, $F^{(0,2)}_{S}$ corresponds to the lowest bosonic component
of an $\mathcal{N}=1$ chiral superfield ${\bf F}_{S}^{(0,2)}$ whose remaining
components can be determined from the explicit off-shell transformations presented in Appendix \ref{8DACTION}.
Summarizing the end result of Appendix \ref{8DACTION}, the partially twisted action in eight dimensions consists
of contributions from $\bar{\delta}$ trivial terms and an F-term:
\begin{equation}\label{IS}
I_{S} = \underset{\mathbb{R}^{3,1}\times S}{\int}d^{4}x d^{2}\overline{\theta}\mathcal{O}
-\underset{\BR^{3,1} \times S}{\int}\!\!d^4 x \,
d^2\theta \> \Tr\Big( {\bf F}_S^{(0,2)} \^ {\mathbf \Phi} \Big)\,,
\end{equation}
where $\mathcal{O}$ is a gauge invariant operator which is $\bar\delta$ trivial.
Here and throughout, `$\Tr$' generally denotes a negative-definite
invariant quadratic form on the Lie algebra of the gauge group $G_{S}$.
As standard, we normalize `$\Tr$' such that for ${G_{S} = SU(n)}$ this form
corresponds to the trace in the fundamental representation.  Also,
$d^4 x$ denotes the standard measure on $\BR^{3,1}$.

The organization of the rest of this subsection is as follows.  We first
derive the BPS equations of motion for the fields of the partially twisted eight-dimensional theory.
The massless spectrum of particles correspond to
zero mode solutions on $S$ in the presence of a non-trivial supersymmetric
background gauge field configuration.  These zero modes are represented
by elements of an appropriate bundle valued Dolbeault cohomology group.
After this analysis, we show how the four-dimensional superpotential descends from the partially twisted
eight-dimensional theory. When $S$ is a Hirzebruch or del Pezzo surface, a general vanishing theorem for
cohomology groups forbids any classical Yukawa couplings.  In subsequent
sections we resolve this issue by treating a more general class of F-theory
compactifications.

\subsubsection{BPS Equations}\label{BPSEOM}

To begin our analysis of four-dimensional supersymmetric vacua, we first present
the BPS equations of motion for the eight-dimensional fields of the partially twisted theory.
In a supersymmetric vacuum, the variations of all of the eight-dimensional fermions must vanish.
These conditions determine the BPS equations of motion.  From the viewpoint of the four-dimensional effective theory,
these equations follow from the requirement that in a supersymmetric vacuum the corresponding effective potential is both
D- and F-flat. In this subsection we present these conditions in the
regime of large volume for $S$.  Whereas the D-flatness
conditions will in general receive perturbative corrections away from
the regime of large volume, up to non-perturbative effects,
the equations of motion derived from the F-terms will remain unchanged.

The on-shell supersymmetry transformations of the eight-dimensional fields of the partially
twisted theory may be found in Appendix \ref{TWISTEDSUSYS}.  From the
variations of $\eta_\beta$ and $\bar\eta_{\dot\beta}$, we see that both
the self-dual and anti-self-dual components of the curvature $F_{\mu \nu}$ on
$\BR^{3,1}$ must vanish, so that
\begin{equation}\label{4dFIELDSTRENGTH}
F_{\mu\nu} \,=\, 0\,.
\end{equation}
From the variations of $\psi_{\beta\, \bar m}$, $\chi_{\beta\, m n}$
and their conjugates, we also see that the mixed components of the
curvature along ${\BR^{3,1} \times S}$ vanish,
\begin{equation}
F_{\mu m} \,=\, F_{\mu \bar m} = 0\,.
\end{equation}
Similarly, both $\varphi$ and $\bar\varphi$ satisfy
\begin{equation}
D_\mu \varphi \,=\, D_\mu \bar\varphi \,=\, 0\,.
\end{equation}

The more interesting BPS equations involve the behavior of the gauge
field and the twisted scalars $(\varphi\,, \bar\varphi)$ on $S$.
First, from the variations of the fermions $\chi_\alpha$ and
$\bar\chi{}_{\dot\alpha}$ in \eqref{FSUSY}, we see that the
components of the curvature on $S$ of holomorphic/anti-holomorphic
type $(2,0)$ and $(0,2)$ both vanish,
\begin{equation}\label{HOLP}
F{}_S^{(0,2)} \,=\, F{}_S^{(2,0)} \,=\, 0\,.
\end{equation}
Hence the $(0,1)$ component $\bar\partial_A$ of the covariant
derivative associated to the gauge field on $S$ endows the
$G_{S}$-bundle $P$ with a holomorphic structure, since
${\bar\partial{}_A^2 = F{}_S^{(0,2)} = 0}$.  The variation of the
fermions in \eqref{FSUSY} also implies a holomorphy condition on
$\varphi$ as well as an anti-holomorphy condition on $\bar\varphi$,
\begin{equation}\label{HOLPHI}
\bar\partial{}_A \varphi \,=\, \partial_A \bar\varphi \,=\, 0\,.
\end{equation}
Finally, from the variations of $\eta_\alpha$ and
$\bar\eta{}_{\dot\alpha}$, the $(1,1)$-component of the curvature on
$S$ must be related to the commutator of $\varphi$ and $\bar\varphi$
as:
\begin{equation}\label{VAFWITTDTERM}
\omega \^ F_S \,+\, {i\over 2}\,[\varphi\,,\bar\varphi] \,=\, 0\,,
\end{equation}
where $\omega$ is the K\"{a}hler form on $S$.

Now, as the discussion
in Appendix \ref{8DACTION} shows, the holomorphy conditions in
\eqref{HOLP} and \eqref{HOLPHI} are both associated to the vanishing
of F-terms in the language of ${\CN=1}$ supersymmetry.  Neither of
these F-term conditions involves the K\"ahler metric on $S$, and we do
not expect them to receive corrections, at least in perturbation theory.
On the other hand, the supersymmetry condition in
\eqref{VAFWITTDTERM} involves the vanishing of a D-term, and this
condition manifestly depends upon the K\"ahler metric on $S$.  In
general, the D-term condition in \eqref{VAFWITTDTERM} can therefore
receive perturbative corrections as we integrate out massive
Kaluza-Klein modes on $S$.  However, if the volume of $S$ is
sufficiently large, we still expect solutions to the classical D-term
equation in \eqref{VAFWITTDTERM} to determine supersymmetric
configurations for $A$, $\varphi$, and $\bar\varphi$ on $S$.

In much of this paper we shall primarily consider vacua such that
$\varphi=0$. In this case, equations \eqref{HOLP} and \eqref{VAFWITTDTERM}
correspond to the usual Hermitian Yang-Mills equations, which are
satisfied when the connection on $P$ is anti-self-dual and the gauge
field describes an instanton on $S$.

We now discuss one immediate consequence of \eqref{VAFWITTDTERM} which is
especially helpful for explicit computations with line bundles.
Given a holomorphic gauge bundle $\mathcal{T}$ transforming in a unitary representation $T$
of some subgroup of $G_{S}$ which we denote by $H_{S}$, we note that the first Chern class
of $\mathcal{T}$ admits a Chern-Weil representation as
${c_1(\mathcal{T}) = {i\over{2\pi}} \Tr_{T}(F_{S})}$.  When $\varphi=0$, tracing over
equation \eqref{VAFWITTDTERM} in the representation $T$ yields the condition:
\begin{equation}\label{PERPC}
\omega \^ c_1(\mathcal{T}) \,=\, 0 \,\in\, H^2(S;\BR)
\end{equation}
where $\omega$ denotes the K\"{a}hler form on $S$.  In the special case where $\mathcal{T}$
is a line bundle and $\varphi=0$, equation \eqref{VAFWITTDTERM} is equivalent to equation \eqref{PERPC}.
Because $c_1(\mathcal{T})$ is also quantized as an element in the integral
cohomology of $S$, equation \eqref{PERPC} can only be solved for generic
$\omega$ if ${c_1(\mathcal{T}) = 0}$.  Conversely, if \eqref{PERPC} is to
admit a solution with ${c_1(\mathcal{T}) \neq 0}$, we must assume that
$\omega$ is non-generic.  Said differently, a supersymmetric
gauge field configuration stabilizes a K\"{a}hler modulus
in the compactification.  Some examples of supersymmetric line bundles
for $S$ a Hirzebruch or del Pezzo surface are presented in Appendix
\ref{VanishAppendix}.

Before closing this subsection, we note that the BPS
equations in \eqref{HOLP}, \eqref{HOLPHI}, and \eqref{VAFWITTDTERM}
have certainly been considered before.  These equations on $S$ are
precisely\footnote{Due to the fact that we follow
  slightly different conventions from \cite{VafaWitten}, a factor of
  `$i/2$' appears in \eqref{VAFWITTDTERM} which is otherwise absent in
  \cite{VafaWitten}.}
the equations which arise from the topological twist of
four-dimensional, ${\CN=4}$ supersymmetric Yang-Mills theory studied
in \cite{VafaWitten} when that theory is specialized to a K\"ahler
four-manifold.  Indeed, the partial twisting of the eight-dimensional
theory we are considering is identical to the theory of
\cite{VafaWitten} in the K\"ahler case.

\subsubsection{Bulk Spectrum of Massless Particles\label{SPECTRUM}}

In direct analogy with heterotic Calabi-Yau compactifications,
the massless spectrum of the seven-brane theory wrapping $S$ originates
from a potentially non-trivial background gauge field configuration on $S$ with
values in some subgroup $H_{S}$.  More precisely,
the spectrum in four dimensions is determined by bundle valued Dolbeault cohomology groups
$H_{\bar{\partial}}^{i}$ on $S$ where $i=0,1,2$.  This fact allows us to express
the total massless spectrum in a given representation of the unbroken gauge group in terms of a
topological index formula.  The end result of this analysis is that
the number of generations minus anti-generations in a representation $\tau$ is:
\begin{equation}\label{GENANTIGEN}
n_{\tau}-n_{\tau^{\ast}}  =-\underset{S}{\int}c_{1}(\mathcal{T})c_{1}(S)
\end{equation}
where $\tau^{\ast}$ denotes the dual representation to $\tau$.  Letting
$\Gamma_{S}$ denote the maximal subgroup of $G_{S}$ such that $G_{S}\supset \Gamma_{S}\times H_{S}$,
in the above, $\mathcal{T}$ denotes a bundle transforming in a representation $T$ of $H_{S}$
such that the decomposition of the adjoint representation of $G_{S}$ to $\Gamma_{S}\times H_{S}$
contains the representation $(\tau,T)$.

We now describe in further detail the massless spectrum.  Our starting point is a supersymmetric
background configuration for the gauge field and the twisted scalars $(\varphi,\bar\varphi)$ on $S$.
For our applications, we shall primarily be interested in configurations with ${\varphi = 0}$
or more generally configurations such that $[\varphi,\bar{\varphi}]=0$.
Under these assumptions, the BPS equations state that the principal $G_{S}$-bundle $P$
carries an anti-self-dual connection, corresponding to an instanton on $S$. While the
case of non-vanishing commutator for the internal four dimensional $\mathcal{N}=4$ twisted theory on $S$ has
been studied in \cite{VafaWitten}, it would be interesting to interpret such vacua from the perspective of the
seven-brane theory.\footnote{When this commutator does not vanish, it is tempting to speculate that the seven-brane
experiences an analogue of the Myers effect.  We thank M. Wijnholt for suggesting this possibility.}

The representation content of the particle spectrum is fixed by decomposing the adjoint representation of $G_{S}$
into irreducible representations of $\Gamma_{S}\times H_{S}$:
\begin{equation}\label{IRREP}
\ad(G_{S}) \,\cong\, \bigoplus_{j} \, \tau_{j} \otimes T_{j}\text{.}
\end{equation}
A similar decomposition holds for the bundle $\ad(P)$.  In the obvious notation, we let $\mathcal{T}_{j}$
denote the corresponding bundle which transforms as a representation $T_{j}$ of $H_{S}$.

Given an instanton on $S$ with structure group $H_{S}$, the unbroken low energy gauge group is given by
the commutant subgroup of $H_{S}$ in $G_{S}$.  We note that in some cases such as when $H_{S}$ contains
semi-simple $U(1)$ factors, $\Gamma_{S}$ corresponds to a proper subgroup of the four-dimensional gauge group.  For example,
when $G_{S}=E_{6}$ and $H_{S}=U(1)$, the commutant is $G_{4d}=SO(10)\times U(1)$. On the other hand,
in many cases this $U(1)$ factor is anomalous and therefore decouples from the low energy dynamics via
the Green-Schwarz mechanism.

The massless spectrum on $\BR^{3,1}$ is given by zero mode solutions on $S$ in
the presence of a potentially non-trivial background gauge field configuration.
Because the partial twisting on $S$ automatically produces a covariantly constant supercurrent along $S$,
each fermionic zero mode possesses a bosonic counterpart.  It therefore suffices to specify
the spectrum of massless fermions in the four-dimensional effective theory.  By inspection of the topological action,
in a background with ${\varphi = \bar\varphi = 0}$, the equations of
motion for the zero modes state that the fermions $\bar\eta{}_{\dot\alpha}$,
$\psi_\alpha$, and $\bar\chi{}_{\dot\alpha}$ in \eqref{TWF} are
annihilated by the operators $\bar\partial{}_A$ and
$\bar\partial{}_A^\dagger$ and hence by the Laplacian
${\triangle^{}_{\bar\partial} \,=\, \bar\partial{}^\dagger_A \bar\partial^{}_A
  \,+\, \bar\partial^{}_A \bar\partial{}^\dagger_A}$.

Via the standard Hodge isomorphism, we then obtain massless fermions
on $\BR^{3,1}$ from the following Dolbeault cohomology groups on $S$,
\begin{equation}\label{DOLBH}
H^0_{\bar\partial}\big(S\,,\ad(P)\big) \,\oplus\,
H^1_{\bar\partial}\big(S\,, \ad(P)\big) \,\oplus\,
H^2_{\bar\partial}\big(S\,, \ad(P)\big)\,
\end{equation}
where the massless modes for $\bar\eta{}_{\dot\alpha}$, $\psi_\alpha$ and $\bar\chi{}_{\dot\alpha}$
are associated to elements of $H_{\overline{\partial}}^0$, $H_{\overline{\partial}}^1$ and
$H_{\overline{\partial}}^2$, respectively.  Note that the natural even/odd
grading on the cohomology correlates with the the chirality of the corresponding
massless fermions on $\BR^{3,1}$.  Labelling the representation content of a four-dimensional massless chiral field by
an additional subscript, the decomposition of $\ad(P)$ in the analogue of equation \eqref{IRREP} implies
that the zero modes in a representation $\tau_{j}$ of $\Gamma_{S}$ belong to the $\mathcal{T}_{j}$ bundle valued cohomology groups:
\begin{align}\label{DOLBHII}
\bar\eta{}_{\dot\alpha \tau_{j}} & \in H_{\bar\partial}^{0}(S,\mathcal{T}_{j})\\
\psi_{\alpha \tau_{j}} & \in H_{\bar\partial}^{1}(S,\mathcal{T}_{j})\\
\bar\chi{}_{\dot\alpha \tau_{j}} & \in H_{\bar\partial}^{2}(S,\mathcal{T}_{j})\text{.}
\end{align}
Further, when $\tau_{j}$ is a complex representation, the CPT conjugates of
$\bar\eta{}_{\dot\alpha \tau_{j}}$ and $\bar\chi{}_{\dot\alpha \tau_{j}}$ correspond to chiral spinors
in the complex conjugate representation $\tau^{\ast}_{j}$.

The same remarks hold for the fermions $\eta_\alpha$,
$\bar\psi{}_{\dot\alpha}$, and $\chi_\alpha$ in \eqref{TWFC} which are related to
the zero modes described above by CPT conjugation.  Because
the zero-mode wavefunctions for $\eta_\alpha$, $\bar\psi{}_{\dot\alpha}$, and
$\chi_\alpha$ are naturally anti-holomorphic on $S$, we identify those
wavefunctions with elements in the conjugate to \eqref{DOLBH}:
\begin{equation}\label{BARDOLBH}
\bar{H^0_{\bar\partial}\big(S\,,\ad(P)\big)} \,\oplus\,
\bar{H^1_{\bar\partial}\big(S\,, \ad(P)\big)} \,\oplus\,
\bar{H^2_{\bar\partial}\big(S\,, \ad(P)\big)}\,.
\end{equation}

Because algebraic geometry typically deals with
holomorphic objects, this is slightly awkward.  A
holomorphic description for the zero modes described by
\eqref{BARDOLBH} is obtained using the isomorphism of vector spaces\footnote{This
isomorphism is obtained as follows. We first observe that the K\"ahler metric on $S$
and the invariant form `$\Tr$' on the Lie algebra of $G_{S}$ together
define the obvious inner product between elements in \eqref{DOLBH} and
\eqref{BARDOLBH}.  The $p$-form indices are contracted along $S$
using the K\"ahler metric with all Lie algebra indices contracted using
the form `$\Tr$'.}:
\begin{equation}\label{Dualizing}
\bar{H^p_{\bar\partial}\big(S\,, \ad(P)\big)} \,\cong\,
  H^p_{\bar\partial}\big(S\,,\ad(P)\big){}^{*}\,.
\end{equation}

Physically, dualization corresponds to CPT conjugation.  It is now immediate that the chiral
and anti-chiral spectrum for fields in a representation $\tau_{j}$ of $\Gamma_{S}$ is given by the cohomology groups:
\begin{align}\label{SPECI}
\text{chiral}  & \text{: }H^0_{\bar\partial}\big(S\,, \mathcal{T}_{j}^{*}\big){}^{*} \oplus
H^1_{\bar\partial}\big(S\,, \mathcal{T}_{j}\big) \oplus
H^2_{\bar\partial}\big(S\,, \mathcal{T}_{j}^{*}\big){}^{*}\\
\text{anti-chiral}  & \text{: }H^0_{\bar\partial}\big(S\,, \mathcal{T}_{j}\big) \oplus
H^1_{\bar\partial}\big(S\,, \mathcal{T}_{j}^{*}\big){}^{*} \oplus
H^2_{\bar\partial}\big(S\,, \mathcal{T}_{j}\big).
\end{align}
Note in particular that the resulting spectrum is automatically
CPT-invariant.

In principle, the spectrum is now determined entirely in terms of the
Dolbeault cohomology groups of \eqref{SPECI} for the bundle
$\mathcal{T}_{j}$ associated with a representation $\tau_{j}$
which appear in the decomposition of $\ad(P)$.  In general, these cohomology
groups are not topological objects because they depend
holomorphically upon the complex structure moduli for $S$ and the vector bundles
$\mathcal{T}_{j}$ and $\mathcal{T}_{j}^{*}$.  Nevertheless, returning to \eqref{SPECI},
the analogous expression for $\tau^{\ast}_{j}$ interchanges all instances of $\mathcal{T}_{j}$
and $\mathcal{T}_{j}^{*}$.  The number of massless four-dimensional chiral fields in the representation $\tau_{j}$
minus the number in $\tau^{\ast}_{j}$ is therefore:
\begin{align}\label{NETNUMBER}
n_{\tau_{j}}-n_{\tau^{\ast}_{j}}  & =h^{0}(S,\mathcal{T}_{j}^{*})+h^{1}(S,\mathcal{T}_{j})+h^{2}(S,\mathcal{T}_{j}^{*})\\
& -(h^{0}(S,\mathcal{T}_{j})+h^{1}(S,\mathcal{T}_{j}^{*})+h^{2}(S,\mathcal{T}_{j}))\\
& =\chi(S,\mathcal{T}_{j}^{*})-\chi(S,\mathcal{T}_{j})
\end{align}
where as usual, $h^{i}=\text{dim}_{\mathbb{C}}H^{i}$ and the Euler character $\chi=h^{0}-h^{1}+h^{2}$ is a topological invariant.

The explicit values of the above Euler characters are determined by an index formula:
\begin{align}
\chi(S,\mathcal{T}) &  =\underset{S}{\int}{ch(\mathcal{T})Td(S)}\label{INDEX}\\
&  =\underset{S}{\int}\left(  \frac{\rk(\mathcal{T})}{12}\left[  c_{1}(S)^{2}+c_{2}(S)\right]
+\frac{1}{2}c_{1}(\mathcal{T})c_{1}(S)+\frac{1}{2}\left[  c_{1}(\mathcal{T})^{2}-2c_{2}%
(\mathcal{T})\right]  \right)  .\label{EXPANDO}%
\end{align}
Because $c_{1}(\mathcal{T})=-c_{1}(\mathcal{T}^{*})$ and $c_{2}(\mathcal{T})=c_{2}(\mathcal{T}^{*})$,
the number of chiral generations minus anti-generations in a representation $\tau_{j}$ is:
\begin{equation}
n_{\tau_{j}}-n_{\tau^{\ast}_{j}}  =-\underset{S}{\int}c_{1}(\mathcal{T}_{j})c_{1}(S).
\end{equation}

In the rigid case where $S$ is either a Hirzebruch of del Pezzo surface, the number
of generations and anti-generations are each computed by a
\textit{distinct} index. This is a consequence of the general vanishing theorem
established in Appendix \ref{VanishAppendix} which shows that for an
arbitrary supersymmetric gauge field configuration, $H_{\bar{\partial}}^{2}(S,\mathcal{T}_{j})=0$.\footnote{A related vanishing theorem is as follows.  Note that
because $\varphi$ transforms as a section of $K_{S}\otimes \ad(P)$,
a zero mode solution for $\varphi$ would imply that
$H_{\bar{\partial}}^{i}(S,K_{S}^{n})$ is non-zero for some $n>0$.
Because $K_S$ is a strictly negative line bundle, the Kodaira
vanishing theorem establishes the absence of such solutions.}
Moreover, when the holomorphic bundle $\mathcal{T}_{j}$ is irreducible
and non-trivial (meaning ${{\mathcal T}_j \neq \CO_S}$), the discussion
in Appendix \ref{VanishAppendix} also shows that
$H_{\bar{\partial}}^{0}(S,\mathcal{T}_{j})=0$.  Combining these facts yields:
\begin{align}
n_{\tau_{j}} & = h^{1}(S,\mathcal{T}_{j})=-\chi(S,\mathcal{T}_{j})\label{SPECIALNGEN}\\
n_{\tau^{\ast}_{j}} & =h^{1}(S,\mathcal{T}_{j}^{*})=-\chi(S,\mathcal{T}_{j}^{*}).
\end{align}

\subsubsection{Bulk Yukawa Couplings\label{BULKYUKAWAS}}

In this section we summarize the Yukawa couplings of the four-dimensional effective theory for a
seven-brane wrapping a general complex surface $S$.  We also find that when
$S$ is either a Hirzebruch or del Pezzo surface, \textit{all} of the classical Yukawa couplings vanish.

When $S$ is a general complex surface,
the four-dimensional Yukawa couplings for the zero mode solutions catalogued by \eqref{SPECI}
originate from the last term in the action of equation \eqref{IS}.  Labelling the
zero mode solutions by $\alpha$, $\beta$ and $\gamma$, the Yukawa couplings of the
four-dimensional effective are therefore determined by the superpotential:
\begin{equation}\label{8DYUK}
d_{\alpha \beta \gamma} = - \underset{S}{\int} h_{ijk} {\bf A}^{\beta,i} \wedge {\bf A}^{\gamma,j} \wedge {\bf \Phi}^{\alpha,k}\text{,}
\end{equation}
where $i,j,k$ and $h_{ijk}$ respectively denote group theory indices and the structure constants of $G_{S}$
associated with breaking $G_{S}$ to the subgroup $\Gamma_{S} \times H_{S}$.

Using the isomorphism of equation \eqref{Dualizing}, a non-trivial Yukawa
coupling between three representations $\tau_{1},\tau_{2},\tau_{3}$ of $\Gamma_{S}$ corresponds to a tri-linear map:
\begin{equation}
H_{\bar{\partial}}^{1}(S,\mathcal{T}_{1})\otimes H_{\bar{\partial}}^{1}(S,\mathcal{T}_{2})\otimes H_{\bar{\partial}}^{2}(S,\mathcal{T}_{3}^{\ast})^{\ast}\rightarrow \mathbb{C}\text{.}
\end{equation}
Note in particular that when $S$ is either a Hirzebruch or del Pezzo surface, the vanishing
theorem used in the previous section to constrain the zero mode content
also requires all Yukawa couplings to vanish.  This is
unacceptable from the viewpoint of phenomenology.  In subsequent
sections we shall rectify this deficiency by treating a more general class of F-theory
compactifications which couple the partially twisted theory to a six-dimensional defect
theory with matter localized along Riemann surfaces in $S$.

\subsection{Bulk Toy Models}

Using the above results, we now present some toy models which illustrate how
both non-chiral and chiral matter can originate from a supersymmetric
gauge field configuration on $S$. Initially, we take $S$ to be a general del Pezzo surface.
We first show that instanton solutions on $S$ with structure group $SU(n)$ can generate
a non-trivial massless spectrum in four dimensions.  Nevertheless, we find that in such cases the resulting
spectrum is always non-chiral because the corresponding instanton solutions have vanishing
first Chern class. We next present an $SO(10)$ model with three chiral generations
transforming in the ${\bf 16}$ spinor representation.

To generate zero modes on $S$ which can descend to massless fields in four dimensions, we introduce an
instanton solution with structure group $SU(n)\subset G_{S}$ on $S$ with instanton number $k$.
As an example, suppose that the decomposition of the adjoint of $G_{S}$ to $\Gamma_{S}\times SU(n)$
contains the representation $(\tau,n)$.  Assuming that a non-trivial instanton solution on $S$ exists,
the resulting number of massless chiral four-dimensional fields in the representation $\tau$ is:
\begin{equation}\label{NUMR}
n_{\tau}=k-n\text{.}%
\end{equation}
The existence of zero mode solutions thus imposes the condition $k\geq n$.

While the explicit parameterization of such an instanton solution is quite non-trivial,
it is enough to check that the moduli space $\mathcal{M}_{k}$ of solutions with instanton
number $k$ is non-trivial. The dimension of $\mathcal{M}_{k}$ is:%
\begin{equation}
\dim\mathcal{M}_{k}=4kn-(n^{2}-1)\frac{(\chi
+\sigma)}{2}%
\end{equation}
where $\chi$ and $\sigma$ respectively denote the Euler character and signature of
$S$. \ When $S$ is a del Pezzo surface, this becomes:%
\begin{equation}
\dim\mathcal{M}_{k}=4kn-2(n^{2}-1)\text{.}%
\end{equation}
The condition $n_{\tau}\geq0$ implies:%
\begin{equation}
\dim\mathcal{M}_{k}=4kn-2(n^{2}-1)\geq2n^{2}+2>0
\end{equation}
so that there exists a moduli space of solutions in this case which can generate
massless chiral four-dimensional fields in the representation $\tau$ of $\Gamma_{S}$.  A similar analysis
shows that when $\tau$ is a complex representation, the number of massless chiral four-dimensional fields
in the complex conjugate representation $\tau^{\ast}$ precisely matches the number transforming
in $\tau$, so that the resulting spectrum is non-chiral.

In fact, we now show that for arbitrary $S$, an $SU(n)$ instanton solution cannot generate a chiral spectrum in four dimensions.  Indeed, because the adjoint
of $G_{S}$ is a real representation, the decomposition to $\Gamma_{S}\times SU(n)$ will contain
the representations $(\tau_{j},T_{j})$ and $(\tau^{\ast}_{j},T^{\ast}_{j})$ when the representation $\tau_{j}$ is complex.
Next recall from equation \eqref{GENANTIGEN} that the net number of generations in $\tau_{j}$ minus the number in $\tau^{\ast}_{j}$ is
$-c_1(S)\cdot c_{1}(\mathcal{T}_{j})$. Because all unitary representations of $SU(n)$ are traceless, $c_{1}(\mathcal{T}_{j})$ vanishes.

We now present an explicit toy model with chiral matter spectrum induced by a
supersymmetric gauge field configuration such that tensor powers of the gauge bundle have non-zero first Chern class.
\ To this end, we take a seven-brane wrapping the Hirzebruch surface $S=\mathbb{F}_{1}$ with bulk gauge group
$G_{S}=E_{6}$. \ By turning on an appropriate supersymmetric $U(1)$ gauge
field configuration defined by a line bundle $\mathcal{L}$ on $S$, $E_{6}$ will break to
$SO(10)\times U(1)$. \ To determine the chiral spectrum of this theory, we
note that the adjoint of $E_{6}$ decomposes as:%
\begin{align}
E_{6} &  \supset SO(10)\times U(1)\\
{\bf 78} &  \rightarrow{\bf 45}_{0}+{\bf 1}_{0}+{\bf 16}_{-3}+\overline{\bf 16}_{+3}%
\end{align}
where we have chosen an integral normalization of the $U(1)$ charge.
\ Invoking the vanishing theorem proved in Appendix
\ref{VanishAppendix}, the zero modes of the Dolbeault operator for
each representation are classified by:%
\begin{align}
{\bf 16}_{-3} &  \in H_{\bar{\partial}}^{1}(S,\mathcal{L}^{-3})\\
\overline{\bf 16}_{+3} &  \in H_{\bar{\partial}}^{1}(S,\mathcal{L}^{+3})\text{.}%
\end{align}
All supersymmetric line bundles of the Hirzebruch surfaces are
classified in Appendix \ref{VanishAppendix}. \ The result is that a
line bundle is supersymmetric for some K\"{a}hler class if and only if
there exist integers $a$ and $b$ such that $ab<0$ so that
$\mathcal{L}$ is given by:%
\begin{equation}
\mathcal{L}=\mathcal{O}_{\mathbb{F}_{1}}(af+b\sigma)\text{,}%
\end{equation}
with $f$ and $\sigma$ as in Appendix \ref{APPDELPEZZO}.  Using the fact
that $c_{1}(\mathbb{F}_1)=3f+2\sigma$, the number of four-dimensional massless fields in the ${\bf 16}$ now
follows from equations \eqref{EXPANDO} and \eqref{SPECIALNGEN}:%
\begin{align}
n_{\bf 16} &  =-\left(  \rk(\mathcal{L}^{-3})+\frac{1}{2}c_{1}(\mathcal{L}^{-3})\cdot(3f
+2\sigma)+\frac{1}{2}c_{1}(\mathcal{L}^{-3})\cdot c_{1}(\mathcal{L}^{-3})\right)  \\
&  =-\left(  1-\frac{3}{2}\left(  2a+b\right)  +\frac{9}{2}\left(
2ab-b^{2}\right)  \right)
\end{align}
where in the above we have used the fact that the Todd genus of $S$ is unity for a Hirzebruch surface.
A similar computation for the number of $\overline{\bf 16}$'s yields:%
\begin{equation}
n_{\overline{\bf 16}}=-\left(  1+\frac{3}{2}\left(  2a+b\right)  +\frac{9}%
{2}\left(  2ab-b^{2}\right)  \right)  \text{.}%
\end{equation}
It is amusing that the net number of ${\bf 16}$'s is always a multiple of three:%
\begin{equation}
n_{\bf 16}-n_{\overline{\bf 16}}=3\left(  2a+b\right)
\end{equation}
so that when $a=1$ and $b=-1$, we have precisely three generations in the $\bf{16}$
of $SO(10)$.

Let us now explicitly check that the zero mode content contains no contribution from $H_{\bar{\partial}}^{0}$ and $H_{\bar{\partial}}^{2}$.
Because $f$ and $\sigma$ generate the cone of all effective classes in $H_{2}(S,\mathbb{Z})$,
when $a$ and $b$ have opposite signs, any purported global section of $\mathcal{L}$
will have a pole along either $\sigma$ or $f$. \ This implies $H_{\bar{\partial}}^{0}(S,\mathcal{L}^{n})$
is trivial for all $n$ a non-trivial integer. \ Moreover, by Serre duality we have:%
\begin{equation}
H_{\bar{\partial}}^{2}(S,\mathcal{O}_{\mathbb{F}_{1}}(naf+nb\sigma)\simeq H_{\bar{\partial}}^{0}(S,\mathcal{O}%
_{\mathbb{F}_{1}}(-(na+3)f-(nb+2)\sigma)^{*}\text{.}%
\end{equation}
In order for $H_{\bar{\partial}}^{2}(S,\mathcal{L}^{n})$ to be non-trivial, both $na$ and $nb$ must be
negative. \ Because this is not possible, it follows that the massless spectrum is completely characterized by $H_{\bar{\partial}}^{1}$.

\section{Intersecting Seven-Branes in F-Theory}\label{COUPLINGDEFECT}

So far, we have analyzed the low energy effective description for
F-theory on a class of very special elliptic Calabi-Yau fourfolds $X$ given by
canonical $ADE$ singularities over a smooth, compact K\"ahler surface
$S$.  We began by considering such models not because they are
particularly realistic, but because they are particularly simple:
F-theory on the local Calabi-Yau fourfold $X$ is described at
low energies by the partially twisted Yang-Mills theory which lives on the
worldvolume of the seven-brane wrapping ${\BR^{3,1} \times S}$.

In this Section, our goal is to study the low-energy effective
description for $F$-theory on a much more general class of local
elliptic Calabi-Yau fourfolds.  From a physical perspective, these new
F-theory backgrounds will contain intersecting seven-branes of various
$ADE$-types.  From a mathematical perspective, we will allow the
singular locus ${\Delta \subset B}$ of the elliptic fibration of $X$
to be reducible, so that $\Delta$ consists of several components of
colliding singularities in the threefold $B$.

For the sake of concreteness, we start in Section \ref{COLLIDE} by
presenting simple local models for elliptic Calabi-Yau fourfolds
associated to intersecting seven-branes in F-theory.  Of course, as
noted long ago \cite{KatzVafa} in the context of F-theory on
elliptic Calabi-Yau threefolds, when two seven-branes intersect along
a common six-dimensional subspace, one finds additional light degrees
of freedom localized at the intersection and charged under the
worldvolume gauge group carried by each seven-brane, analogous to
the bifundamental matter that arises at the intersection of ordinary
D-branes.  Somewhat surprisingly, the effective description of this
light charged matter seems never to have been worked out for ${\CN=1}$
supersymmetric F-theory models.

To remedy that gap, we consider in Section \ref{DEFECT} the case that
our original seven-brane wrapping the surface $S$ intersects another
seven-brane wrapping a surface $S'$ transversely along a smooth
complex curve ${\Sigma = S \cap S'}$.  As we explain, the
effective dynamics of the light, charged degrees of freedom living on
the intersection ${\BR^{3,1} \times \Sigma}$ are captured by a
partially twisted six-dimensional defect theory which is jointly
coupled to the bulk, eight-dimensional Yang-Mills theories living on
the two seven-branes. If $S'$ is non-compact, as will often be the
case in our local models, then the fields in the twisted Yang-Mills
theory on ${\BR^{3,1}\times S'}$ are non-dynamical background fields, and we simply consider a
twisted defect theory on ${\BR^{3,1} \times \Sigma}$ coupled
to the twisted Yang-Mills theory on ${\BR^{3,1} \times S}$ introduced
in Section \ref{SEVENBRANE}.

We next present evidence that this more involved theory accurately describes the corresponding
compactification in F-theory.  To this end, in Section \ref{UNFOLDING} we match the additional moduli of
the defect theory on $\Sigma$ to the moduli of F-theory on $X$.  This
match is quite interesting, since it relies upon the interpretation
of the defect theory on $\Sigma$ as inducing a surface operator for
the bulk gauge theory on $S$.

Having argued that the partially twisted theory does indeed describe a
more genral class of F-theory compactifications, in Section \ref{6DMATTER}
we study the resulting spectrum of massless charged matter and
effective superpotential in four dimensions.  Once again, the spectrum
of massless charged matter is determined by certain Dolbeault
cohomology groups on $\Sigma$, and the net chirality in four
dimensions is given by a corresponding index.  We also find
generically non-zero Yukawa couplings involving two chiral superfields
derived from $\Sigma$ and one from the bulk theory on $S$.

\subsection{Colliding Singularities and Intersecting Seven-Branes}\label{COLLIDE}

In this Section we present geometries for a more general class of F-theory vacua which
contain intersecting seven-branes.  To describe these intersections geometrically in terms of
colliding singularities, we begin by generalizing the local model for the
elliptically-fibered Calabi-Yau fourfold $X$ given in Section \ref{LOCALMODEL}.

Once again, $X$ will be the total space of an elliptic $K3$-fibration
over the compact surface $S$,
\begin{equation}
\begin{matrix}
&Y \,\longrightarrow\,X\cr
&\mskip 65mu\Big\downarrow\lower 0.5ex\hbox{$^\pi$}\cr
&\mskip 55mu S\cr
\end{matrix}\,,
\end{equation}
where $Y$ remains a local, elliptically-fibered $K3$ given as a
hypersurface in $\BC^3$.  However, we now allow the singularities in
$Y$ to vary from point to point over $S$.  At points where the generic
singularity on $S$ enhances, two seven-branes intersect.

\bigskip\noindent{\it Intersecting Seven-Branes of A-Type}\smallskip

As a first example, we exhibit a local model for $X$ that describes a
compact 7-brane of type $A_n$ which wraps $S$ and intersects
another, non-compact seven-brane of type $A_m$ along an effective
divisor ${\Sigma\subset S}$.  The fibering of $Y$ over $S$ is
necessarily more complicated in this situation, so we generalize
\eqref{RKTHV} by allowing the affine coordinates $(x,y,z)$ on $\BC^3$
to transform over $S$ as sections of the rank three bundle:
\begin{equation}\label{RKTHVII}
V \,=\, L \oplus L \oplus K_S\,.
\end{equation}
At the moment, $L$ is an arbitrary line bundle over $S$, and the
motivation behind this particular ansatz for $V$ will be clear
momentarily.

We now take $Y$ to be the hypersurface in $V$ given by
\begin{equation}\label{AMNY}
y^2 \,=\, x^2 \,+\, \alpha^{m+1}\, z^{n+1}\,,\qquad\qquad
\alpha\,\in\,H^0_{\bar\partial}\big(S, \CO_S(\Sigma)\big)\,.
\end{equation}
Here $\alpha$ is a holomorphic section of the line bundle
$\CO_S(\Sigma)$ associated to the divisor $\Sigma$, along which we
assume $\alpha$ vanishes to first-order.  The equation \eqref{AMNY} is
homogeneous if we take ${L^2= \CO_S\big((m+1)\, \Sigma\big) \otimes
  K_S^{(n+1)}}$.  Note that the holomorphic two-form $(1/y) \, dx\^dz$ on $Y$
automatically transforms as a section of $K_S$, so that $X$ is
Calabi-Yau.\footnote{As in \eqref{ADESIII}, the potentially
  half-integral nature of $L$ is an artifact associated
to our particular parametrization for the $A_n$ singularity.  If we
instead take ${xy = \alpha^{m+1} z^{n+1}}$, no square-roots are necessary.}

Away from $\Sigma$, $\alpha$ is non-vanishing, so \eqref{AMNY}
describes a generic $A_n$ singularity along $S$, identified as the
zero-section in $V$.  Therefore we have a seven-brane of type $A_n$
wrapping $S$.  On the other hand, away from $S$ itself, the normal
coordinate $z$ to ${S \subset B}$ is non-vanishing.  Because
$\alpha$  vanishes to first-order along $\Sigma$, we still find a
generic $A_m$ singularity in the fiber of $K_S$ over $\Sigma$.  Thus
we also have a seven-brane of type $A_m$ extending in the fiber over
$\Sigma$.

Over ${\Sigma \subset S}$, the $A_m$ and $A_n$ singularities collide
and enhance to an $A_{m + n + 1}$ singularity.  That is, we have a
transversal intesection of one seven-brane carrying the worldvolume gauge
group ${SU(n+1)}$ with another seven-brane carrying the worldvolume
gauge group ${SU(m+1)}$.  As usual, we expect to find
bifundamental matter localized along $\Sigma$ and transforming under
${SU(n+1) \times SU(m+1)}$ as
\begin{equation}\label{BIFUND}
\big({\bf n+1}\,, \bar{\bf m+1}\big) \oplus \big(\bar{\bf n+1}\,, {\bf
  m+1}\big)\,.
\end{equation}

The local model for $X$ determined by \eqref{AMNY} is only the
simplest in a wide class of examples.  For instance, given holomorphic
sections
\begin{align}
t_k \,&\in\, H^0_{\bar\partial}\big(S,
K_S\big)\,,& &k=1,\ldots,n+1\,, & &\sum_{k=1}^{n+1} \, t_k \,=\, 0\,,\cr
u_\ell \,&\in\, H^0_{\bar\partial}\big(S, \CO_S(\Sigma)\big)\,,&
&\ell=1,\ldots,m+1\,, & &\sum_{\ell=1}^{m+1} \, u_\ell \,=\, 0\,,
\end{align}
we can deform \eqref{AMNY} to
\begin{equation}\label{AMNYII}
y^2 \,=\, x^2 \,+\, \prod_{\ell=1}^{m+1} \left(\alpha \,+\,
  u_\ell\right) \, \prod_{k=1}^{n+1} \left(z \,+\, t_k\right)\,.
\end{equation}
The `unfolding' of the respective $A_n$ and $A_m$ singularities in equation
\eqref{AMNYII} can be interpreted physically as a separation of
the seven-branes in the original stacks wrapping $S$ and the fiber
over $\Sigma$.  Indeed, equation \eqref{AMNYII} describes
seven-branes which now wrap the divisors in the threefold $B$ given by
${z + t_k = 0}$ and ${\alpha +  u_\ell = 0}$.

Of course, if the canonical bundle $K_S$ has no holomorphic sections,
or if $\alpha$ is the only holomorphic section of $\CO_S(\Sigma)$ up
to scale, then the local model for $X$ in \eqref{AMNY} cannot be
deformed as in \eqref{AMNYII}.  However, we can still generalize
\eqref{AMNYII} as follows.

To simplify the notation, let us focus on the case ${m=0}$, for which
we expand \eqref{AMNYII} as
\begin{align}\label{AMNYIII}
y^2 \,&=\, x^2 \,+\, \alpha \, \prod_{k=1}^{n+1} \left(z \,+\,
  t_k\right)\,,\cr
&=\, x^2 \,+\, \alpha\,\Big[z^{n+1} \,+\, s_2 \, z^{n-1} \,+\,
  \cdots \,+\, s_{n+1}\Big]\,.
\end{align}
where $s_k$ denotes the elementary symmetric polynomial of degree $k$ in the
$n+1$ variables $t_{1},\cdots ,t_{n+1}$. If $m=0$ as in \eqref{AMNYIII}, then the hypersurface
$Y$ is actually non-singular along the locus where $\alpha$ vanishes,
but the corresponding elliptic fibration has a singular fiber of
Kodaira type ${\rm I}_1$.  A singularity of type ${\rm I}_1$ is the
generic singularity in an elliptic fibration, and corresponds to a
seven-brane with worldvolume gauge group $U(1)$.

Each coefficient ${\alpha \cdot s_k}$ in \eqref{AMNYIII} transforms
as a section of the line bundle $\CO_S(\Sigma) \otimes K_S^{k}$.
Even if $K_S$ itself has no holomorphic sections, such as
occurs when $K_S$ is strictly negative (the del Pezzo case), the
tensor product ${\CO_S(\Sigma) \otimes K_S^{k}}$ might admit
holomorphic sections, depending upon our choices for $\Sigma$ and $k$.
Thus, we can further generalize \eqref{AMNYIII} by taking
\begin{align}\label{BMNY}
y^2 \,&=\, x^2 \,+\, \alpha \, z^{n+1} \, +\, \widehat{\alpha}_2 \, z^{n-1} \,+\,
\cdots \,+\, \widehat{\alpha}_{n+1}\,,\cr
\widehat{\alpha}_k \,&\in\, H^0_{\bar\partial}\big(S\,, \CO_S(\Sigma) \otimes
K_S^{k}\big)\,
\end{align}
where the $\widehat{\alpha}_{k}$ are the generalizations of the deformation
parameters $\alpha_{k}$ given in \eqref{ADESDF} in the special case where $\alpha$ transforms as a section
of a trivial bundle.  As we discuss in subsection \ref{UNFOLDING}, the low-energy physics of F-theory on
the local elliptic Calabi-Yau fourfold determined by \eqref{BMNY} turns out to be quite rich, despite the fact that
\eqref{BMNY} describes the unfolding of only an $A_n$ singularity.  In fact, from
equation \eqref{AMNYIII}, we notice that the Calabi-Yau geometry remains regular even when the $s_i$ have
first order poles along the locus $\alpha=0$.  We will interpret
this effect as the coupling of certain light modes living on $\Sigma$ which source
the bulk field $\varphi$.  In particular, the residue of $\varphi$ along the
$\alpha =0$ pole will be identified with a condensate of these modes on $\Sigma$.

\bigskip\noindent{\it Intersecting Seven-Branes of Arbitrary ADE-Type}\smallskip

Since this paper is primarily concerned with exceptional singularities in F-theory,
we must also consider intersecting seven-branes of arbitrary $ADE$-type.
By analogy to the unfolding of the $A_n$ singularity over
$S$ in \eqref{BMNY}, we thus consider local models for $X$ which are
based on the unfolding of the general $ADE$-singularity.

In all cases, the essential
feature of these models is that the coefficients corresponding
to the analogues of the $\widehat{\alpha}_{k}$ in equation \eqref{BMNY}
are also holomorphic sections of appropriate line bundles
over $S$.  As these coefficients vary over $S$, the singularity
along $S$ can enhance, at which point another seven-brane intersects
the seven-brane wrapping $S$.

We now present our local models for $X$ which are based on the general
unfolding of the $D$- and $E$-type singularities.  Again, some important
numerology will be at play in the various line bundles which appear in the
constructions below.

In the case that $X$ is associated to an unfolding of a $D$-type
singularity, we take the coordinates $(x,y,z)$ on $Y$ to transform as
sections of the rank three bundle
\begin{equation}\label{RKTHVD}
V \,=\, (L \otimes K_S^{-1}) \oplus L \oplus K_S^2\,.
\end{equation}
Here $L$ is an arbitrary line bundle over $S$.  This ansatz for $V$
has been taken so that the equation for the $D_n$ singularity will be
homogeneous in $(x,y,z)$ and so that the Calabi-Yau
condition on $X$ will be satisfied.

With this ansatz for $V$, we then consider an equation for the
hypersurface $Y$ of the form
\begin{equation}\label{DX}
D_n:\qquad y^2 \,=\, - x^2 \, z \,+\, \alpha^2\, z^{n-1}\,,\qquad\qquad \alpha
\,\in\, H^0_{\bar\partial}\!\left(S,\, L \otimes K_S^{-(n-1)}\right)\,.
\end{equation}
In this equation, $\alpha$ is an arbitrary holomorphic section of ${L
  \otimes K_S^{-(n-1)}}$ which we introduce as part of the defining
data for the local Calabi-Yau fourfold $X$.   If $L$ is sufficiently
positive, then a non-trivial $\alpha$ always exists.  Of course, we
have made a special choice in assuming that $\alpha$ appears
quadratically in \eqref{DX}.  This choice is related to a `splitness'
condition for the singularity discussed in \cite{BershadskyPLUS}.  We
will motivate our choice from the perspective of topological field
theory later, but for now it is simply part of the model for $X$.

As in the preceding discussion of intersecting $A_n$ singularities,
we let $\Sigma$ be the effective divisor in $S$ along which $\alpha$
vanishes, so that
\begin{equation}
\CO_S(\Sigma) \,=\, L \otimes K_S^{-(n-1)}\,.
\end{equation}
Away from $\Sigma$, the equation in \eqref{DX} thus describes
a $D_n$ singularity over $S$.  In precise analogy to \eqref{BMNY}, we
now unfold the $D_n$ singularity in \eqref{DX} as
\begin{equation}\label{EXDN}
D_n:\qquad y^2 \,=\, - x^2 \, z \,+\, \alpha^2 \, z^{n-1} \,+\, \sum_{k=1}^{n-1}
\widehat{\delta}_{2k} \, z^{n-k-1} \,-\, 2  \, \widehat{\gamma}_n \, x\,,
\end{equation}
where:
\begin{equation}\label{EXDNII}
\widehat{\delta}_{2k} \,\in\, H^0_{\bar\partial}\big(S,\,\CO_S(2\Sigma) \otimes
K_S^{2k}\big)\,,\qquad\qquad
\widehat{\gamma}_n \,\in\, H^0_{\bar\partial}\big(S,\,\CO_S(\Sigma)\otimes
K_S^n\big)\,.
\end{equation}
Like $\alpha$, the sections $\widehat{\delta}_{2k}$ and $\widehat{\gamma}_n$ are
parameters which define the local model for $X$, and the line bundles
appearing in \eqref{EXDNII} are just determined by homogeneity
of \eqref{EXDN}.

Similarly, for the $E$-type singularities, we take $(x,y,z)$ to transform
in the rank three bundle:
\begin{equation}
V \,=\, L^2 \oplus L^3 \oplus (L \otimes K_S)\,,
\end{equation}
where $L$ is again an arbitrary line bundle on $S$.  As before, this
ansatz for $V$ is fixed by homogeneity and the Calabi-Yau condition on
$X$.

To describe a generic $E_{6,7,8}$ singularity over $S$, we consider
the hypersurface equations for $Y$,
\begin{align}
E_{6} &  \text{:}\text{ \ \ \ }y^{2}=x^{3}+\alpha^{2}z^{4}\text{
\ \ \ \ \ \ \ \ \ \ \ }\alpha\in H_{\overline{\partial}}^{0}\left(  S,L\otimes
K_{S}^{-2}\right)  \label{EX}\\
E_{7} &  \text{:}\text{ \ \ \ }y^{2}=-x^{3}+16\alpha xz^{3}\text{
\ \ \ \ \ }\alpha\in H_{\overline{\partial}}^{0}\left(  S,L\otimes K_{S}%
^{-3}\right)  \\
E_{8} &  \text{:}\text{ \ \ \ }y^{2}=x^{3}-\alpha z^{5}\text{
\ \ \ \ \ \ \ \ \ \ \ \ \ }\alpha\in H_{\overline{\partial}}^{0}\left(
S,L\otimes K_{S}^{-5}\right)  \text{.}%
\end{align}
As explained previously, the particular choice of numerical coefficients
have been chosen for ease of comparison with the results of \cite{KatzMorrison}.
In each case, away from the divisor $\Sigma$ on which $\alpha$
vanishes, $Y$ has an exceptional singularity along $S$.  In the case
of the $E_6$ singularity, we have again made a special choice
associated to the quadratic appearance of $\alpha$ in \eqref{EX} and
related to the splitness condition in \cite{BershadskyPLUS}.

We now unfold each exceptional singularity as:
\begin{align}\label{EXUNFOLD}
&E_6:\qquad y^2 \,=\, x^3 \,+\, \alpha^{2} z^4 \,+\, \widehat{\varepsilon}_2 \, x z^2
  \,+\, \widehat{\varepsilon}_5 \, x z \,+\, \widehat{\varepsilon}_6 \, z^2 \,+\, \widehat{\varepsilon}_8 \,
  x \,+\, \widehat{\varepsilon}_9 \, z \,+\, \widehat{\varepsilon}_{12}\,,\cr\cr
&E_7:\qquad y^2 \,=\, - x^3 \,+\, 16\alpha x\,z^3 \,+\, \widehat{\varepsilon}_2 \, x^2 z
  \,+\, \widehat{\varepsilon}_6 \, x^2 \,+\, \widehat{\varepsilon}_8 \, x z \,+\,\cr
&\qquad\qquad\qquad\,+\, \widehat{\varepsilon}_{10} \, z^3 \,+\, \widehat{\varepsilon}_{12} \, x
  \,+\, \widehat{\varepsilon}_{14} z \,+\, \widehat{\varepsilon}_{18}\,,\cr\cr
&E_8:\qquad y^2 \,=\, x^3 \,-\, \alpha z^5 \,+\, \widehat{\varepsilon}_2 \, x z^3
  \,+\, \widehat{\varepsilon}_8 \, x z^2 \,+\, \widehat{\varepsilon}_{12} \, z^3 \,+\,
  \widehat{\varepsilon}_{14} \, x z \,+\,\cr
&\qquad\qquad\qquad\,+\, \widehat{\varepsilon}_{18} \, z^2 \,+\,
\widehat{\varepsilon}_{20}\, x \,+\, \widehat{\varepsilon}_{24} \, z \,+\, \widehat{\varepsilon}_{30}\,.
\end{align}
Here the coefficients $\widehat{\varepsilon}_k$ transform as holomorphic
sections of various line bundles which are determined by the
homogeneity of the equations in \eqref{EXUNFOLD}.  We collect the bundle assignments for
these deformation parameters in the following tables:
\begin{equation}%
\begin{tabular}
[t]{|l|l|l|l|}\hline
$E_{6}$ &  &  & \\\hline
& $\widehat{\varepsilon}_{2}\in\mathcal{O}_{S}(2\Sigma)\otimes K_{S}^{2}$ &
$\widehat{\varepsilon}_{5}\in\mathcal{O}_{S}(3\Sigma)\otimes K_{S}^{5}$ &
$\widehat{\varepsilon}_{6}\in\mathcal{O}_{S}(4\Sigma)\otimes K_{S}^{6}%
$\\\hline
& $\widehat{\varepsilon}_{8}\in\mathcal{O}_{S}(4\Sigma)\otimes K_{S}^{8}$ &
$\widehat{\varepsilon}_{9}\in\mathcal{O}_{S}(5\Sigma)\otimes K_{S}^{9}$ &
$\widehat{\varepsilon}_{12}\in\mathcal{O}_{S}(6\Sigma)\otimes K_{S}^{12}%
$\\\hline
\end{tabular}
\label{E6UNFOLD}%
\end{equation}%
\begin{equation}%
\begin{tabular}
[c]{|l|l|l|l|}\hline
$E_{7}$ &  &  & \\\hline
& $\widehat{\varepsilon}_{2}\in\mathcal{O}_{S}(\Sigma)\otimes K_{S}^{2}$ &
$\widehat{\varepsilon}_{6}\in\mathcal{O}_{S}(2\Sigma)\otimes K_{S}^{6}$ &
$\widehat{\varepsilon}_{8}\in\mathcal{O}_{S}(3\Sigma)\otimes K_{S}^{8}%
$\\\hline
& $\widehat{\varepsilon}_{10}\in\mathcal{O}_{S}(3\Sigma)\otimes K_{S}^{10}$ &
$\widehat{\varepsilon}_{12}\in\mathcal{O}_{S}(4\Sigma)\otimes K_{S}^{12}$ &
$\widehat{\varepsilon}_{14}\in\mathcal{O}_{S}(5\Sigma)\otimes K_{S}^{14}%
$\\\hline
& $\widehat{\varepsilon}_{18}\in\mathcal{O}_{S}(6\Sigma)\otimes K_{S}^{18}$ &
& \\\hline
\end{tabular}
\label{E7UNFOLD}%
\end{equation}%
\begin{equation}%
\begin{tabular}
[c]{|l|l|l|l|}\hline
$E_{8}$ &  &  & \\\hline
& $\widehat{\varepsilon}_{2}\in\mathcal{O}_{S}(\Sigma)\otimes K_{S}^{2}$ &
$\widehat{\varepsilon}_{8}\in\mathcal{O}_{S}(2\Sigma)\otimes K_{S}^{8}$ &
$\widehat{\varepsilon}_{12}\in\mathcal{O}_{S}(3\Sigma)\otimes K_{S}^{12}%
$\\\hline
& $\widehat{\varepsilon}_{14}\in\mathcal{O}_{S}(3\Sigma)\otimes K_{S}^{14}$ &
$\widehat{\varepsilon}_{18}\in\mathcal{O}_{S}(3\Sigma)\otimes K_{S}^{18}$ &
$\widehat{\varepsilon}_{20}\in\mathcal{O}_{S}(4\Sigma)\otimes K_{S}^{20}%
$\\\hline
& $\widehat{\varepsilon}_{24}\in\mathcal{O}_{S}(5\Sigma)\otimes K_{S}^{24}$ &
$\widehat{\varepsilon}_{30}\in\mathcal{O}_{S}(6\Sigma)\otimes K_{S}^{30}$ &
\\\hline
\end{tabular}
\label{E8UNFOLD}%
\end{equation}
A non-trivial feature of the above bundle assignments
is that the power of $K_{S}$ appearing in each entry
\textit{again} precisely matches to the degrees of the primitive Casimir invariant.  The discrepancy by an overall factor of $\mathcal{O}_{S}(n \Sigma)$ anticipates the beautiful match we shall find between condensates of fields localized on $\Sigma$ and the geometric unfolding of the singularity.

The class of local, elliptic Calabi-Yau fourfolds described by equations
\eqref{BMNY}, \eqref{EXDN}, and \eqref{EXUNFOLD} is quite broad, and
serves as a significant extension over the very restricted class of
models studied in Section \ref{SEVENBRANE}.  Given such a local
Calabi-Yau fourfold $X$, one of our primary goals in the rest of the
paper will be to determine the low-energy effective description for F-theory
compactified on $X$.

\subsection{Topological Field Theory on a Defect}\label{DEFECT}

The local geometries for the elliptically-fibered Calabi-Yau fourfold
$X$ introduced in Section \ref{COLLIDE} correspond in F-theory to
rather complicated configurations of intersecting seven-branes inside
the threefold $B$ sitting at the base of $X$.  So if we wish to
analyze F-theory on $X$, we clearly need to know something about the
low-energy effective description for intersecting seven-branes in
F-theory.

For concreteness, let us consider two seven-branes wrapping smooth
complex surfaces $S$ and $S'$ in $B$ such that $S$ and $S'$ intersect
transversely along a complex curve ${\Sigma = S \cap S'}$.  In
general, $\Sigma$ might be reducible and consist of several components
which themselves intersect at points inside $S$.  However, to keep
matters simple, we suppose that $\Sigma$ is an irreducible, smooth
curve.  In Section \ref{MULTINT}, we consider what happens when
$\Sigma$ becomes reducible.

As explained in Section \ref{SEVENBRANE}, on each of $S$ and $S'$
is a twisted Yang-Mills theory which captures the effective dynamics
for the low-energy degrees of freedom living on a seven-brane in
F-theory.  But just as for ordinary D7-branes, we now expect
additional light degrees of freedom to be localized along the subspace
${\BR^{3,1}\times \Sigma}$ where the seven-branes wrapping $S$ and
$S'$ intersect.  These light degrees of freedom on ${\BR^{3,1} \times
  \Sigma}$ are then described by an effective defect theory coupled to
the bulk Yang-Mills theories on ${\BR^{3,1} \times S}$ and ${\BR^{3,1}
  \times S'}$.

If the defect theory on ${\BR^{3,1} \times \Sigma}$ is to preserve
${\CN=1}$ supersymmetry in four dimensions, the defect theory must be
twisted along $\Sigma$ for the same reason that the bulk Yang-Mills
theory on ${\BR^{3,1} \times S}$ is twisted along $S$.  Though the
corresponding defect theories that live on the intersections of
D-branes in Minkowski and Anti-de Sitter space have been studied for example in
\cite{DefectKirsch,DefectKapustinSethi,DefectOoguriFreedman}, little attention
has been paid to their twisted relatives.  For this reason, we now establish some
basic facts about the structure of the partially twisted theory on ${\BR^{3,1}\times\Sigma}$.

\subsubsection{Topological Twist on $\Sigma$}

We first identify the light defect degrees of freedom that
propagate on ${\BR^{3,1} \times \Sigma}$.  The simplest case to
consider is ${\Sigma = \BC}$, corresponding to F-theory compactified
to $\BR^{5,1}$ on an elliptically-fibered Calabi-Yau threefold.  Such
an F-theory background preserves ${\CN=2}$ supersymmetry in four
dimensions, and the massless charged matter arising from intersecting
seven-branes in that situation was analyzed in \cite{KatzVafa}.

To recall the result of \cite{KatzVafa}, we suppose that the
seven-brane wrapping $S$ carries a worldvolume gauge group $G_S$, and
the seven-brane wrapping $S'$ carries a worldvolume gauge group
$G_{S'}$.  Both $G_S$ and $G_{S'}$ are simply-laced
Lie groups associated to $ADE$ singularities along $S$ and $S'$.  We
also allow the possibility that either $G_S$ or $G_{S'}$ is $U(1)$,
corresponding to a Kodaira fiber of type ${\rm I}_1$ over $S$ or
$S'$.  At the intersection of $S$ and $S'$ over $\Sigma$, the
singularities generically enhance to a new singularity associated to a
simply-laced Lie group $G_\Sigma$, where
\begin{equation}\label{GSIGMA}
G_S \times G_{S'} \,\subset\, G_\Sigma\,.
\end{equation}
According to \cite{KatzVafa}, a massless hypermultiplet charged under
${G_S \times G_{S'}}$ then propagates along the seven-brane intersection.

In the case ${G_S = SU(n+1)}$, ${G_{S'}=SU(m+1)}$, and ${G_\Sigma =
  SU(m+n+2)}$, the charged hypermultiplet decomposes into a pair of
${\CN=1}$ chiral multiplets which transform as bifundamentals
of ${G_S \times G_{S'}}$.  To extend the notion of
``bifundamental'' matter to the general $ADE$ setting, we decompose the
adjoint representation of $G_\Sigma$ under ${G_S \times
  G_{S'}}$ as
\begin{equation}\label{BIFUNDT}
\ad(G_\Sigma) \,=\, \ad(G_S) \oplus \ad(G_{S'}) \oplus
\left[\bigoplus_j \, U^{}_j \otimes U'_j\right].
\end{equation}
Here $U^{}_j$ and $U'_j$ are irreducible representations
of $G_S$ and $G_{S'}$, and `$j$' is a dummy index running over
whatever summands appear in the decomposition above.  Under the
decomposition of the hypermultiplet into ${\CN=1}$ chiral
multiplets, the light ``bifundamental'' matter localized
along $\Sigma$ then transforms in the representation of ${G_S \times
  G_{S'}}$ given by the non-adjoint summand of \eqref{BIFUNDT}, namely
\begin{equation}\label{BIFUNDII}
\bigoplus_j \, U^{}_j \otimes U'_j\,.
\end{equation}

If $\Sigma$ is not $\BC$ but rather a compact complex curve, we apply the
adiabatic argument as in Section \ref{SEVENBRANE} to conclude that the
massless fields which propagate on ${\BR^{3,1} \times \Sigma}$
correspond to a twisted version of the charged hypermultiplet propagating
on $\BR^{5,1}$.  Once again, the requirement of ${\CN=1}$
supersymmetry in four dimensions leaves us with no choice about how to
twist the hypermultiplet.

To specify the twist, we recall that the untwisted
hypermultiplet on $\BR^{5,1}$ contains a pair of complex bosons
$(\sigma\,, \bar\sigma^c)$ and a negative-chirality\footnote{Our choice of
chirality here may seem a bit strange, but it makes certain
conventions about holomorphy more natural later.} Weyl fermion,
which transforms in the ${\bf 4}'$ of $SO(5,1)$.  Under the gauge
group ${G_S \times G_{S'}}$ in \eqref{GSIGMA}, we take the bosons and
fermions in the hypermultiplet to transform in the representation
${U\otimes U'}$, corresponding to one of the summands in \eqref{BIFUNDII}.
If all kinetic terms are canonical, the hypermultiplet also respects a
global $SU(2)_R$ symmetry under which $(\sigma\,,\bar\sigma^c)$
transform as a doublet, and the Weyl fermion transforms as a singlet.
Finally, the supersymmetry generator $\epsilon$ transforms under
${SO(5,1) \times SU(2)_R}$ in the representation ${{\bf 4}' \otimes {\bf 2}}$.\footnote{The generator
$\epsilon$ also obeys a reality condition which will play no role in the discussion to follow.}

To twist the hypermultiplet on $\Sigma$, we reduce the global
$SO(5,1)$ symmetry to ${SO(3,1) \times U(1)}$, where the $U(1)$ factor
is to be identified with the structure group of the tangent bundle on
$\Sigma$.  Under the reduction to ${SO(3,1) \times U(1)}$, the
negative-chirality spinor of $SO(5,1)$ decomposes as
\begin{equation}\label{DECOM4}
{\bf 4}' \,\longmapsto\, \left({\bf 2}\,,{\bf 1}\,, -\ha\right) \oplus
\left({\bf 1}\,,{\bf 2}\,,+\ha\right)\,.
\end{equation}

To specify the bundle assignments of the twisted theory, note that
the ${\bf 2}$ of $SU(2)_R$ decomposes to the Cartan $U(1)_R$ subgroup as:
\begin{equation}
{\bf 2} \,\longmapsto\, {\bf 1}_{+1} \oplus {\bf 1}_{-1}\,.
\end{equation}
The subscripts above indicate the charges under the generator $R$ of
$U(1)_R$.  Equivalently, the distinguished $U(1)_R$ subgroup of
$SU(2)_R$ can be identified with the R-symmetry used to partially twist the
bulk eight-dimensional Yang-Mills theory described in Section \ref{SEVENBRANE}.

The twist of the hypermultiplet is now specified by a homomorphism
from $U(1)_R$ to the $U(1)$ factor in ${SO(3,1) \times U(1)}$.  Let
$J$ be the generator of the $U(1)$ factor in ${SO(3,1) \times U(1)}$
normalized according to \eqref{DECOM4}.  In order to preserve
${\CN=1}$ supersymmetry on $\BR^{3,1}$, half of the original eight
supersymmetries generated by $\epsilon$ must transform as scalars on
$\Sigma$ once we twist.  As one can check, this requirement implies
that the generator $J_{top}$ of the twisted $U(1)$ must be
\begin{equation}
J_{top} \,=\, J \,\pm\, \ha R\,.
\end{equation}
Either choice of sign above leads to an isomorphic twist, so we take
${J_{top} = J - \ha R}$ without loss of generality.

Because the fermions in the hypermultiplet transform trivially
under $SU(2)_R$, the twist by ${U(1)_R \subset SU(2)_R}$ does not
alter their geometric interpretation on $\Sigma$.  According to
\eqref{DECOM4}, the twisted hypermultiplet on
${\BR^{3,1}\times\Sigma}$ therefore contains fermions
$(\lambda_\alpha\,,\bar\lambda{}^c_{\dot\alpha})$ transforming as
spinors on the curve $\Sigma$,
\begin{align}\label{HYPF}
&\lambda^{}_\alpha\qquad \hbox{section of }\, K^{1/2}_\Sigma
\otimes \SU \otimes \SU'\,,\cr
&\bar\lambda{}^c_{\dot\alpha}\qquad \hbox{section of }\, \bar K^{1/2}_\Sigma
\otimes \SU \otimes \SU'\,.
\end{align}

In \eqref{HYPF}, $K^{1/2}_\Sigma$ denotes a square-root of the
canonical bundle $K_\Sigma$ on $\Sigma$.  In general, the choice of a
square-root for $K_\Sigma$ is not unique.  If $\Sigma$ has genus $g$,
then $\Sigma$ admits $2^{2g}$ distinct spin structures, and we must
identify precisely which one we pick to define $K^{1/2}_\Sigma$.  As we
explain in Section \ref{COSMIC}, $\Sigma$ inherits a distinguished
spin structure from its embedding in the surface $S$ (or equally well
from its embedding in $S'$), and the distinguished spin structure
defines $K^{1/2}_\Sigma$.  Also, $\SU$ and $\SU'$ are vector bundles on
$\Sigma$ associated to the representations $U$ and $U'$.  As one
expects, $\SU$ and $\SU'$ are determined by the restriction to
$\Sigma$ of principal bundles $P$ and $P'$ over the respective
surfaces $S$ and $S'$.

Even though the fermions in the hypermultiplet are not affected by the
twist on $\Sigma$, the bosons in the hypermultiplet are.  Under the
generator $J_{top}$, $\sigma$ and $\bar\sigma^c$ carry respective
charges $\mp \ha$, inherited from their charges under $U(1)_R$.  As a
result, $\sigma$ and $\bar\sigma^c$ also transform as spinors on $\Sigma$,
\begin{align}\label{HYPB}
&\sigma\qquad\, \hbox{section of }\, K^{1/2}_\Sigma
\otimes \SU \otimes \SU'\,,\cr
&\bar\sigma^c\qquad \hbox{section of }\, \bar K^{1/2}_\Sigma
\otimes \SU \otimes \SU'\,.
\end{align}

Finally, as is often useful when we discuss the hypermultiplet in the
language of ${\CN=1}$ supersymmetry, we introduce the complex boson
$\sigma^c$ and the fermion $\lambda_\alpha^c$ which are the CPT
conjugates of $\bar\sigma^c$ and $\bar\lambda{}^c_{\dot\alpha}$,
\begin{align}\label{HYPC}
&\sigma^c\qquad \hbox{section of }\, K^{1/2}_\Sigma
\otimes \SU^* \otimes (\SU')^*\,,\cr
&\lambda^c_\alpha\qquad \hbox{section of }\, K^{1/2}_\Sigma
\otimes \SU^* \otimes (\SU')^*\,.
\end{align}
Here $\SU^*$ and $(\SU')^*$ are the bundles
associated to the dual representations $U^*$ and $(U')^*$.  Of course,
if ${U \otimes U'}$ appears as a summand in \eqref{BIFUNDII}, then so
too does ${U^* \otimes (U')^*}$, since the representation in
\eqref{BIFUNDII} is necessarily real.

\subsubsection{On the Supersymmetric Defect Action}

We are still left to determine the supersymmetric action for the
twisted defect theory on ${\BR^{3,1} \times S}$.  In particular, we
need to consider how the defect fields on $\Sigma$ couple to the bulk
Yang-Mills fields on $S$ and $S'$.  As will be clear, supersymmetry
and gauge-invariance leave us little choice as to how the coupling can
be done.

Like the partially twisted Yang-Mills theory discussed in Section
\ref{SEVENBRANE}, the twisted fields in the hypermultiplet on
${\BR^{3,1} \times \Sigma}$ naturally arrange themselves into the
standard representations of the ${\CN=1}$ supersymmetry algebra on
$\BR^{3,1}$.  Thus from the field content in \eqref{HYPF},
\eqref{HYPB}, and \eqref{HYPC}, we clearly obtain ${\CN=1}$ chiral
multiplets
\begin{equation}\label{BIGS}
\left(\sigma,\,\lambda^{}_\alpha\right),\qquad\qquad
\left(\sigma^c,\,\lambda^c_\alpha\right),
\end{equation}
along with the CPT-conjugate ${\CN=1}$ anti-chiral multiplets, which will
not play a role in the following.

We are now in a slightly unusual situation.  In the most common
topological field theories, the on-shell supersymmetries of the
ten-dimensional Yang-Mills theory immediately determine the on-shell supersymmetries of the
lower-dimensional twisted theory.  From the on-shell twisted
supersymmetries, one then tries to write an appropriately twisted
version of the super Yang-Mills action.

But in the case at hand, we do not know {\it a priori} how the on-shell
supersymmetries should act on the chiral multiplets in \eqref{BIGS},
since we do not yet know precisely how the defect theory will couple to
the bulk Yang-Mills theories on $S$ and $S'$.  However, we do have an
(essentially) off-shell formulation for the ${\CN=1}$ supersymmetry
algebra in our problem.  Inverting the usual order of analysis, in
Appendix \ref{6DACTION} we therefore construct the most general
supersymmetric off-shell action for the six-dimensional defect
theory.  We then integrate out auxiliary fields to determine the
on-shell supersymmetry transformations and associated BPS equations
for the system.  The details of the analysis in Appendix
\ref{6DACTION} are mostly unimportant, but the analysis does yield two
important results.

\bigskip\noindent{\it The Defect Superpotential}\smallskip

First, the off-shell defect action, including couplings to the bulk
Yang-Mills fields on $S$ and $S'$, can be written concisely in
superspace as
\begin{equation}\label{IDEFECT}
I_\Sigma \,=\, \underset{\BR^{3,1} \times \Sigma}{\int} \!\! d^4x \,
d^2\bar\theta \,\, \CO \ + \ \underset{\BR^{3,1} \times \Sigma}{\int}
\!\! d^4x \, d^2\theta \; W_\Sigma\,,
\end{equation}
where
\begin{equation}\label{WDEFECT}
W_\Sigma \,=\, \left\langle{\mathbf \Lambda}^c \,, \bar\partial_{{\bf A}
  \,+\, {\bf A}'} {\mathbf \Lambda}\right\rangle\text{.}
\end{equation}
The notation of the above expression is explained below. In
\eqref{IDEFECT}, $\CO$
is a gauge-invariant operator whose particular form does not much matter once we restrict attention to the
cohomology of $\bar Q{}_{\dot\alpha}$.  As we demonstrate in Appendix
\ref{6DACTION}, the terms in $I_\Sigma$ derived from $\CO$ are the standard
kinetic terms on $\BR^{3,1}$ for the defect fields, along with
couplings conjugate to those derived from the superpotential in
\eqref{IDEFECT}.

The superpotential $W_\Sigma$ is clearly the object of interest in
$I_\Sigma$.  To explain our notation, ${\mathbf \Lambda}$ and
${\mathbf \Lambda}^c$ are chiral superfields associated to the chiral
multiplets  in \eqref{BIGS}, so that
\begin{align}\label{BIGLAMBDA}
{\mathbf \Lambda} \,&=\, \sigma \,+\, \sqrt{2} \theta \lambda \,+\, \theta\theta \CK \,+\, \cdots\,,\cr
{\mathbf \Lambda}^c \,&=\, \sigma^c \,+\, \sqrt{2} \theta \lambda^c \,+\, \theta\theta \CK^c \,+\,\cdots\,.
\end{align}
Here $\CK$ and $\CK^c$ are auxiliary bosonic fields that transform
like $\sigma$ and $\sigma^c$ as sections of $K^{1/2}_\Sigma \otimes
\SU \otimes \SU'$ and $K^{1/2}_\Sigma \otimes \SU^* \otimes (\SU')^*$,
and the `$\cdots$' indicate additional terms involving
$\bar\theta{}^{\dot\alpha}$ which do not play a role in our discussion
of $W_\Sigma$.  The pairing $\langle\,\cdot\,,\,\cdot\,\rangle$
between ${\mathbf \Lambda}$ and ${\mathbf \Lambda}^c$ in $W_\Sigma$ is
the canonical pairing between sections of $\SU \otimes \SU'$ and the dual.
 Recall that from the four-dimensional perspective, the
$(0,1)$ component of the gauge field along $S$ appears as the lowest
bosonic component of the chiral superfield described in Section
\ref{SEVENBRANE},
\begin{equation}\label{BIGA}
{\bf A}_{\bar m} \,=\, A_{\bar m} \,+\, \sqrt{2} \theta \psi_{\bar m}
\,+\, \theta\theta \CG_{\bar m} \,+\, \cdots\,,
\end{equation}
where $\CG_{\bar m}$ is an auxiliary bosonic field transforming on $S$ as a
section of ${\bar\Omega{}^1_S \otimes \ad(P)}$.  Finally, the covariant
derivative $\bar\partial_{{\bf A} + {\bf A}'}$ appearing in
\eqref{IDEFECT} is meant to be interpreted literally in superspace as
\begin{align}\label{BIGDA}
\bar\partial_{{\bf A} + {\bf A}'} \,&=\, \bar\partial \,+\, {\bf A}
\,+\, {\bf A}'\,,\cr
&=\, \bar\partial_{A + A'} \,+\, \sqrt{2} \theta \!\left(\psi \,+\, \psi'\right) \,+\, \theta\theta \!\left(\CG \,+\, \CG'\right)\,+\, \cdots\,.
\end{align}

Given \eqref{BIGLAMBDA}, \eqref{BIGA}, and \eqref{BIGDA}, we can
immediately work out the component expansion for the superpotential in
$I_\Sigma$.  However, before we do so, let us note that $W_\Sigma$ is
quite special.  Due to the twisting of ${\mathbf \Lambda}$ and
${\mathbf \Lambda}^c$ as spinors on $\Sigma$, the superpotential in
\eqref{WDEFECT} itself transforms as a differential form of type
$(1,1)$ on $\Sigma$.  Hence $W_\Sigma$ can be naturally integrated
over $\Sigma$, without reference to a metric on the curve.  As a
result, when we compactify the defect theory to $\BR^{3,1}$ in Section
\ref{6DMATTER}, $W_\Sigma$ leads immediately to effective Yukawa
couplings in four dimensions which involve only holomorphic data on
$\Sigma$.

\bigskip\noindent{\it BPS Equations}\smallskip

The second important result obtained in Appendix \ref{6DACTION} is a derivation
of the conditions for unbroken supersymmetry when the defect theory on
${\BR^{3,1} \times \Sigma}$ is coupled to the bulk Yang-Mills theories on
${\BR^{3,1} \times S}$ and ${\BR^{3,1} \times S'}$ via the action in
\eqref{IDEFECT}.

From the perspective of the present paper, the most interesting BPS
equations are the F-term equations associated to the auxiliary fields
which enter $W_\Sigma$.  Explicitly, we expand in components
\begin{align}\label{COMPW}
 \underset{\BR^{3,1} \times \Sigma}{\int}
\!\! d^4x \, d^2\theta \; W_\Sigma \; &=\,
\underset{\BR^{3,1}\times\Sigma}{\int} d^4 x\,\Big[\left\langle
  \mathcal{K}^{c},\overline{\partial}_{A+A^{\prime}}%
\sigma\right\rangle+\left\langle \sigma^{c},\overline{\partial
}_{A+A^{\prime}}\mathcal{K}\right\rangle+\left\langle \sigma
^{c},(\mathcal{G}+\mathcal{G}^{\prime})\cdot\sigma\right\rangle\cr
&\qquad\,-\,\left\langle\lambda^c,\bar\partial_{A+A'}\lambda
\right\rangle \,-\, \left\langle\sigma^c,(\psi+\psi')\cdot\lambda\right\rangle
\,-\, \left\langle\lambda^c,(\psi+\psi')\cdot\sigma\right\rangle\Big]\,.
\end{align}
Here in expressions such as $\CG\cdot\sigma$, we indicate the linear
action by elements in the Lie algebra of the group $G_S$ on the
representation $U$, and similarly for the action of $G_{S'}$ on the
representation $U'$.

Because $\CK$ and $\CK^c$ are the auxiliary bosonic fields appearing
in the defect chiral superfields ${\mathbf \Lambda}$ and ${\mathbf
  \Lambda}^c$, the linear terms involving $\CK$ and $\CK^c$ in
\eqref{COMPW} immediately imply the F-term supersymmetry conditions
\begin{equation}\label{HOLSIG}
\bar\partial{}_{A + A'}\,\sigma \,=\, \bar\partial{}_{A + A'} \, \sigma^c
\,=\, 0\,.
\end{equation}
Thus $\sigma$ and $\sigma^c$ must be holomorphic as sections of the
respective bundles $K^{1/2}_\Sigma \otimes \SU \otimes \SU'$ and
$K^{1/2}_\Sigma \otimes \SU^* \otimes (\SU')^*$ on $\Sigma$.

The essentially new ingredient in \eqref{COMPW} is the coupling of the
auxiliary bulk fields $\CG$ and $\CG'$ appearing in ${\bf A}$ and ${\bf
  A}'$ to the defect fields $\sigma$ and $\sigma^c$.  In the absence
of the defect, the F-term supersymmetry condition associated to $\CG$
(and similarly to $\CG'$) merely states that
\begin{equation}\label{ORIGHOLP}
\bar\partial{}_A \varphi \,=\, 0\,,
\end{equation}
so that $\varphi$ is holomorphic as a section of
${K_S\otimes\ad(P)}$ on $S$.

However, in the presence of the defect, the linear coupling to $\CG$
in equation \eqref{COMPW} induces a source term for the BPS equation
in \eqref{ORIGHOLP}, so that the new supersymmetry condition on
$\varphi$ becomes
\begin{equation}\label{MODBPSI}
\bar\partial_A \varphi \,=\, \delta_\Sigma \,
\llangle\sigma^{c},\sigma\rrangle_{\ad(P)}\,.
\end{equation}
Here $\delta_\Sigma$ is a two-form on $S$ with delta-function support
along $\Sigma$ which represents the Poincar\'e dual of $\Sigma$.  In
particular, because $\Sigma$ is a holomorphic curve, $\delta_\Sigma$ has
holomorphic/anti-holomorphic type $(1,1)$ on $S$, so that both sides
of \eqref{MODBPSI} are associated to differential forms of type
$(2,1)$ on $S$.

We also introduce in \eqref{MODBPSI} the natural
`outer-product' determined by the action of $G$ on $U$,
\begin{equation}
\llangle\,\cdot\,,\,\cdot\,\rrangle_{\ad(P)}: \big[\SU^* \otimes
  (\SU')^*\big] \otimes \big[\SU \otimes \SU'\big]
\,\longrightarrow\, \ad(P)\big|_\Sigma\,,
\end{equation}
obtained from the individual pairings
\begin{equation}\label{PAIRII}
\SU^* \otimes \SU \,\longrightarrow\, \ad(P)\big|_\Sigma\,,\qquad\qquad (\SU')^*
\otimes \SU' \,\longrightarrow\, \CO_\Sigma\,.
\end{equation}
Explicitly, if $(T^I)^a_{a'}$ for ${I=1,\ldots,\dim(G_S)}$ represent the
generators of $G_S$ acting on $U$ in a given basis $\{u^a\}$, then locally
${\llangle\sigma^c,\sigma\rrangle_{\ad(P)} \,=\,\sigma^c_a \,
  (T^I)^a_{a'} \, \sigma^{a'}}$.

As will be very important in Section \ref{UNFOLDING}, configurations
of $\varphi$
which satisfy equation \eqref{MODBPSI} have a very natural interpretation in algebraic
geometry.  Namely, the delta-function source appearing on the right in equation \eqref{MODBPSI} implies
that $\varphi$ is now a \textit{meromorphic} section of $K_S \otimes
\ad(P)$ which has a first-order pole along $\Sigma$, with residue
given by $\llangle\sigma^{c},\sigma\rrangle_{\ad(P)}$. Thus, if $\Sigma$
is determined by the vanishing of $\alpha$ as in Section
\ref{COLLIDE}, then $\varphi$ appears locally near $\Sigma$ as
\begin{equation}\label{POLEPHI}
\varphi = \,\frac{\llangle\sigma^c,\sigma\rrangle_{\ad(P)} \, ds^1\^ds^2}{\alpha}
+ \cdots\,,\end{equation}
where $(s^1,s^2)$ are local holomorphic coordinates near
${\Sigma\subset S}$, and the `$\cdots$' indicate terms which are
regular in $\varphi$.

We will have quite a bit more to say about equation \eqref{MODBPSI} in Section
\ref{UNFOLDING}, but for now let us record the rest of the BPS
equations in the coupled defect and Yang-Mills theories.  The presence
of the defect on $\Sigma$ does not change the conditions
\begin{equation}\label{OLDHOLP}
F_S^{(0,2)} = F_S^{(2,0)} = 0\text{.}
\end{equation}
Thus the gauge field on $S$ still endows the principal $G_S$-bundle
$P$ with a holomorphic structure in the presence of the defect.
However, the defect does induce source terms in the D-term equation
for the gauge field, so that the new D-term equation becomes
\begin{equation}\label{MODDTERM}
\omega\wedge F_{S}+\frac{i}{2}[\varphi,\overline{\varphi}] \,=\,
\ha \, \omega\^\delta_{\Sigma } \, \Big[\mu(\bar\sigma\,, \sigma) \,-\,
\mu(\bar\sigma^c\,,\sigma^c)\Big]\,.
\end{equation}

In \eqref{MODDTERM}, we have introduced the moment map $\mu$
associated to the action of $G_S$ on the representation $U$.
Thus,
\begin{equation}
\mu(\cdot,\cdot):\, \Big[\overline{\SU} \otimes \overline{\SU'}\Big]
\otimes \Big[\SU \otimes \SU'\Big] \,\longrightarrow\,
\ad(P)\big|_\Sigma\,,
\end{equation}
and dually when $\mu$ is evaluated on $\SU^* \otimes (\SU')^*$.  The
moment map $\mu$ is closely related to the canonical outer-product
$\llangle\,\cdot\,,\,\cdot\,\rrangle_{\ad(P)}$ that we introduced earlier, but
$\mu$ is defined using a hermitian metric on the bundles $\SU$ and
$\SU'$.  This hermitian metric is also used to define the kinetric terms
on $\BR^{3,1}$ for $\sigma$ and $\sigma^c$, as in Appendix \ref{6DACTION}.
Explicitly, in terms of the local generators $(T^I)^a_{a'}$ which we
introduced to describe the outer-product
$\llangle\,\cdot\,,\,\cdot\,\rrangle_{\ad(P)}$, the moment map is
given as usual by ${\mu(\bar\sigma,\sigma)=\bar\sigma{}^{\bar a}
  (T^I)_{\bar a \, a} \sigma^a}$.

Finally, for each BPS equation involving the gauge field $A$ or scalar
$\varphi$ in \eqref{MODBPSI}, \eqref{OLDHOLP}, and \eqref{MODDTERM},
we obtain completely parallel BPS equations for the gauge field
$A'$ and the scalar $\varphi'$ on the surface $S'$.

\subsubsection{The Defect as a Cosmic String}\label{COSMIC}

So far, we have used indirect arguments to determine the structure of
the defect theory on ${\BR^{3,1} \times \Sigma}$.  Those arguments
were based upon the adiabatic extension of older ideas \cite{KatzVafa}
about the charged matter present in F-theory compactifications
on Calabi-Yau threefolds, along with the requirement of ${\CN=1}$
supersymmetry in four dimensions.  However, with a bit more work, the
structure of the defect theory on ${\BR^{3,1}\times\Sigma}$ can also
be determined directly from the partially twisted Yang-Mills theory on
${\BR^{3,1}\times S}$ itself.  Furthermore, as we explain, the
analysis from the perspective of $S$ identifies the proper spin
structure on $\Sigma$ with which to define the bundle $K_\Sigma^{1/2}$.

The analysis we are about to perform is not really new.  In the
context of F-theory compactifications on Calabi-Yau threefolds,
precisely the same arguments were originally used in \cite{KatzVafa}
to deduce the presence of charged hypermultiplets on $\BR^{5,1}$, a
result which is the basic ingredient in our previous adiabatic
analysis.  Essentially the same ideas have also appeared in \cite{Witten4M},
where a twisted version of four-dimensional, ${\CN=1}$ supersymmetric
Yang-Mills theory on a K\"ahler manifold is considered.  Nonetheless,
the following observations do serve to illuminate the relationship
between the bulk Yang-Mills theory on $S$ and the defect theory on
$\Sigma$, and therefore they seem worthwhile to review.

To start, let us assume that the holomorphic surface $S$ satisfies
$h^{2,0}(S) \neq 0$, so that $K_S$ admits holomorphic sections and
${S \subset B}$ is {\em not} rigid.  In this case, if we start with a
seven-brane with worldvolume gauge group $G_S$ which wraps $S$, then to
obtain a configuration of intersecting seven-branes in $B$, we can
simply turn on a holomorphic expectation value for the twisted scalar field
$\varphi$.

Specifically, we assume that $\varphi$ takes the abelian form
\begin{equation}\label{ABPHI}
\varphi \,=\, \varphi_0\, t\,,\qquad\qquad \varphi_0
\,\in\,H^0_{\bar\partial}\big(S, K_S\big)\,,\qquad\qquad t\,\in\,\ad(G_S)\,.
\end{equation}
Here $\varphi_0$ is a non-trivial element in
$H^0_{\bar\partial}\big(S, K_S\big)$, and $t$ is a fixed generator in
the Lie algebra of $G_S$.  In this very simple background, we also take
the connection on $P$ to be trivial, so that ${F_S = 0}$ and the
D-term equation on $S$ is automatically satisfied.

Simple though the background described by \eqref{ABPHI} may be, it
still describes a configuration of intersecting seven-branes in $B$.
The expectation value in \eqref{ABPHI} generically breaks the
simply-laced group $G_S$ to a product
\begin{equation}\label{HIGGSI}
\Gamma_S \times U(1)\subset G_S\,,
\end{equation}
where the $U(1)$ factor is generated by $t$, and $\Gamma_S$ is
generated by the other elements in the Lie algebra of $G_S$ which commute
with $t$.  Equivalently, from the perspective of the local
Calabi-Yau fourfold $X$ introduced in Section \ref{SEVENBRANE} to
describe the original seven-brane wrapping ${\BR^{3,1} \times S}$ with
gauge group $G_S$, the expectation value for $\varphi$ determines a
deformation of $X$ as in \eqref{ADESDF}, such that the generic
singularity along $S$ unfolds to the singularity corresponding to the
simply-laced group $\Gamma_S$.  Finally, from the physical
perspective, the expectation value for $\varphi$ implies that some
seven-branes have been moved away from the original stack of
seven-branes wrapping $S$, such that the gauge group $G_S$ is
generically broken as in \eqref{HIGGSI}.

The qualifier ``generic'' in the preceding statement is essential.
Unless $K_S$ is trivial, $\varphi$ vanishes over a curve ${\Sigma \subset S}$
representing the canonical divisor of $S$.  At points in $\Sigma$
where $\varphi$ vanishes, the gauge group $G_S$ is unbroken, the
generic singularity along $S$ enhances, and the seven-brane
represented by the $U(1)$ factor in \eqref{HIGGSI} intersects the
other seven-branes associated to the generic singularity of type
$\Gamma_S$ along $S$.  Hence a configuration of intersecting
seven-branes with respective worldvolume gauge groups $\Gamma_S$
and $U(1)$ can be described purely in terms of the twisted Yang-Mills
theory on ${\BR^{3,1} \times S}$ with gauge group $G_S$, such that
$\varphi$ has the expectation value in \eqref{ABPHI}.

To determine the structure of the defect theory along $\Sigma$, we
therefore want to consider what happens in the eight-dimensional
Yang-Mills theory when $\varphi$ has a non-trivial, holomorphic
expectation value.  Certainly away from the curve ${\Sigma \subset S}$
where $\varphi$ vanishes, the usual Higgs mechanism operates, and the
bifundamental components of the Yang-Mills multiplet on $S$ are
massive.  Here we use ``bifundamental'' in the general sense that it is
defined in \eqref{BIFUNDT} and \eqref{BIFUNDII}.

Over $\Sigma$, something more interesting happens.  As discussed in
\cite{Witten4M}, $\Sigma$ appears from the four-dimensional
perspective of $S$ as a kind of global cosmic string associated to
the vanishing of the holomorphic mass term induced by $\varphi$ for the
bifundamental components of the Yang-Mills multiplet.  In precisely this
situation, one expects to find bifundamental boson and fermion zero-modes
which are trapped along $\Sigma$ and which lead to massless, charged
matter on $\BR^{3,1}$.

We naturally want to identify the massless modes trapped along
$\Sigma$ with the defect degrees of freedom that we
initially introduced by hand on $\Sigma$.  To do so, let us
identify which fermion zero-modes in the twisted Yang-Mills theory are
actually trapped along $\Sigma$.  Of course, once we identify which
fermionic zero-modes are trapped along $\Sigma$, supersymmetry determines
which bosonic zero-modes are trapped.

For the following analysis, we need only work in a small neighborhood
of ${\Sigma \subset S}$, which we parametrize with local holomorphic
coordinates $(s{}^1,s{}^2)$.  Since $\varphi$ vanishes to first-order on
$\Sigma$ by assumption, $\varphi$ takes the local form
\begin{equation}\label{EXPP}
\varphi \,=\, t \, s^2 \, ds^1 \^ ds^2\,.
\end{equation}
According to \eqref{EXPP}, $\Sigma$ corresponds to the locus ${s^2 = 0}$.
Hence $s^1$ is a coordinate along $\Sigma$, and $s^2$ is a coordinate
on the normal bundle $N_{\Sigma/S}$ to $\Sigma$ inside $S$.
Similarly, the conjugate $\bar\varphi$ is given by
\begin{equation}\label{EXPPC}
\bar\varphi \,=\, t \, \bar s{}^{\bar 2} \, d\bar s^{\bar 1} \^ d\bar
s^{\bar 2}\,.
\end{equation}

We now want to solve the equations of motion for the twisted fermions
locally near $\Sigma$ in the background described by \eqref{EXPP}.
According to Appendix \ref{8DACTION}, the relevant terms in the
twisted Yang-Mills action which determine the fermionic equations
of motion on $S$ are
\begin{align}\label{TOYI}
I_S \,&= \underset{\BR^{3,1} \times S}{\int}\!\!d^4x \,
\Tr\Big( \chi^\alpha \^ \bar\partial{}_A \psi_\alpha \,+\,
\bar\chi{}_{\dot\alpha} \^ \partial_A \bar\psi{}^{\dot\alpha} \,+\, 2
i \sqrt{2} \, \omega \^ \partial_A \eta^\alpha \^ \psi_\alpha \,-\, 2
i \sqrt{2} \, \omega \^ \bar\partial{}_A \bar\eta{}_{\dot\alpha} \^
\bar\psi{}^{\dot\alpha}\cr
&\qquad\,-\, \frac{1}{2}\,\bar\psi{}_{\dot\alpha}
\big[\bar\varphi, \bar\psi{}^{\dot\alpha}\big] \,+\,
\frac{1}{2}\,\psi^\alpha \big[\varphi, \psi_\alpha\big] \,+\, \sqrt{2}\,
\eta^\alpha \big[\bar\varphi, \chi_\alpha\big] \,+\, \sqrt{2}\,
\bar\eta{}_{\dot\alpha} \big[\varphi,
\bar\chi{}^{\dot\alpha}\big]\Big)\,+\,\cdots\,.
\end{align}
Hence the fermionic wavefunctions near $\Sigma$ satisfy
\begin{align}\label{DIRACI}
\bar\partial{}_A \psi^\alpha \,-\, \sqrt{2} \, \big[\bar\varphi,
\eta^\alpha]\,&=\,0\,,& \partial_A \bar\psi{}^{\dot\alpha} \,+\,
\sqrt{2} \, \big[\varphi, \bar\eta{}^{\dot\alpha}\big]\,&=\, 0\,,\cr
\omega\^\bar\partial{}_A \bar\eta{}^{\dot\alpha} \,+\,
\frac{i}{2\sqrt{2}}\,\big[\bar\varphi, \bar\psi{}^{\dot\alpha}\big] \,&=\,0\,,&
\omega\^\partial_A \eta^\alpha \,+\, \frac{i}{2\sqrt{2}}\,\big[\varphi,
\psi^\alpha\big] \,&=\,0\,,
\end{align}
as well as
\begin{align}\label{DIRACII}
\omega\^\partial_A \psi^\alpha \,+\ \frac{i}{2}\, \big[\bar\varphi,
\chi^\alpha\big] \,&=\,0\,,&\omega\^\bar\partial{}_A
\bar\psi{}^{\dot\alpha} \,-\, \frac{i}{2}\, \big[\varphi,
\bar\chi{}^{\dot\alpha}\big]\,&=\,0\,,\cr
\bar\partial{}_A \chi^\alpha \,-\, \big[\varphi,
\psi^\alpha\big]\,&=\,0\,,& \partial_A \bar\chi{}^{\dot\alpha}\,-\,
\big[\bar\varphi, \bar\psi{}^{\dot\alpha}\big]\,&=\,0\,.
\end{align}

To look for fermion zero-modes trapped along $\Sigma$, we take all
derivatives in \eqref{DIRACI} and \eqref{DIRACII} to act in the
direction normal to $\Sigma$.  We also assume that $\omega$ takes the
canonical Euclidean form near $\Sigma$.  As a result, in
\eqref{DIRACI} only the components of $\psi^\alpha$ and $\bar\psi{}^{\dot\alpha}$
which are tangent to $\Sigma$ appear, and in \eqref{DIRACII} only the
components of  $\psi^\alpha$ and $\bar\psi{}^{\dot\alpha}$ which are
normal to $\Sigma$ appear.

If $\Psi$ and $\wt\Psi$ represent any pair of fermions appearing in
\eqref{DIRACI} and \eqref{DIRACII}, then $\Psi$ and $\wt\Psi$ will
have a zero-mode trapped along $\Sigma$ if these fermions satisfy
classical equations of the schematic form (we ignore irrelevant constants)
\begin{align}
\frac{\partial \Psi}{\partial s^{2}} \, \,+\, \bar s{}^{\bar 2}
\, \wt\Psi \,&=\, 0\,,\cr
\frac{\bar\partial \wt\Psi}{\bar\partial \bar s^{\bar 2}} \,+\, s^2 \,
\Psi \,&=\, 0,
\end{align}
implying that $\Psi$ and $\wt\Psi$ behave near $\Sigma$ as
$\exp(-|s^2|^2)$.  Given the local expressions for $\varphi$ and
$\bar\varphi$ in \eqref{EXPP} and \eqref{EXPPC}, one can then
check that the fermions in \eqref{DIRACI} do {\em not} have
zero-modes localized along $\Sigma$.  In contrast, each bifundamental
fermion in \eqref{DIRACII} has precisely one such zero-mode, with
Gaussian decay along the normal direction to $\Sigma$.  In fact, since
we are free to scale the K\"ahler metric near $\Sigma$ as we wish, the
Gaussian decay away from $\Sigma$ can be made arbitrarily fast.

We thus obtain massless bifundamental matter localized along $\Sigma$
and associated to the following pairs of twisted fermions,
\begin{equation}\label{NORMOM}
\begin{pmatrix}
\psi^\alpha_{\bar 2} \, d\bar s^{\bar 2} \\
\chi{}^\alpha_{1 2} \, ds^1 \^ ds^2
\end{pmatrix}\,,\qquad\qquad
\begin{pmatrix}
\bar\psi{}^{\dot\alpha}_2 \, ds^2 \\
\bar\chi{}^{\dot\alpha}_{\bar 1 \bar 2} \, d\bar s^{\bar 1} \^ d\bar s^{\bar 2}
\end{pmatrix}\,.
\end{equation}
By appropriately raising or lowering indices using the K\"ahler metric
on $S$, we naturally regard the fermion zero-modes derived
from ${(\psi^\alpha_{\bar 2}, \bar\chi{}^{\dot\alpha}_{\bar 1 \bar 2})
  \equiv (\psi^{\alpha \, 2}, \bar\chi{}^{\dot\alpha\, 2}_{\bar 1})}$ as
transforming along $\Sigma$ in the bundles $N_{\Sigma / S}$ and
$\bar\Omega{}^1_\Sigma \otimes N_{\Sigma / S}$, and similarly
for the CPT-conjugates in \eqref{NORMOM}.

At first glance, this observation presents a small puzzle, since the
corresponding fermions $\lambda_\alpha$ and
$\bar\lambda{}^c_{\dot\alpha}$ in the defect theory transform as
sections of $K^{1/2}_\Sigma$ and $\bar K^{1/2}_\Sigma$.  However, we
now recall that $\Sigma$ is itself defined by the
vanishing of $\varphi_0$, a section of $K_S$.  Thus the normal bundle
${N_{\Sigma / S}}$ is isomorphic to $K_S\big|_\Sigma$,
\begin{equation}\label{NSIG}
N_{\Sigma / S} \,=\, K_S\big|_\Sigma\,.
\end{equation}
We also recall that the adjunction formula implies
\begin{equation}\label{ADJSIG}
K_\Sigma \,=\, K_S\big|_\Sigma \otimes N_{\Sigma / S}\,.
\end{equation}
Hence by \eqref{NSIG} and \eqref{ADJSIG},
\begin{equation}\label{SPINOR}
N_{\Sigma / S} = K^{1/2}_\Sigma\,.
\end{equation}
As a result, the massless fermions localized along $\Sigma$
in the eight-dimensional Yang-Mills theory on $S$ can be identified
with the defect fermions $\lambda_\alpha$ and
$\bar\lambda{}^c_{\dot\alpha}$ on $\Sigma$ which we initially introduced
by hand.

Another important consequence of the isomorphism in \eqref{SPINOR} is
that it establishes a canonical choice of spin structure on $\Sigma$.
Namely, we use the spin structure associated to the normal bundle
$N_{\Sigma / S}$ to define the defect fields on $\Sigma$.

\subsection{Unfolding Singularities via Surface Operators\label{UNFOLDING}}

As shown in Section \ref{SEVENBRANE}, when the singularity type remains
constant over all of $S$, the primitive Casimir invariants constructed from
the vevs of $\varphi$ exactly match all possible ways that a singularity of
general $ADE$ type can unfold. \ In this Section we demonstrate that the gauge
theory degrees of freedom of the intersecting seven-brane theory match to the
possible unfoldings of a more general class of F-theory compactifications
where the singularity type enhances along a real codimension two subspace of $S$
corresponding to a matter curve.

Given the precise match between gauge theory and geometry in the absence of a
matter curve, it is natural to suspect that the primitive Casimir invariants
of $\varphi$ still describe all possible ways to unfold the singularity type
for more general F-theory compactifications. \ Indeed, as seen in Section
\ref{COLLIDE}, the possible deformations are locally identical to the case of
the pure seven-brane theory. An immediate objection to this proposal follows from
inspection of tables \ref{E6UNFOLD}, \ref{E7UNFOLD} and \ref{E8UNFOLD} as well
as the analogous results for the $A$- and $D$- type singularities. \ Indeed,
the bundle assignments for the primitive Casimir invariants only match to
those of the deformation parameters of the geometry up to tensoring by $\mathcal{O}_{S}(n \Sigma)$ for some $n > 0$.
In this Section we give an explanation for this apparent mismatch.  The essential ingredient in this analysis
is that a non-zero vev for a six-dimensional field localized along the matter curve $\Sigma$ can
source a surface operator in the eight-dimensional partially twisted theory.
In particular, as shown in equation \eqref{POLEPHI} the bulk field $\varphi$ develops a pole
 along $\Sigma$ whose residue is given by the condensate of massless fields living on $\Sigma$:
\begin{equation}\label{POLEPHITWO}
\varphi = \frac{\llangle\sigma^{c},\sigma\rrangle_{\ad(P)} \,
  ds^1\^ds^2}{\alpha} + \cdots\,.
\end{equation}
 This will effectively shift the bundles which the casimirs of $\varphi$ belong to.
 Moreover, even though we will not be using it in this paper,
it follows from equation (\ref{MODDTERM}) that\ generic values of $\sigma$ and
$\sigma^{c}$ can also cause the bulk gauge field to develop singularities
along $\Sigma$. \

We interpret these singularities as a surface operator. \ As
noted in \cite{BershadskyFOURD}, the reduction of equations (\ref{MODBPSI})
and (\ref{MODDTERM}) in the directions normal to $\Sigma$ in $S$ corresponds to
Hitchin's equations in the presence of a source term. \ This is quite similar
to the operative definition of surface operators given in
\cite{GukovWittenDefect}. \ While the topological twist of the
four-dimensional gauge theory in \cite{GukovWittenDefect} does not
contain a $(2,0)$ form, an analogous singularity develops in the
one-forms of that twisted theory. \ In the present context, a surface
operator corresponds to a pole in $\varphi$ along the curve
$(\alpha=0)$. \ Because the pole in $\varphi$ arises from an F-term
supersymmetry condition, the path integral over arbitrary field
configurations for $\varphi$ on $S$ automatically localizes onto those
configurations of the form \eqref{POLEPHITWO}, so we obtain a disorder
operator associated to ${\Sigma \subset S}$.

In the rest of this subsection we further elaborate on the connection between
surface operators of the partially twisted theory and the unfolding of
singularities in an F-theory compactification which includes intersecting
seven-branes. \ As a warmup, we first show how surface operators in the partially
twisted eight-dimensional theory coupled to a six-dimensional defect describe seven-brane recombination in
both perturbatively and non-perturbatively realized compactifications. \ Next,
we show that when the corresponding complex deformation exists, there is an
\textit{exact} match between meromorphic vevs of $\varphi$ with a simple pole
structure and an arbitrary unfolding of all $ADE$ singularities other than
$E_{8}$. \ In the $E_{8}$ case, the matter curve, where the $E_{8}$ singularity
enhances to a higher singularity, will lead to exotic matter structure, which
has not been analyzed.\footnote{In appendix \ref{EXOTICA} we will discuss some related
questions.}  We thus will not have anything to say about the unfolding of the $E_{8}$
singularity because this requires further knowledge of the exotic physics living
on the ``matter curve''.

\subsubsection{Brane Recombination}

\begin{figure}
[ptb]
\begin{center}
\includegraphics[
height=3.1479in,
width=2.4803in
]%
{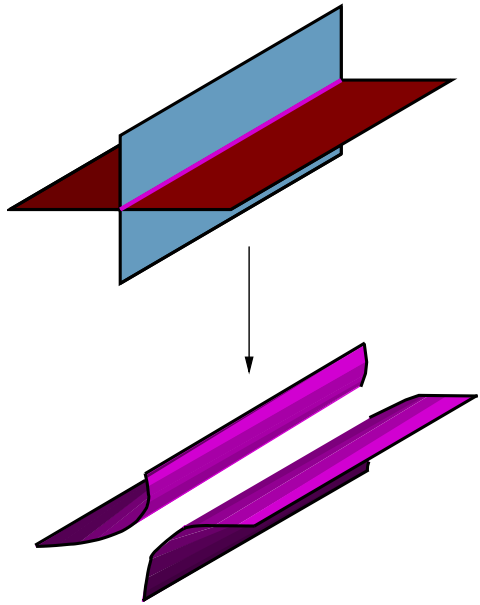}%
\caption{Prior to brane recombination, the common locus of two stacks of
intersecting seven-branes lead to additional light degrees of freedom which
propagate along a six-dimensional defect theory (top). \ When these light
degrees of freedom condense, the branes recombine (bottom).}%
\label{newbranerecombination}%
\end{center}
\end{figure}

In this Section we demonstrate that surface operators in both perturbative and
non-perturbatively realized intersecting brane configurations correspond to
brane recombination in the compactification. \ In the context of
perturbatively realized intersecting D-brane configurations, it is well-known
that vevs for bifundamental matter trigger brane recombination. \ For example,
a four-dimensional gauge theory with gauge group $U(1)\times U(1)$ and two Higgs fields with
opposite $U(1)$ charges will break to the diagonal $U(1)$ subgroup for
appropriate Higgs vevs. \ In string theory this change in rank is interpreted
as brane recombination.  See figure \ref{newbranerecombination} for a depiction of
this process.

To interpret brane recombination in terms of surface operators, first recall
that the transverse intersection of $n+1$ D7-branes along $(z=0)$ with another
stack of $m+1$ D7-branes along $(\alpha=0)$ is given by equation
(\ref{AMNYII}) with all $u_{l}$ and $t_{k}$ zero. \ For simplicity, we take
the divisor $(\alpha=0)$ in the threefold base to be non-compact so that we
may effectively treat the fields of this theory as non-dynamical constants.
\ Returning to equation (\ref{MODBPSI}), note that the vev $\llangle
\sigma^{c},\sigma\rrangle _{\ad(P)}$ corresponds to an $(n+1)\times(n+1)$
matrix of rank:%
\begin{equation}
l\leq\min(m+1,n+1)\text{.}\label{rankbound}%
\end{equation}
By a shift of coordinates, we may assume without loss of generality that the
deformation in the $t_{i}$ is:%
\begin{equation}
\overrightarrow{t}=\left(  \frac{b_{1}}{\alpha}+O(\alpha^{0}),...,\frac{b_{l}%
}{\alpha}+O(\alpha^{0}),0^{n+1-l}\right)
\end{equation}
so that the geometry now changes as:%
\begin{equation}
y^{2} = x^{2} + \alpha^{m+1}z^{n+1}\mapsto y^{2}=x^{2} + \alpha^{m+1}z^{n+1-l}%
\underset{i=1}{\overset{l}{%
{\displaystyle\prod}
}}\left(  z+\frac{b_{i}}{\alpha}\right)
\end{equation}
or,
\begin{equation}
y^{2} = x^{2}+\alpha^{m+1-l}z^{n+1-l}%
\underset{i=1}{\overset{l}{%
{\displaystyle\prod}
}}\left(  \alpha z+b_{i}\right)  \text{.}%
\end{equation}
which corresponds to brane recombination of $l$ branes of the $(z=0)$ stack
with $l$ branes of the $(\alpha=0)$ stack to form up to $l$ distinct stacks at
$(\alpha z+b_{i}=0)$ for $i=1,...,l$. Note that here $b_i$ is the $i^{\text{th}}$ eigenvalue
of the matrix $\llangle \sigma^{c} ,\sigma\rrangle_{\ad(P)}$.
Physically, the bound of
(\ref{rankbound}) corresponds to the fact that once all of the branes of a
stack have combined with other branes, there are none left over.

Starting from the local model:
\begin{equation}
y^{2}=x^{2}+\alpha z^{n+1}\text{,}%
\end{equation}
a similar analysis establishes that the geometry:%
\begin{equation}
y^{2}=x^{2}+\alpha z^{n+1}+\underset{i=2}{\overset{n}{\sum}}\widehat{\alpha}_{i}z^{n-i}%
\end{equation}
\ can be interpreted as a sequence of successive brane recombinations.

Because a similar analysis holds for brane recombination for $D$-type
singularities, we now proceed to examples involving $E$-type branes.\  In
fact, it is in principle possible that a similar analysis of $A$- and $D$-type
singularities will not be possible in the $E$-type case because this is
an intrinsically non-perturbative feature of F-theory. \ Nevertheless, we now
show that a sequential Higgsing of fields on the matter curve still produces a
general deformation of lower singularity type.

In the next subsection we will present a more general analysis of unfolding of
singularities in terms of surface operators, so for now we confine our remarks
to some representative examples. The geometry:%
\begin{equation}
y^{2}=x^{3}+\alpha^{2}z^{4}%
\end{equation}
corresponds to an $E_{6}$ singularity at $z=0$ and a non-compact $A_{2}$
singularity at $\alpha=0$. \ With conventions for deformations of $E$-type
singularities as in \cite{KatzMorrison,KatzVafa}, breaking to $SO(10)$ will
occur when the vev of $\varphi$ lies in the direction $(t,-2t,t,t,t,t)$ of the
Cartan subalgebra of $E_{6}$. \ After a suitable change of variables, the
geometry is given by rescaling the result in \cite{KatzVafa} by factors of
$\alpha$:%
\begin{equation}
y^{2}=\alpha^{2}z^{4}-(4\alpha zt-x)(2\alpha zt+x)^{2}\text{.}%
\end{equation}
We note that when $t$ develops a first order pole along $\alpha$ with residue
$b$, the geometry is now given by:%
\begin{equation}
y^{2}=\alpha^{2}z^{4}-(4bz-x)(2bz+x)^{2}%
\end{equation}
so that the resulting matter in the ${\bf 16}$ of $SO(10)$ is localized at $(b=0)$.

As a final example of brane recombination, the local geometry:%
\begin{equation}
y^{2}=x^{3}+\alpha xz^{3}%
\end{equation}
corresponds to an $E_{7}$ singularity at $z=0$ and a non-compact $A_{1}$
singularity at $\alpha=0$. \ Breaking to $E_{6}$ will occur when the vev of
$\varphi$ lies in the direction $(0,0,0,0,0,t,0)$ of the Cartan subalgebra.
\ Rescaling the result of \cite{KatzVafa} by powers of $\alpha$ yields:%
\begin{equation}
y^{2}=x^{3}+\alpha\left(  xz^{3}+\alpha t^{2}z^{4}\right)
\end{equation}
where we have performed a shift in the $y$ and $z$ coordinates in order to
cast the geometry in the above form. \ Note that when  $t$ develops a first
order pole along $\alpha=0$ with residue $b$, the resulting geometry is:%
\begin{equation}
y^{2}=x^{3}+\alpha xz^{3}+b^{2}z^{4}\text{.}%
\end{equation}
Proceeding step by step, a general geometry can be realized by further
Higgsing the matter localized on Riemann surfaces.

\subsubsection{General Unfolding}\label{GENUNFOLDING}

In the previous Section we presented a number of examples showing that general
deformations of colliding singularities can be understood in terms of a
sequence of brane recombinations induced by surface operators in the eight-dimensional
partially twisted theory. \ In this Section we show that the general unfolding
of a singularity again precisely matches to properties of the surface
operator. \ To present this more general analysis, we shall assume throughout
that the resulting deformation corresponds to a holomorphic section of an
appropriate bundle on $S$. \ For example, when $S$ corresponds to a del Pezzo
surface, $K_{S}$ is a strictly negative line bundle, so $K_{S}^{n}$ admits no
holomorphic sections for $n>0$. \ Under these assumptions, we now match
meromorphic vevs of $\varphi$ with a simple pole structure to an arbitrary
unfolding of all $ADE$ singularities other than $E_{8}$. \

We begin by describing the unfolding of a general $A_{n}$ singularity of the
form:%
\begin{equation}
y^{2}=x^{2}+\alpha z^{n+1}\text{.}%
\end{equation}
When the vev of $\varphi$ lies in a general direction of the Cartan subalgebra
of $A_{n}$, the singularity deforms to equation (\ref{AMNYIII}) which we
reproduce here:%
\begin{equation}
y^{2}=x^{2}+\alpha\left[  z^{n+1}+s_{2}z^{n-1}+\cdot\cdot\cdot+s_{n+1}\right]
\text{.}%
\end{equation}
By definition of the $s_{i}$, when one of the $t_{i}$'s develops a simple pole
along $\alpha=0$, each $s_{i}$ will also contain a simple pole. \ In
particular, this implies that each $\alpha \cdot s_{i}$ is a regular section which
does not have to vanish at $\alpha=0$. \ Solving for a general deformation in terms of
these regular sections, we see that the vevs of $\varphi$ indeed match to
a general deformation of an $A_{n}$ singularity.

Similarly, a general unfolding of the $D_{n}$ singularity:%
\begin{equation}
y^{2}=x^{2}z+\alpha^{2}z^{n-1}%
\end{equation}
by the vev of $\varphi$ in a general direction of the Cartan subalgebra is the same as in
\cite{KatzMorrison} up to non-trivial powers of $\alpha$ which are required in order to preserve
the overall homogeneity of the defining hypersurface equation:
\begin{equation}
y^{2}= - x^{2}z+\alpha^{2}\frac{\underset{i=1}{\overset{n}{%
{\displaystyle\prod}
}}(z+t_{i}^{2})-\underset{i=1}{\overset{n}{%
{\displaystyle\prod}
}}t_{i}^{2}}{z}+2\alpha x\underset{i=1}{\overset{n}{%
{\displaystyle\prod}
}}t_{i}\text{.}%
\end{equation}
We note that when a single $t_{i}$ develops a pole along $\alpha=0$, the
overall $\alpha$ dependence of all lower order deformations again cancels out.
\ In this way, the vevs of $\varphi$ again match to the general class of
deformations presented in equation (\ref{EXDN}).

While the above match between geometry and surface operators in the eight-dimensional
partially twisted theory is already non-trivial, we now show that this
connection persists for the unfolding of $E_{6}$ and $E_{7}$ singularities.
\ Consider first the unfolding of an $E_{6}$ singularity of the form:%
\begin{equation}
y^{2}=x^{3}+\alpha^{2}z^{4}%
\end{equation}
by a vev of $\varphi$ in an arbitrary direction $(t_{1},...,t_{6})$ of the
Cartan subalgebra of $E_{6}$. The end result of this deformation is the same as in \cite{KatzMorrison} up to
scaling by powers of $\alpha$:
\begin{align}
y^{2} &  =x^{3}+\alpha^{2}z^{4}+\varepsilon_{2}(2\alpha)^{2}xz^{2}%
+\varepsilon_{5}(2\alpha)^{3}xz+\varepsilon_{6}(2\alpha)^{4}z^{2}%
\label{E6def}\\
&  +\varepsilon_{8}(2\alpha)^{4}x+\varepsilon_{9}(2\alpha)^{5}z+\varepsilon
_{12}(2\alpha)^{6}%
\end{align}
where the explicit functions $\varepsilon_{i}$ as functions of the $t_{i}$ are
defined in Appendix 1 of \cite{KatzMorrison} as functions of the elementary
symmetric polynomials $s_{i}$ introduced earlier.  The particular numerical
coefficients have been chosen to conform with the conventions of \cite{KatzMorrison}.

Our expectation is that when the ${\bf 27}$ localized along the matter curve in $S$
develops a vev, a pole in $\varphi$ will remove the overall $\alpha$
dependence of each type of deformation. \ In the present case, however, it
follows from Appendix 1 of \cite{KatzMorrison} that the leading order behavior
of the $\varepsilon_{j}$ is:
\begin{equation}
\varepsilon_{j}\propto\left(  s_{1}\right)  ^{j}+O(t_{i}^{j-1})
\end{equation}
where $O(t_{i}^{j-1})$ denotes contributions of lower degree in each parameter
$t_{i}$. \ In particular, when a single $t_{i}$ develops a pole along
$\alpha=0$, $\varepsilon_{j}$ has a pole of order $j$ along $\alpha=0$.
\ Returning to equation (\ref{E6def}), it would at first appear that this
branch does not match to a geometric unfolding of the singularity due to the
presence of uncancelled poles in $\alpha$.

In fact, this is an artifact of the choice of coordinates used to present the
most general possible unfolding. \ Returning to equation (\ref{E6def}), we
present the general deformation in terms of coordinates $X\equiv x$ and
$Y\equiv y$ and:
\begin{equation}
Z\equiv z-\frac{4}{27}\alpha(s_{1})^{3}+\frac{1}{3}\alpha s_{1}s_{2}\text{,}%
\end{equation}
the definining hypersurface equation for a general deformation may now be
written as:%
\begin{align}
Y^{2} &  =X^{3}+\alpha^{2}Z^{4}+\beta_{2}\alpha^{2}XZ^{2}+\beta_{3}\alpha
^{3}Z^{3}+\beta_{5}\alpha^{3}XZ\\
&  +\beta_{6}\alpha^{4}Z^{2}+\beta_{8}\alpha^{4}X+\beta_{9}\alpha^{5}%
Z+\beta_{12}\alpha^{6}%
\end{align}
where the subscripts of the $\beta_{i}$ denote the degree in terms of
polynomials in the $t_{i}$. \ Using the explicit expressions for the
$\varepsilon_{i}$ given in Appendix 1 of \cite{KatzMorrison}, we find that the
leading order behavior of the coefficients as polynomials in the $t_{i}$ is:%
\begin{align}
\beta_{2}\alpha^{2} &  =\alpha^{2}\left(  -\frac{4}{3}\left(  s_{1}\right)
^{2}+O(t_{i})\right)  \\
\beta_{3}\alpha^{3} &  =\alpha^{3}\left(  \frac{16}{27}(s_{1})^{3}+O(t_{i}%
^{2})\right)  \\
\beta_{5}\alpha^{3} &  =\alpha^{3}\left(  -\frac{4}{3}s_{1}(s_{2})^{2}%
+\frac{8}{3}\left(  s_{1}\right)  ^{2}s_{3}+O(t_{i}^{2})\right)  \\
\beta_{6}\alpha^{4} &  =\alpha^{4}\left(  \frac{8}{9}\left(  s_{1}\right)
^{2}(s_{2})^{2}-\frac{16}{9}(s_{1})^{3}s_{3}+O(t_{i}^{3})\right)  \\
\beta_{8}\alpha^{4} &  =\alpha^{4}\left(  -\frac{1}{3}(s_{2})^{4}+\frac{4}%
{3}s_{1}(s_{2})^{2}s_{3}-\frac{4}{3}(s_{1})^{2}(s_{3})^{2}+O(t_{i}%
^{3})\right)  \\
\beta_{9}\alpha^{5} &  =\alpha^{5}\left(  \frac{4}{9}s_{1}(s_{2})^{4}%
-\frac{16}{9}(s_{1})^{2}(s_{2})^{2}s_{3}+\frac{16}{9}(s_{1})^{3}(s_{3}%
)^{2}+O(t_{i}^{4})\right)  \\
\beta_{12}\alpha^{6} &  =\alpha^{6}\left(
\begin{array}
[c]{l}%
\frac{2}{27}(s_{2})^{6}-\frac{4}{9}s_{1}(s_{2})^{4}s_{3}+\frac{8}{9}%
(s_{1})^{2}(s_{2})^{2}(s_{3})^{2}\\
-\frac{16}{27}(s_{1})^{3}(s_{3})^{3}+O(t_{i}^{5})
\end{array}
\right)  \text{.}%
\end{align}
When a single $t_{i}$ develops a simple pole, each $s_i$ also
develops a simple pole along $\alpha =0$.  By inspection of
the above result, we thus conclude that each product $\beta_i \alpha^k$ above is still
regular and has no leading order dependence on $\alpha$. As for the
$A$- and $D$-type singularities, this implies that an arbitrary
unfolding of the singularity matches to some choice of eigenvalues for
$\varphi$. \ This is a highly non-trivial match between possible deformations
of the singularity and the partially twisted seven-brane theory!

In a similar fashion, we now consider unfolding an $E_{7}$ singularity of the
form:%
\begin{equation}
y^{2}=x^{3}+\alpha xz^{3}%
\end{equation}
by a vev of $\varphi$ in an arbitrary direction $(t_{1},...,t_{7})$ of the
Cartan subalgebra of $E_{7}$. Up to powers of $\alpha$ introduced to preserve homogeneity,
the unfolding of $E_{7}$ in \cite{KatzMorrison} is:
\begin{align}
y^{2} &  =-x^{3}+16\alpha xz^{3}+\varepsilon_{2}\alpha x^{2}z+\varepsilon
_{6}\alpha^{2}x^{2}+\varepsilon_{8}\alpha^{3}xz\\
&  +\varepsilon_{10}\alpha^{4}z^{2}+\varepsilon_{12}\alpha^{4}x+\varepsilon
_{14}\alpha^{5}z+\varepsilon_{18}\alpha^{6}%
\end{align}
where the $\varepsilon_{i}$ as functions of the $t_{j}$ are defined in
Appendix 2 of \cite{KatzMorrison}. \ As before, the leading order behavior of
$\varepsilon_{i}\propto\left(  s_{1}\right)  ^{i}$ so that a simple pole in
one of the $t$'s~would appear to not admit a geometric interpretion. \ As in
the case of the $E_{6}$ singularity, we now show that this is an artifact of
the choice of coordinates.

To this end, we first shift the $x$ coordinate via the substitution:%
\begin{equation}
x\mapsto x+\frac{1}{3}\left(  \varepsilon_{2}\alpha z+\varepsilon_{6}%
\alpha^{2}\right)  \text{.}%
\end{equation}
Setting $Y\equiv y$ and $X\equiv x$, we also define:%
\begin{equation}
Z\equiv z-\frac{1}{16}\alpha(s_{1})^{4}+\frac{1}{6}\alpha(s_{1})^{2}s_{2}-\frac
{1}{6}\alpha s_{1}s_{3}%
\end{equation}
so that the defining hypersurface equation for a general deformation may now
be written as:%
\begin{align}
Y^{2} &  =-X^{3}+16\alpha XZ^{3}+\beta_{2}\alpha^{2}Z^{4}+\beta_{4}\alpha
^{2}XZ^{2}+\beta_{6}\alpha^{3}Z^{3}\label{ESEVDEFGEN}\\
&  +\beta_{8}\alpha^{3}XZ+\beta_{10}\alpha^{4}Z^{2}+\beta_{12}\alpha
^{4}X+\beta_{14}\alpha^{5}Z+\beta_{18}\alpha^{6}%
\end{align}
where as before, the subscripts indicate the degree of each $\beta_{i}$ in
terms of polynomials in the $t_{i}$. \ The explicit leading order behavior of
each $\beta_{i}$ as a polynomial in the $t_{i}$ may be found in Appendix
\ref{Explicit}.

It follows from the results in Appendix \ref{Explicit} that a simple pole in
one of the $t_{i}$'s exactly cancels the overall $\alpha$ dependence of each
deformation parameter. \ In particular, this implies that the degrees of
freedom of the partially twisted $E_{7}$ theory with a defect exactly matches
the unfolding of the singularity.

\subsection{Chiral Matter and Yukawa Couplings from $\Sigma$}\label{6DMATTER}

Having determined how the partially twisted seven-brane theory couples to a six-dimensional
defect, we now determine some basic properties of supersymmetric vacua in
the associated four-dimensional effective theory. Consider first the trivial case where
the matter curve is given by a flat $T^{2}$. In this case, ${\bf \Lambda}$ and ${\bf \Lambda}^{c}$
correspond to two four-dimensional $\mathcal{N}=1$ chiral superfields transforming in complex conjugate
representations. \ Indeed, these chiral multiplets determine an $\mathcal{N}=2$
hypermultiplet. \ In the special case where $U$ is a real representation of $G_{S}$,
the six-dimensional fields can also organize into a half-hypermultiplet.  This reduces to a single
four-dimensional $\mathcal{N}=1$ chiral multiplet.

In the presence of potentially non-trivial background gauge field configurations on $S$ and $S'$,
the resulting zero mode spectrum is determined by bundle valued Dolbeault
cohomology groups with support on $\Sigma$. \ The net chirality of the
spectrum is then given by a topological invariant which is uniquely fixed
by the representation and gauge bundle content of the six-dimensional fields localized on $\Sigma$.
Because much of this discussion parallels a similar treatment given for the pure
seven-brane theory, our discussion will be brief.
As opposed to the case of the pure seven-brane theory wrapping a Hirzebruch or del Pezzo surface,
we find that the non-trivial coupling between bulk gauge fields propagating on a compact surface
and six-dimensional fields localized along $\Sigma$ induces non-trivial Yukawa couplings among
the zero modes.  We conclude this Section by presenting
some toy models which further explicate these results.

\subsubsection{Massless Spectrum}

In this subsection we determine the massless particle spectrum in four dimensions of fields
localized along a Riemann surface $\Sigma$ in $S$ in the presence of a potentially
non-trivial background gauge field configuration. \ We begin with an analysis
of the relevant group theory. Proceeding in a parallel fashion to the case
of the pure seven-brane theory, non-trivial background gauge field configurations
on $S$ and $S^{\prime}$ which take values in subgroups $H_{S}\subset G_{S}$
and ${H_{S^{\prime}}}\subset G_{S^{\prime}}$ will break $G_{S}\times G_{S^{\prime}}$
to the commutant subgroup. In many applications, $S^{\prime}$ is non-compact so that
the associated gauge group factor is non-dynamical. As before, we let $\Gamma_{S}$ denote
the maximal subgroup of $G_{S}$ such that $G_{S}\supset \Gamma_{S}\times H_{S}$, with
similar notation for $\Gamma_{S^{\prime}}$. Letting $\Gamma= \Gamma_{S}\times \Gamma_{S^{\prime}}$ and
$H= H_{S}\times H_{S^{\prime}}$, decomposing $U\times U^{\prime}$ into irreducible
representations of $\Gamma\times H$ yields:%
\begin{equation}
U\otimes U^{\prime} =\underset{j}{%
{\displaystyle\bigoplus}
}(\nu_{j},V_{j})\text{.}%
\end{equation}
A similar decomposition holds for the bundle $\mathcal{U}\otimes \mathcal{U}^{\prime}$.
In the obvious notation, we let $\mathcal{V}_{j}$ denote the corresponding bundle which
transforms as a representation $V_{j}$ of $H$.

Because the supercurrent is covariantly constant in the partially twisted six-dimensional theory, it is enough
to specify the massless spectrum of fermions. \ Taking
into account the additional twist by an ambient line bundle on $\Sigma$ so
that the resulting fermions transform as zero- and one-forms on $\Sigma$, it
follows that the fermions $\lambda_{\alpha\nu_{j}}$ and $\lambda
_{\alpha\nu^{\ast}_{j}}^{c}$ are both annihilated by $\overline{\partial
}_{A + A^{\prime}}$ and $\overline{\partial}_{A + A^{\prime}}^{\dag}$ and therefore also by the
Laplacian $\Delta_{\overline{\partial}}=\overline{\partial}_{A + A^{\prime}}\overline
{\partial}_{A + A^{\prime}}^{\dag}+\overline{\partial}_{A + A^{\prime}}^{\dag}\overline{\partial}_{A + A^{\prime}}$.
The chiral spectrum is therefore:%
\begin{align}
\lambda_{\alpha\nu_{j}} &  \in H_{\overline{\partial}}^{0}(\Sigma,K_{\Sigma
}^{1/2}\otimes\mathcal{V}_{j})\label{firsthyper}\\
\lambda_{\alpha\nu^{\ast}_{j}}^{c} &  \in H_{\overline{\partial}}%
^{0}(\Sigma,K_{\Sigma}^{1/2}\otimes\mathcal{V}_{j}^{*})\simeq H_{\overline
{\partial}}^{1}(\Sigma,K_{\Sigma}^{1/2}\otimes\mathcal{V}_{j})^{*}%
\text{.}\label{secondhyper}%
\end{align}
Using the analogue on $\Sigma$ of the isomorphism in \eqref{Dualizing}, the
anti-chiral spectrum is:%
\begin{align}
\overline{\lambda}_{\dot{\alpha}\nu^{\ast}_{j}} &  \in\overline{H_{\overline
{\partial}}^{0}(\Sigma,K_{\Sigma}^{1/2}\otimes\mathcal{V}_{j})}\simeq
H_{\overline{\partial}}^{0}(\Sigma,K_{\Sigma}^{1/2}\otimes\mathcal{V}%
_{j})^{*}\label{CPTfirsthyper}\\
\overline{\lambda^{c}}_{\dot{\alpha}\nu_{j}} &  \in\overline{H_{\overline
{\partial}}^{0}(\Sigma,K_{\Sigma}^{1/2}\otimes\mathcal{V}_{j}^{*})}\simeq
H_{\overline{\partial}}^{1}(\Sigma,K_{\Sigma}^{1/2}\otimes\mathcal{V}%
_{j})\label{CPTsecondhyper}%
\end{align}
where in the above we have used Serre duality. \ Comparing equations
(\ref{firsthyper}) and (\ref{secondhyper}) with (\ref{CPTfirsthyper}) and
(\ref{CPTsecondhyper}), we observe that the resulting spectrum is manifestly
CPT-invariant. \ It now follows that the net number of generations minus
anti-generations transforming in the representation $\nu_{j}$ is:%
\begin{equation}\label{SIGMAGEN}
n_{\nu_{j}}-n_{\nu^{\ast}_{j}}=h^{0}(\Sigma,K_{\Sigma}^{1/2}%
\otimes\mathcal{V}_{j})-h^{1}(\Sigma,K_{\Sigma}^{1/2}\otimes\mathcal{V}%
_{j})=\chi(\Sigma,K_{\Sigma}^{1/2}\otimes\mathcal{V}_{j})\text{.}%
\end{equation}

The Euler character $\chi(\Sigma,K_{\Sigma}^{1/2}\otimes\mathcal{V}_{j})$ is now
given by the analogue of the index theorem found in equation \eqref{INDEX} for
the Riemann surface $\Sigma$.  Returning to equation \eqref{SIGMAGEN}, we therefore find:
\begin{equation}\label{SIGMAINDEX}
n_{\nu_{j}}-n_{\nu_{j}^{\ast}}=\left(  1-g\right)  \rk\left(
K_{\Sigma}^{1/2}\otimes\mathcal{V}_{j}\right)  +\underset{\Sigma}{\int}%
c_{1}\left(  K_{\Sigma}^{1/2}\otimes\mathcal{V}_{j}\right)
\end{equation}
where $g$ denotes the genus of the Riemann surface $\Sigma$.

\subsubsection{More Yukawa Couplings}\label{MOREYUKAWA}

In this Section we demonstrate that as opposed to the pure seven-brane theory, non-trivial Yukawa
couplings can originate from the coupling of the bulk gauge fields to the
matter curve. As explained near equation \eqref{IDEFECT}, it is convenient
to treat ${\bf \Lambda}$ and ${\bf \Lambda}^{c}$ as a collection of four-dimensional fields labelled by points on $\Sigma$.
Labelling all zero mode solutions of the bulk and defect theory by $\alpha$, $\beta$ and $\gamma$,
the resulting cubic superpotential term now follows from equation \eqref{IDEFECT}:
\begin{equation}\label{WRIEMANNSURFACE}
d_{\alpha \beta \gamma}=\underset{\Sigma}{\int} c_{ijk}\left(  {\bf \Lambda}
^{c,\alpha,i}\left({\bf A}^{\beta,j} + {\bf A}^{\prime \beta^{\prime},j} \right){\bf \Lambda}^{\gamma,k} \right)%
\end{equation}
where the indices $i,j,k$ are group indices in $G_{\Sigma} \supset G_{S}\times G_{S^{\prime}}$ and $c_{ijk}$ is a structure constant
associated with the decomposition to $\Gamma_{S}\times \Gamma_{S^{\prime}}$.

Before closing this subsection, we now elaborate on the geometric content of the
coupling between two fields localized on $\Sigma$ in representations $\nu_{1}$
and $\nu_{{2}}$ and a bulk field transforming in a representation $\tau$ with
associated bundles $\mathcal{V}_{1}$, $\mathcal{V}_{2}$ and $\mathcal{T}$ which
transform in representations of the structure group of the instanton configuration.
A non-trivial Yukawa coupling corresponds to a tri-linear map:
\begin{equation}
H_{\overline{\partial}}^{0}(  \Sigma,K_{\Sigma}^{1/2}\otimes
\mathcal{V}_{1})  \otimes H_{\overline{\partial}}^{0}(
\Sigma,K_{\Sigma}^{1/2}\otimes\mathcal{V}_{2})  \otimes H_{\overline
{\partial}}^{1}(  \Sigma,\mathcal{T}_{\Sigma})  \rightarrow%
\mathbb{C}
\end{equation}
where $\mathcal{T}_{\Sigma}$ denotes the restriction of $\mathcal{T}$ to
$\Sigma$. \ This map is natural in the sense that $H_{\overline{\partial}}%
^{1}(  \Sigma,\mathcal{T}_{\Sigma})$ canonically pairs with
$H_{\overline{\partial}}^{0}(\Sigma,K_{\Sigma}^{1/2}\otimes
\mathcal{V}_{1})  \otimes H_{\overline{\partial}}^{0}(
\Sigma,K_{\Sigma}^{1/2}\otimes\mathcal{V}_{2})$ because of Serre
duality on $\Sigma$:%
\begin{equation}
H_{\overline{\partial}}^{1}(  \Sigma,\mathcal{T}_{\Sigma})  \simeq
H_{\overline{\partial}}^{0}(  \Sigma,K_{\Sigma}\otimes\mathcal{T}%
_{\Sigma}^{\ast})  ^{\ast}=H_{\overline{\partial}}^{0}(
\Sigma,K_{\Sigma}\otimes\mathcal{V}_{1}\otimes\mathcal{V}_{2})  ^{\ast}%
\end{equation}
where in the final equality we have used the fact that the corresponding
F-term only transforms as a gauge invariant singlet provided
$\mathcal{T}_{\Sigma}=\mathcal{V}_{1}^{\ast}\otimes\mathcal{V}_{2}^{\ast}$.
\ In this final form for $H_{\overline{\partial}}^{1}(  \Sigma
,\mathcal{T}_{\Sigma})  $, we see that the two fields on $\Sigma$ indeed
pair naturally with the dual cohomology group which describes a component of
the bulk gauge field.

\subsubsection{A Refined Toy Model}

Although we shall defer the construction of more realistic GUTs to future work in \cite{BHV}, we now explain how
to generate a three generation $SO(10)$ toy model with non-trivial Yukawa
couplings. \ We find it encouraging that many of the broadest features of $SO(10)$
GUTs can be achieved simply using supersymmetric gauge field configurations corresponding to line bundles on $S$.
Consider a seven-brane wrapping a del Pezzo $3$ surface with $G_{S}=SO(12)$ coupled to a defect.
The charged matter localized along the defect is given by a six-dimensional half
hypermultiplet in the $32$ spinor representation. \ We assume that the matter
curve is given by an exceptional curve $\Sigma$ with homology class $E_{1}$ in $H_{2}(dP_{3},%
\mathbb{Z}
)$. \ While it is important to describe the explicit geometry of the
F-theory compactification, in order to focus on the most salient features of the toy model
we shall defer such issues to future investigations.

The bulk gauge group $G_{S}=SO(12)$ breaks to $SO(10)\times U(1)$ in the
presence of a supersymmetric gauge field configuration which takes non-trivial
values in the $U(1)$ factor. This corresponds to a supersymmetric line
bundle $\mathcal{L}$ on the surface $S$. \ The adjoint $\bf{66}$ and
spinor $\bf{32}$ decompose under this subgroup as:%
\begin{align}
SO(12)  &  \supset SO(10)\times U(1)\\
{\bf 66}  &  \rightarrow{\bf 45}_{0}+{\bf 1}_{0}+{\bf 10}_{2}+{\bf 10}_{-2}\\
{\bf 32}  &  \rightarrow{\bf 16}_{1}+\overline{\bf 16}_{-1}%
\end{align}
so that the resulting matter content in four dimensions is:%
\begin{align}
{\bf 10}_{2}  &  \in H_{\bar{\partial}}^{1}(S,\mathcal{L}^{2})\\
{\bf 10}_{-2}  &  \in H_{\bar{\partial}}^{1}(S,\mathcal{L}^{-2})\\
{\bf 16}_{1}  &  \in H_{\bar{\partial}}^{0}(\Sigma,K_{\Sigma}^{1/2}\otimes \mathcal{L}_{\Sigma})\label{NUM16}\\
\overline{\bf 16}_{-1}  &  \in H_{\bar{\partial}}^{0}(\Sigma,K_{\Sigma}^{1/2}\otimes
\mathcal{L}_{\Sigma}^{-1})\text{,}\label{NUM16BAR}%
\end{align}
where $\mathcal{L}_{\Sigma}$ denotes the restriction of $L$ to the matter curve $\Sigma$.

In Appendix \ref{VanishAppendix} we show that there exist a family of
K\"{a}hler classes such that the line bundle:%
\begin{equation}
\mathcal{L}=\mathcal{O}_{S}(a_{1}E_{1}+a_{2}E_{2}+a_{3}E_{3})
\end{equation}
is supersymmetric provided $a_{i}a_{j}<0$ for some $i\neq j$.

To achieve three ${\bf 16}$'s with a minimal number of net ${\bf 10}$'s in the
bulk, we take $a_{1}=-3$, $a_{2}=2$, $a_{3}=2$ so that:%
\begin{align}
\mathcal{L}  &  =\mathcal{O}_{S}(-3E_{1}+2E_{2}+2E_{3})\\
\mathcal{L}_{\Sigma}  &  =\mathcal{O}_{\Sigma}(+3)\text{.}%
\end{align}
Because they are bulk fields, the net number of massless chiral fields transforming in the $\bf{10}$ follows
from equations \eqref{EXPANDO} and \eqref{SPECIALNGEN}:%
\begin{align}
n_{{\bf 10}_{2}}  &  =32\\
n_{{\bf 10}_{-2}}  &  =34\\
n_{{\bf 10}_{-2}}-n_{{\bf 10}_{2}}  &  =2\text{.}%
\end{align}
Next consider the six-dimensional half hypermultiplet localized along the matter curve.
\ The zero modes for the ${\bf 16}$'s and $\bar{\bf 16}$'s are respectively given by harmonic
representatives in the cohomology groups \eqref{NUM16} and \eqref{NUM16BAR}.  The total
number transforming in each representation is:
\begin{align}
n_{\bf 16}  &  =\dim H_{\bar{\partial}}^{0}(\Sigma,K_{\Sigma}^{1/2}\otimes \mathcal{L}_{\Sigma})=\dim
H_{\bar{\partial}}^{0}(\Sigma,\mathcal{O}_{\Sigma}(2))=3\\
n_{\overline{\bf 16}}  &  =\dim H_{\bar{\partial}}^{0}(\Sigma,K_{\Sigma}^{1/2}\otimes \mathcal{L}_{\Sigma
}^{-1})=\dim H_{\bar{\partial}}^{0}(\Sigma,\mathcal{O}_{\Sigma}(-4))=0\text{.}%
\end{align}

Labelling the three zero modes on $\Sigma$ as ${\bf 16}^{(i)}$ for $i=1,2,3$, it follows from
equation \eqref{WRIEMANNSURFACE} that the contribution to the superpotential
from the bulk to surface interaction term is schematically of the form:
\begin{equation}\label{schematicW}
W\supset\lambda_{ij}{\bf 16}^{(i)}\times {\bf 16}^{(j)}\times {\bf 10}^{(S)}%
\end{equation}
where ${\bf 10}^{(S)}$ is shorthand for all possible contributions from
the large number of zero mode solutions transforming in this
representation of the $SO(10)$ GUT.  Note that in this particular case,
the structure constants of $G_{S}$ only allow the ${\bf 10}_{-2}$'s to couple
to the ${\bf 16}$'s because of the $U(1)$ charges of all fields in the
decomposition of $G_{S}$.

While we defer a full discussion of Yukawa couplings to \cite{BHV}, we now briefly
comment on some general features of the flavor structure associated with coupling
two matter fields on $\Sigma$ to a bulk field in $S$ for models in which such
interaction terms are the sole contribution to the four-dimensional effective
superpotential.  While in the example just presented we have localized all three
generations on a single matter curve, it is in principle possible to localize
different generations on distinct matter curves, or to allow some generations to
descend from bulk fields.  This can in principle be used to induce non-trivial texture zeroes in the
Yukawa matrices.  As an example, suppose we have three ${\bf 16}$'s localized on
three distinct matter curves in $S$ with some number of bulk ${\bf 10}^{(S)}$'s.  Labelling the
three generations as ${\bf 16}^{(i)}$ for $i=1,2,3$, the Yukawa couplings are still
described by equation \eqref{schematicW} where now the Yukawa matrix
takes the schematic form:
\begin{equation}
\lambda_{ij}\sim\left(
\begin{array}
[c]{ccc}%
\lambda_{11} & 0 & 0\\
0 & \lambda_{22} & 0\\
0 & 0 & \lambda_{33}%
\end{array}
\right)\text{.}
\end{equation}
As another example, suppose that we have constructed a
consistent three generation $SO(10)$ GUT model with one ${\bf 16}$ and a ${\bf 10}$
coming from bulk zero modes on $S$ with two generations localized along
a single matter curve. Denoting the bulk ${\bf 16}$ by ${\bf 16}^{(1)}$, we
let ${\bf 16}^{(2)}$ and ${\bf 16}^{(3)}$ denote the two generations localized on the matter curve.
In this case, the analogue of equation \eqref{schematicW} implies that the Yukawa matrix takes the form:%
\begin{equation}
\lambda_{ij} \sim \left(
\begin{array}{ccc}
0 & 0 & 0 \\
0 & \lambda_{22} & \lambda_{23} \\
0 & \lambda_{23} & \lambda_{33}
\end{array}
\right)\text{.}
\end{equation}

\section{Multiple Intersections and More Yukawa Couplings\label{MULTINT}}

While the generic singularity type of the elliptic Calabi-Yau fourfold
corresponds to $G_{S}$ over $S$ with corresponding gauge group, simple
genericity arguments imply that there exist subloci of complex codimension one
and two in $S$ along which the rank of $G_{S}$ increases by one and two,
respectively. \ This follows from the dictionary established between the
unfolding of the singularity of the geometry and its interpretation in terms
of fields of the partially twisted theory. For example, consider the case
where $G_{S}=E_{6}$. \ In this case, the local geometry is:%
\begin{equation}
y^{2}=x^{3}+\alpha^{2} z^{4}%
\end{equation}
where $\alpha$ is a section of a suitable bundle over $S$. \ When this bundle
is non-trivial, there is a codimension one locus on $S$ where $\alpha=0$.
\ Although the singularity type appears to degenerate when $\alpha=0$, there
are higher order terms in this defining equation which do not correspond to
light degrees of freedom for $\alpha\neq0$. \ Along this locus, however, the
next relevant term in the singularity is:%
\begin{equation}
y^{2}=x^{3}+\alpha^{2} z^{4}+\beta xz^{3}%
\end{equation}
so that for generic points on the curve $\alpha=0$ the geometry contains an
$E_{7}$ singularity. \ Note that in the case where $S$ is a del Pezzo surface,
$K_{S}^{-1}$ is an ample line bundle so that when $\alpha$ has a section, $\beta$ does as well.
\ In particular, this implies that for generic points of $S$,
$\beta\neq0$. \ The singularity type can enhance to even higher type along the
discrete collection of points in $S$ where $\alpha=\beta=0$. \ Adding in the
next highest term, the model is therefore of the form:%
\begin{equation}
y^{2}=x^{3}+\alpha^{2} z^{4}+\beta xz^{3}+\gamma z^{5}
\end{equation}
so that when $\alpha=\beta=0$ the geometry contains an $E_{8}$ type singularity.

\begin{figure}
[ptb]
\begin{center}
\includegraphics[
height=3.0986in,
width=3.7178in
]%
{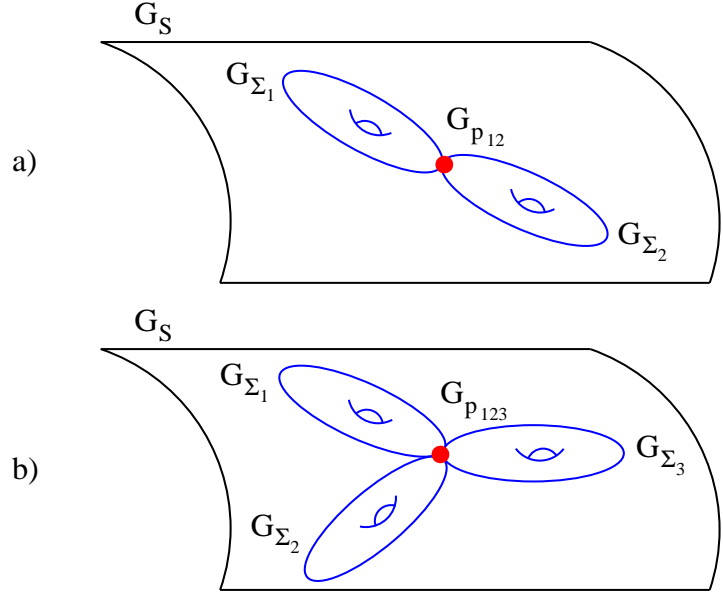}%
\caption{Depiction of the bulk eight-dimensional gauge theory defined by F-theory with
singularity type $G_{S}$ at generic points of the complex surface $S$. \ Along
a codimension one defect, the singularity type can generically enhance to the
higher rank singularity $G_{\Sigma_{j}}$. \ At codimension two defects where
distinct matter curves intersect, the geometry can enhance to the even higher
rank singularity $G_{p_{ij}}$ when two curves intersect (a) and $G_{p_{ijk}}$
when three curves intersect (b). \ Contrary to expectations based on dimension
counting, we find that case (b) is generic for geometries with exceptional type singularities.}
\label{ENHANCETWICE}%
\end{center}
\end{figure}

This example is illustrative of the generic situation. \ Letting $r$
denote the rank of $G_{S}$, along curves $\Sigma_{j}$ the singularity type
$G_{\Sigma_{j}}$ enhances to rank $r+1$. \ Moreover, at $l$ points
$p_{I}^{(k)}$, some collection of curves $\Sigma_{i_{1}}$,..., and
$\Sigma_{i_{n}}$ intersect where $k=1,...,l$ and $I$ denotes a mult-index in
the $n$ variables. \ Along such points, the singularity type $G_{p_{I}^{(k)}}$
has rank $r+2$. \ While arguments based on dimension counting would suggest
that only two matter curves can intersect, we shall argue that for exceptional
type singularities, three curves can also generically meet. \ The above
analysis implies that the singularity types obey the containment relations: \
\begin{equation}
G_{p_{I}^{(k)}}\supset G_{\Sigma_{j}}\times U(1)\supset G_{S}\times U(1)\times
U(1)\text{.}
\end{equation}
See figure \ref{ENHANCETWICE} for a depiction of the enhancement in
singularity type along defects of the bulk theory on $S$.

As discussed in Section \ref{COUPLINGDEFECT}, one potential source of four-dimensional chiral matter localized
along a curve $\Sigma_{j}$ originates from the local enhancement of singularity
type from $G_{S}$ to $G_{\Sigma_{j}}$. \ In this Section we perform a similar
analysis to analyze the last stage of enhancement to a singularity of type
$G_{p_{I}^{(k)}}$.

Before proceeding to a more detailed analysis of rank two enhancement in
singularity type, note that when $G_{S}=E_{7}$, the resulting $G_{p_{ij}%
^{(k)}}$ would appear to have rank nine. \ Because the largest compact
exceptional group has rank eight, we conclude that the resulting physics is
likely to be somewhat more exotic. \ This special case has been discussed in
\cite{MorrisonFlorea} and we defer further discussion to Appendix
\ref{EXOTICA}. \ For now, we will assume that all of the singularities
encountered are of $ADE$ type.

To better understand the physics associated with points $p_{I}$ where the
singularity type enhances twofold, we first treat the case where $G_{p_{I}}$
is an $A$-type singularity. In this case, a perturbative treatment of
the geometry in terms of intersecting D7-branes is available, and it is well-known
that such configurations signal the presence of additional superpotential terms
in the four-dimensional effective theory. \ Re-interpreting these results in terms of the general
philosophy outlined in \cite{KatzVafa} yields a similar result for more general
geometries which contain points of twofold enhancement in the singularity type. We find that multiple
intersections of matter curves are generic for certain F-theory compactifications
and moreover induce additional Yukawa couplings in the four-dimensional effective theory.  To better illustrate
these facts, we present explicit examples of this phenomenon for $E$-type singularities.

\subsection{Enhancement to $A$-Type Singularities}

As a representative example, we first consider the triple intersection of
D7-branes wrapping three distinct divisors in the threefold base described by
the local geometry:%
\begin{equation}
y^{2}=x^{2}+(z-t_{1})^{n_{1}}(z-t_{2})^{n_{2}}(z-t_{3})^{n_{3}}
\label{tripleinter}%
\end{equation}
so that there are six-dimensional bifundamentals of $U(n_{i})\times U(n_{j})$ localized at each
pairwise intersection of divisors.\footnote{In order to emphasize the connection with D-branes, in this Section
we include the explicit $U(1)$ factor of the worldvolume gauge group so that the full gauge group is $U(n)$.} When
the $t_{i}$ are all regular sections, the three stacks of D7-branes wrap compact divisors in the threefold base.
On the other hand, as explained in previous Sections, we can also consider the case where some of the $t_{i}$ are
meromorphic sections so that some of these divisors are non-compact.
Decomposing into four-dimensional $\mathcal{N}=1$ superfields ${\bf \Lambda}$ and ${{\bf \Lambda}}^{c}$,
by abuse of notation we shall often denote possible contributions from either type of
superfield as ${\bf \Lambda}_{i\overline{j}}$ when the context is clear.

To proceed further, note that this situation is well-described by perturbative type IIB
superstrings. \ In this context, we recall that a pair of intersecting
D7-branes gives bifundamental fields ${\bf \Lambda}_{i\overline{j}}$ on the
intersection. \ Further, while triple intersections of D7-branes at a point do
not contribute additional matter fields, the disc diagram amplitude ending on
the three D7-branes generates a superpotential term in the four dimensional effective theory between the
${\bf \Lambda}_{i\overline{j}}$. In fact, the superpotential computation localizes
to constant maps of the disc to the point of the triple intersection. \ The
end result is the superpotential term localized at the triple intersection
point:%
\begin{equation}
W \supset{\bf \Lambda}_{1\overline{2}}{\bf \Lambda}_{2\overline{3}}{\bf \Lambda}_{3\overline{1}%
}|_{p}.
\end{equation}

We will now try to recover this well-known result in the framework of
geometric singularities so that we can apply this methodology to more general
examples and in particular to E-type singularities. \ Specializing to the case
where $n_{1}=n_{2}=1$ and setting $n_{3}=n$ yields:%
\begin{equation}
y^{2}=x^{2}+z^{n}(z^{2}-(t_{1}+t_{2})z+t_{1}t_{2}) \label{specialtripleinter}%
\end{equation}
where in passing from equation (\ref{tripleinter}) to equation
(\ref{specialtripleinter}) we have set $t_{3}=0$ for notational simplicity.
\ At generic points of $S$, the singularity type of the geometry at $z=0$ is
an $A_{n-1}$ singularity. \ In this case, each $\Sigma_{i}$ corresponds to the
locus $(z=0)\cap(z=t_{i})$. \ Along the vanishing locus for each $\Sigma
_{i}$, this appears to enhance to $A_{n}$ and at the discrete collection of
points in $S$ where $t_{1}=t_{2}=0$, the singularity type enhances to
$A_{n+1}$.

We now interpret this system from the perspective of the gauge theory.
\ Recall that the $t_{i}$ denote non-zero vevs for the eigenvalues of the
$\varphi$ form on $S$. \ Along the locus where only one $t_{i}$ vanishes, the
generic $U(n)\times U(1)_{1}\times U(1)_{2}$ gauge group enhances to
$U(n+1)_{i}\times U(1)_{i}$, where the subscript indicates the embedding
inside $U(n+2)$. \ Indeed, given the partially broken gauge group $U(n)\times
U(1)_{1}\times U(1)_{2}$, there are two ways in which such an embedding can
occur. \ We note that there is one additional enhancement when the vevs $\varphi$
are proportional to the identity in the $U(2)$ direction so that
$t_{1}=t_{2}$. \ A six-dimensional field in the bifundamental of two of the unitary group
factors localizes along the defining equations for the corresponding $t$'s.
This additional field lies on another Riemann surface $\Sigma^{\prime}$ which
is not localized in $S$. \ We note that this is consistent with the fact that
the corresponding matter fields on $\Sigma^{\prime}$ are not charged
under the gauge group $G_{S}$. \ Decomposing the adjoint representation of
$U(n+2)$ to $U(n)\times U(1)_{1}\times U(1)_{2}$, we therefore conclude that
there is a six-dimensional field in the $({\bf n},0,+1)$ localized along the matter curve
$(z=0)\cap(z=t_{1})$, another six-dimensional field in the $(\overline{\bf n},-1,0)$ localized
along $(z=0)\cap(z=t_{2})$,\ and a third in the $(\cdot,+1,-1)$ localized
along $(z=t_{1})\cap(z=t_{2})$.

At the common intersection point $t_{1}=t_{2}=0$, the wave functions of the
three matter fields overlap. \ In fact, we can directly generalize the
philosophy in \cite{KatzVafa} to the case at hand and study the theory from
the viewpoint of the gauge theory $G_{p}=U(n+2)$. \ Viewing $t_{1}$ and
$t_{2}$ as scalar fields in the Cartan of $U(2)\subset U(n+2)$, note that in
the limit where the $t_{i}$ vanish, we have an eight-dimensional theory with gauge group
$U(n+2)$. \ Turning on the $t_{i}$'s has the effect of creating fields localized
on each matter curve. \ Note that the eight-dimensional theory with gauge group $U(n+2)$ also
contains a cubic Yukawa coupling between fields localized on matter curves.
\ Evaluating the triple overlap of the three wavefunctions at the common point
of intersection, we find a non-zero contribution to the superpotential in the four dimensional effective theory. We
have thus recovered the perturbatively generated superpotential from those
locations in the geometry where a twofold enhancement in the singularity type occurs.

Note that when $t_{1}$ and $t_{2}$ share a mutual pole, the wave function for
the bifundamental of $U(1)_{1}\times U(1)_{2}$ along the corresponding Riemann
surface can have non-compact support.  In this case, this field appears
as a coupling constant in the compact theory realized on the surface $S$.
\ In the four dimensional effective theory, a non-zero vev for this bifundamental generates
a mass term for the vector-like pair of fields in the fundamental and
anti-fundamental of $U(n)$.

In the above analysis for $A$-type singularities, we have implicitly assumed
that the $t_{i}$ are global sections. \ In general, this may not hold because
only $t_{1}+t_{2}$ and $t_{1}t_{2}$ appear in the local presentation of the
geometry. \ This more general case corresponds to geometries where the $t_{i}$
do not remain invariant under a monodromy in the fiber direction so that the
singularity is not of `split' type. \ As explained in
\cite{BershadskyKachruSadov}, when the singularity is not of split type, the
gauge group is reduced by some outer automorphism which in the perturbative
string setup would correspond to orientifolding an $A_{n}$ singularity.
Because our primary interest in this paper is GUT groups which descend from
an $E$ type singularity, we shall not treat in any detail the action of outer
automorphisms on the corresponding GUT groups.

A similar analysis holds when the highest singularity type of the geometry is
$D_{n}$. In this case, the six-dimensional fields localized on matter curves are bifundamentals
of $SO(2n)\times U(m)$, $SO(2n)\times U(m^{\prime})$ and $U(m)\times U(m^{\prime})$.
As in the case of $A$-type singularities, a cubic coupling corresponds to a mass term
because the last type of bifundamental is non-dynamical in the four-dimensional effective theory.

\subsection{Cubic Couplings from Codimension Two}

Given the above example, we now describe how cubic couplings originate for more general geometries.
As discussed above, we have a sequence of gauge groups%
\begin{equation}
G_{p_{I}}\supset G_{\Sigma_{j}}\times U(1)\supset G_{S}\times U(1)\times
U(1)\text{.}%
\end{equation}
For each $\Sigma_{j}$ there exists a six-dimensional matter field
${\bf \Lambda}_{j}$ which can potentially lead to chiral matter in four dimensions. \ Each
${\bf \Lambda}_{j}$ is part of the adjoint representation of the bigger group
$G_{p_{I}}$.  The interaction term in the theory with gauge group
$G_{p_{I}}$ induces an interaction between the fields ${\bf \Lambda}_{j}^{a}$,
where $a$ denotes a group theory index for $G_{p_{I}}$.
\ Letting $f_{abc}$ denote the structure constants of $G_{p_{I}}$, the
resulting induced interaction term coming from the adjoint interaction for the seven-branes
with gauge group $G_{p_{I}}$ is:
\begin{equation}
f_{abc}{\bf \Lambda}_{j}^{a}{\bf \Lambda}_{k}^{b}{\bf \Lambda}_{l}^{c}\text{.}%
\label{cubicinteraction}%
\end{equation}
We now interpret the geometric content of the above equation.  Recall from the
twisting of the six-dimensional defect theory that each ${\bf \Lambda}_{i}$ transforms as a
section of the bundle $K_{\Sigma_{i}}^{1/2}$.  On the other hand, we will now argue that
there is a canonical identification:
\begin{equation}\label{TRIVIALIZING}
K_{\Sigma_{1}}\otimes K_{\Sigma_{2}}\otimes K_{\Sigma_{3}}|_{p} = {\bf 1}.
\end{equation}
Further, a mild generalization of the discussion in subsection \ref{COSMIC}
near equation \eqref{SPINOR} demonstrates that there is in fact a canonical
choice of square root for each $K_{\Sigma}$ in equation \eqref{TRIVIALIZING}.
This uniquely fixes the overall sign of ${\bf 1}^{1/2}$.  To see equation
\eqref{TRIVIALIZING}, note that equation \eqref{SPINOR} implies:
\begin{equation}
K_{\Sigma_{3}}=N^2_{\Sigma_{3}/S}
\end{equation}
and so we can write:
\begin{equation}
K_{\Sigma_{1}}\otimes K_{\Sigma_{2}}\otimes K_{\Sigma_{3}}|_{p}=(K_{\Sigma_{1}}\otimes N_{\Sigma_{3}/S})|_{p}\otimes
(K_{\Sigma_{2}}\otimes N_{\Sigma_{3}/S})|_{p}= {\bf 1}
\end{equation}
where the last equality follows from the fact that the $\Sigma_i$'s intersect pairwise
inside $S$.

We now describe the form of the resulting superpotential term in four dimensions. \ Letting
${\bf \Lambda}^{\alpha_{j},a}$ denote one of the zero modes of ${\bf \Lambda}_{j}$ on
$\Sigma_{j}$, $\alpha_{j}$ denotes an index which runs over the set of chiral
zero modes and as before, $a$ denotes a group index. \ The resulting cubic
superpotential term between the corresponding chiral fields is therefore given
by:%
\begin{equation}
d_{\alpha_{j},\beta_{k},\gamma_{l}}=f_{abc}{\bf \Lambda}_{j}^{\alpha_{j},a}(p_{jkl}%
){\bf \Lambda}_{k}^{\beta_{k},b}(p_{jkl}){\bf \Lambda}_{l}^{\gamma_{l},c}(p_{jkl})\text{.}%
\end{equation}
with similar conventions for $\beta_{k}$ and $\gamma_{l}$. We now proceed
to some examples of the relevant group theory in the context of $E$-type singularities.

\subsection{Local Cubic Couplings for $E$-Type Singularities}

To illustrate the above formalism, we now show how cubic couplings can
originate from $E$-type singularities. \ First consider the local geometry
mentioned previously where $G_{p}=E_{8}$, $G_{\Sigma}=E_{7}$ and $G_{S}=E_{6}$
so that:%
\begin{equation}
y^{2}=x^{3}+\gamma z^{5}+\beta xz^{3}+\alpha^{2} z^{4}\text{.}%
\end{equation}
Because we may locally trivialize $\gamma$ near the common vanishing locus of
$\alpha$ and $\beta$, we may analyze the matter content of this theory as an
$E_{8}$ gauge theory where $\varphi$ has developed a vev. \ The
deformation in the Cartan which breaks $E_{8}$ to $E_{6}\times U(1)\times
U(1)$ corresponds to the t-direction $\overrightarrow{t} = (t_{1},t_{2},0,0,0,0,0,0)$. Decomposing the
adjoint representation of $E_{8}$ under $E_{6}\times U(1)\times U(1)$, there
are three $\mathbf{27}$'s with $U(1)$ charges $(+1,0)$, $(0,-1)$ and $(-1,+1)$
localized along the matter curves $(t_{1}=0)$, $(t_{2}=0)$ and $(t_{1}%
=-t_{2})$. Note that along each matter curve, the singularity type enhances to
$E_{7}$. As discussed in Section \ref{COUPLINGDEFECT}, the matter curves can
be read off from the geometry by writing $\alpha,\beta,\gamma$ as explicit
functions of the $t_{i}$. \ It now follows that there are three matter curves
in $S$ with the background $U(1)$ charge for each determining the total number
of $\mathbf{27}$'s on each curve. \ The overlap of the three wavefunctions at
the common intersection points $t_{1}=t_{2}=0$ contributes an additional
$\mathbf{27}^{3}$ gauge invariant term to the superpotential. A similar
analysis holds for more general breaking patterns. As another example, when
$G_{p}=E_{7}$, $G_{\Sigma}=E_{6}$ and $G_{S}=SO(10)$, the resulting
superpotential will contain the product ${\bf 16}\times {\bf 16}\times {\bf 10}$. This type of
interaction term can originate from the unfolding:%
\begin{equation}
y^{2}=x^{3}+\gamma xz^{3}+\beta z^{4}+\alpha x^{2}z\text{.}%
\end{equation}

We wish to emphasize that even if the bulk gauge group is of classical $A$- or
$D$-type, a twofold enhancement in rank to an $E$-type singularity can
generate Yukawa couplings which vanish identically in string perturbation
theory. To illustrate this point, we note that while it is certainly
possible to engineer a D-brane construction of an $SU(5)$ GUT model with
three generations of $\overline{\bf 5}$'s and ${\bf 10}$'s as in for example
\cite{Lykkenbraneguts}, a semi-realistic $SU(5)$ GUT must contain
contributions to the superpotential of the form:%
\begin{equation}
W_{SU(5)}\supset\overline{\bf 5}_{H}\times\overline{\bf 5}_{M}\times{\bf 10}_{M}%
\label{pertcoup}%
\end{equation}
and:%
\begin{equation}
W_{SU(5)}\supset{\bf 5}_{H}\times{\bf 10}_{M}\times{\bf 10}_{M}\label{FFTYUK}%
\end{equation}
where the subscripts $H$ and $M$ respectively denote Higgs and matter fields
of the $SU(5)$ GUT. In perturbative type II string theory, the coupling of
(\ref{FFTYUK}) is zero because the group indices only contract in the presence
of a five index $\varepsilon$ tensor so that the purported coupling would
violate the $U(1)$ charge associated with the $U(5)$ gauge symmetry of the
D-brane. \ Note that even if this $U(1)$ lifts via a Green-Schwarz mechanism,
in perturbative string theory it will persist as a global symmetry which can, however,
be violated by non-perturbative effects.

\begin{figure}
[ptb]
\begin{center}
\includegraphics[
height=1.4762in,
width=2.0064in
]%
{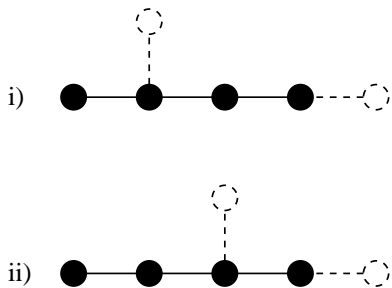}%
\caption{Dynkin diagrams which illustrate how the gauge group $SU(5)$
with corresponding singularity $A_{4}$ can enhance along matter curves
to $A_{5}$ and $D_{5}$ and can undergo a further rank two enhancement
at isolated points to either $D_{6}$ (i) or $E_{6}$
(ii). Whereas the first possibility can generically be realized
in perturbative type II string theory constructions for any rank,
the second case is exceptional and can give rise to Yukawa couplings which
vanish perturbatively.}%
\label{EXCYUK}%
\end{center}
\end{figure}

From the perspective of geometry, we can expect to get ${\bf 5}$'s and $\overline{\bf 5}$'s
from enhancement to an $A_{5}$ singularity and ${\bf 10}$'s from enhancement to a $D_{5}$ singularity.
Whereas the resulting twofold enhancement to a $D$-type singularity
can always be realized in perturbative D-brane constructions, for $SU(5)$ there is another possibility where the
singularity type enhances to $E_{6}$. See figure \ref{EXCYUK} for a depiction of
this enhancement. In such cases, the
$\mathbf{78}^{3}$ interaction of the $E_{6}$ gauge theory will contain a
contribution of the form given by (\ref{FFTYUK}). Note that at other points
of the geometry we can generically expect rank two enhancement to a $D$-type
singularity. These points will lead to superpotential contributions
of the form given by (\ref{pertcoup}). This example illustrates our general
philosophy that even for classical GUT groups, the presence of an $E$-type
singularity in the geometry is important for phenomenology.

\section{A Final Toy Model\label{FINALTOY}}

In this Section we present a final toy model for an $SO(10)$ GUT which
combines many of the ingredients described in the previous Sections. To this
end, we treat the theory of a seven-brane wrapping a del Pezzo 3 surface with
bulk gauge group $G_{S}=SO(12)$. In order to introduce chiral matter with
non-trivial interactions, we analyze a model in which an enhancement by one
rank occurs along some matter curves to both $E_{7}$ and $D_{7}$ type
singularities. To allow a more varied class of interaction terms, we also
require that these matter curves intersect at points such that the singularity
type enhances to $E_{8}$. After introducing a supersymmetric gauge field
configuration to induce a chiral matter spectrum in four dimensions, we
determine the cubic coupling contributions to the superpotential in the
four-dimensional effective theory.

We begin with an analysis of how deformations in the Cartan of the $E_{8}$
singularity localized at some point $p$ descend to a rank 7 singularity along
matter curves and the rank 6 bulk gauge group $SO(12)$. \ The breaking
pattern:%
\begin{equation}
E_{8}\supset SO(12)\times U(1)_{1}\times U(1)_{2}%
\end{equation}
is achieved by noting that $SO(16)$ is a maximal subalgebra of $E_{8}$.
\ Indeed, the adjoint representation of $E_{8}$ decomposes to the adjoint and
spinor representations of $SO(16)$ as:%
\begin{align}
E_{8} &  \supset SO(16)\\
\mathbf{248} &  \rightarrow {\bf 120} + {\bf 128}\text{.}%
\end{align}
Because the adjoint of $E_{8}$ decomposes to $SO(12)$ via:%
\begin{align}
E_{8} &  \supset SO(12)\times SU(2)_{1}\times SU(2)_{2}\\
\mathbf{248} &  \rightarrow(\mathbf{66},\mathbf{1},\mathbf{1})+(\mathbf{1}%
,\mathbf{3},\mathbf{1})+(\mathbf{1},\mathbf{1},\mathbf{3})+(\mathbf{32}%
^{\prime},\mathbf{2},\mathbf{1})+(\mathbf{32},\mathbf{1},\mathbf{2}%
)+(\mathbf{12},\mathbf{2},\mathbf{2})
\end{align}
we can identify four matter curves in $S$ by examining the $U(1)$ charges of
the fields. \ The half hypermultiplet content of each curve $\Sigma_{i}$ for
$i=1,...,4$ is therefore given by a $\bf{32^{\prime}}$ localized along $\Sigma_{1}%
$, a $\bf{32}$ along $\Sigma_{2}$, a $\bf{12}$ along $\Sigma_{3}$ and
another $\bf{12}$ along $\Sigma_{4}$. \ Because the matter content from $\Sigma_{3}$ and $\Sigma_{4}$
is essentially identical, for illustrative purposes we shall assume that the
gauge field configuration has been chosen so that the zero mode content on
$\Sigma_{4}$ is trivial.

We now study vacua with a non-trivial supersymmetric gauge field configuration
in the bulk theory which breaks $SO(12)$ to $SO(10)\times U(1)$. \ The
relevant representations of $SO(12)$ decompose as:%
\begin{align}
\mathbf{66}  &  \rightarrow\mathbf{45}_{0}+\mathbf{10}_{2}+\mathbf{10}%
_{-2}+\mathbf{1}_{0}\\
\mathbf{32}  &  \rightarrow\mathbf{16}_{1}+\overline{\mathbf{16}}_{-1}\\
\mathbf{32}^{\prime}  &  \rightarrow\mathbf{16}_{-1}+\overline{\mathbf{16}%
}_{+1}\\
\mathbf{12}  &  \rightarrow\mathbf{10}_{0}+\mathbf{1}_{2}+\mathbf{1}%
_{-2}\text{.}%
\end{align}

Again invoking the results of Appendix \ref{VanishAppendix}, as an example we
consider supersymmetric gauge field configurations determined by line bundles
of the form:
\begin{equation}
\mathcal{L}=\mathcal{O}_{S}(E_{1}+E_{2}+aE_{3})
\end{equation}
where $a$ is a negative integer.

Labelling the the zero modes corresponding to bulk $\mathbf{10}_{\pm2}$'s by
$\mathbf{10}_{\pm2}^{(S)}$, their representatives are classified by the bundle
valued cohomology groups:%
\begin{equation}
\mathbf{10}_{\pm2}^{(S)}\in H_{\overline{\partial}}^{1}\left(  S,\mathcal{L}%
^{\pm2}\right)  \text{.}%
\end{equation}
The number of massless $\mathbf{10}_{\pm2}^{(S)}$'s in the bulk of $S$ is now
given by the same index computation used in previous toy models:%
\begin{align}
n_{\mathbf{10}_{+2}^{(S)}}  &  =2a^{2}-a+1\\
n_{\mathbf{10}_{-2}^{(S)}}  &  =2a^{2}+a+5\text{.}%
\end{align}

To deduce the matter content along each matter curve, we first specify the
homology class of each $\Sigma_{i}$. \ While it is important to specify the
explicit Calabi-Yau fourfold which realizes such a configuration, for
illustrative purposes we defer such issues to \cite{BHV} and assume that the
effective class in $H_{2}(S,%
\mathbb{Z}
)$ of each matter curve is:%
\begin{align}
\left[  \Sigma_{1}\right]   &  =E_{1}\\
\left[  \Sigma_{2}\right]   &  =H-E_{1}-E_{2}\\
\left[  \Sigma_{3}\right]   &  =3H-E_{1}-E_{2}%
\end{align}
so that $\Sigma_{1}$ and $\Sigma_{2}$ have genus zero and $\Sigma_{3}$ has
genus one.\footnote{The genus $g$ of a smooth curve $C$ in $S$ is given by the
intersection-theoretic formula $C\cdot(C+K_{S})=2g-2$} \ Note that these
classes intersect pairwise as $\left[ \Sigma_{i}\right]\cdot\left[ \Sigma_{j}\right]=+1$
for $i\neq j$. \ Restricting $\mathcal{L}$ to $\Sigma_{1}$ and $\Sigma_{2}$ yields:%
\begin{align}
\mathcal{L}_{\Sigma_{1}}  &  =\mathcal{O}_{\Sigma_{1}}(-1)\\
\mathcal{L}_{\Sigma_{2}}  &  =\mathcal{O}_{\Sigma_{2}}(+2)\text{.}%
\end{align}
For our present purposes, it is enough to note that $\mathcal{L}_{\Sigma_{3}}$
is a degree two line bundle on $\Sigma_{3}$.

We now present the zero mode content localized along each matter curve.
\ Along $\Sigma_{1}$, the number of massless $\mathbf{16}$'s and
$\overline{\mathbf{16}}$'s are classified by the cohomology groups:
\begin{align}
1\times\mathbf{16}_{-1}  &  \in H_{\overline{\partial}}^{0}(\Sigma
_{1},K_{\Sigma_{1}}^{1/2}\otimes\mathcal{O}_{\Sigma_{1}}(+1))=H_{\overline
{\partial}}^{0}(\Sigma_{1},\mathcal{O}_{\Sigma_{1}}(0))\\
0\times\overline{\mathbf{16}}_{+1}  &  \in H_{\overline{\partial}}^{0}%
(\Sigma_{1},K_{\Sigma_{1}}^{1/2}\otimes\mathcal{O}_{\Sigma_{1}}%
(-1))=H_{\overline{\partial}}^{0}(\Sigma_{1},\mathcal{O}_{\Sigma_{1}}(-2))
\end{align}
where we have also indicated the multiplicity of each type of massless mode by
the overall prefactor. \ Summarizing, we find a single $\mathbf{16}$ which
descends from $\Sigma_{1}$.

Along $\Sigma_{2}$, the number of $\mathbf{16}$'s and $\overline{\mathbf{16}}%
$'s are classified by the cohomology groups:%
\begin{align}
2\times\mathbf{16}_{+1}  &  \in H_{\overline{\partial}}^{0}(\Sigma
_{2},K_{\Sigma_{2}}^{1/2}\otimes\mathcal{O}_{\Sigma_{2}}(+2))=H_{\overline
{\partial}}^{0}(\Sigma_{2},\mathcal{O}_{\Sigma_{2}}(1))\\
0\times\overline{\mathbf{16}}_{-1}  &  \in H_{\overline{\partial}}^{0}%
(\Sigma_{2},K_{\Sigma_{2}}^{1/2}\otimes\mathcal{O}_{\Sigma_{2}}%
(-2))=H_{\overline{\partial}}^{0}(\Sigma_{2},\mathcal{O}_{\Sigma_{2}}(-3))
\end{align}
so that two generations transforming in the $\mathbf{16}$ descend from
$\Sigma_{2}$.

To compute the dimensions of the relevant cohomology groups for powers of the
degree two line bundle $\mathcal{L}_{\Sigma_{3}}$ on $\Sigma_{3}$, we first
note that because $K_{\Sigma_{3}}$ is trivial, the Kodaira vanishing theorem
implies:%
\begin{equation}
H_{\overline{\partial}}^{0}(\Sigma_{3},\mathcal{L}_{\Sigma_{3}}^{-n})\simeq
H_{\overline{\partial}}^{1}(\Sigma_{3},\mathcal{L}_{\Sigma_{3}}^{n})^{\ast}=0
\end{equation}
when $n>0$. \ In this particular case, it now follows that either
$h^{0}(\Sigma_{3},\mathcal{L}_{\Sigma_{3}}^{m})$ or $h^{1}(\Sigma
_{3},\mathcal{L}_{\Sigma_{3}}^{m})$ is zero for $m\neq0$ so that the index
theorem of equation (\ref{SIGMAINDEX}) in fact counts the total number of zero
mode solutions. \ Along $\Sigma_{3}$, the number of $\mathbf{1}_{\pm2}$'s are
classified by the cohomology groups:%
\begin{align}
4\times\mathbf{1}_{2}  &  \in H_{\overline{\partial}}^{0}(\Sigma_{3}%
,K_{\Sigma_{3}}^{1/2}\otimes\mathcal{L}_{\Sigma_{3}}^{2})=H_{\overline
{\partial}}^{0}(\Sigma_{1},\mathcal{L}_{\Sigma_{3}}^{2})\\
0\times\mathbf{1}_{-2}  &  \in H_{\overline{\partial}}^{0}(\Sigma
_{1},K_{\Sigma_{1}}^{1/2}\otimes\mathcal{L}_{\Sigma_{3}}^{-2})=H_{\overline
{\partial}}^{0}(\Sigma_{1},\mathcal{L}_{\Sigma_{3}}^{-2})=0
\end{align}
where in computing the multiplicities we have used the fact that:%
\begin{equation}
\underset{\Sigma_{3}}{\int}c_{1}\left(\mathcal{L}_{\Sigma_{3}}\right) =\deg\mathcal{L}_{\Sigma_{3}} = 2\text{.}%
\end{equation}
Even though the $\mathbf{10}_{0}$'s are uncharged under the background gauge
field on $\Sigma_{3}$, because $\Sigma_{3}$ is a genus one curve, we also find
non-trivial zero mode solutions on $\Sigma_{3}$ which are classified by the
cohomology group:%
\begin{equation}
1\times\mathbf{10}_{0}\in H_{\overline{\partial}}^{0}(\Sigma_{3}%
,\mathcal{O}_{\Sigma_{3}})\simeq H_{\overline{\partial}}^{1}(\Sigma
_{3},\mathcal{O}_{\Sigma_{3}})^{\ast}\simeq%
\mathbb{C}
\text{.}%
\end{equation}
In the present context, because the $\mathbf{10}_{0}$ descends from a single
half hypermultiplet on $\Sigma_{3}$, $H_{\overline{\partial}}^{0}(\Sigma
_{3},\mathcal{O}_{\Sigma_{3}})$ and $H_{\overline{\partial}}^{1}(\Sigma
_{3},\mathcal{O}_{\Sigma_{3}})$ classify CPT conjugate particles so that the
\textquotedblleft chiral matter\textquotedblright\ is computed by
$H_{\overline{\partial}}^{0}(\Sigma_{3},\mathcal{O}_{\Sigma_{3}})$. To
summarize, the chiral matter content on $\Sigma_{3}$ is given by four singlets
and a single $\mathbf{10}$.

We now summarize the contributions to the superpotential of the
four-dimensional effective theory:
\begin{equation}
W_{tot}=W_{\Sigma\Sigma S}+W_{\Sigma\Sigma\Sigma}%
\end{equation}
where $W_{\Sigma\Sigma S}$ denotes the contribution from couplings between two
fields on a Riemann surface and a bulk field in $S$, and $W_{\Sigma
\Sigma\Sigma}$ denotes the contribution from the triple overlap of zero modes
localized at the intersection point. \ The contribution to $W_{\Sigma\Sigma
S}$ is:%
\begin{align}
W_{\Sigma\Sigma S}  &  =\lambda^{(1)}\mathbf{16}_{-1}\times\mathbf{16}%
_{-1}\times\mathbf{10}_{+2}^{(S)}\\
&  +\lambda_{ij}^{(2)}\mathbf{16}_{+1}^{(i)}\times\mathbf{16}_{+1}^{(j)}%
\times\mathbf{10}_{-2}^{(S)}\\
&  +\lambda_{k}^{(3)}{\bf 1}_{+2}^{(k)}\times\mathbf{10}_{0}\times\mathbf{10}%
_{-2}^{(S)}%
\end{align}
where the superscript on the $\lambda$'s labels the contribution from each
matter curve, and $i,j=1,2$ label the two zero modes in the $\mathbf{16}$
localized along $\Sigma_{2}$ and $k=1,...,4$ labels the four zero modes in the
singlet localized along $\Sigma_{3}$. \ The contribution to $W_{\Sigma
\Sigma\Sigma}$ is:%
\begin{equation}
W_{\Sigma\Sigma\Sigma}=\alpha_{i}\mathbf{16}_{-1}\times\mathbf{16}_{+1}%
^{(i)}\times\mathbf{10}_{0}%
\end{equation}
where in the above equation, all wave functions for the zero modes are
evaluated at the point $p$ and $\alpha_{i}$ includes the contribution from the
structure constants of $E_{8}$ restricted to $SO(10)$.

\section{Conclusions\label{CONCLUDE}}

In this paper we have developed a framework for constructing GUT models from
local compactifications of F-theory in terms of a partially twisted eight-dimensional theory
with codimension one and two defects. \ In the presence of a non-trivial gauge
bundle, both the pure eight-dimensional theory and theory with defects admit a chiral matter
spectrum. \ When the K\"{a}hler surface of the partially twisted theory is a del Pezzo
or Hirzebruch surface, non-zero interaction terms require the presence of codimension one
defects. \ In the absence of codimension two defects, the possible interaction
terms are greatly limited because two fields localized along a single matter
curve must participate in each such Yukawa coupling. \ More general
couplings are possible in the presence of codimension two defects. \ In
this case interaction terms can arise from the triple overlap of matter curves
at a single point of the bulk gauge theory. \ In addition to providing
explicit examples of various model building ingredients, we have also shown
that the degrees of freedom of the partially twisted eight-dimensional theory precisely match
to the unfolding of the singularity type of the F-theory geometry. \ In
particular, we have demonstrated that brane recombination for both
perturbative and non-perturbatively realized gauge theories in F-theory admits
both a consistent gauge theory and geometric interpretation in terms of
field vevs with a given pole structure. \ In the remainder of this Section we
discuss some possible applications and extensions of the above work.

Although our ultimate goal is the construction of semi-realistic GUT models,
the present investigation has also established a beautiful connection between
a partially twisted eight-dimensional theory and the unfolding of geometric singularities in
F-theory. \ Along these lines, the appearance of
fields localized along surface operators, and the presence of point-like \textquotedblleft
defects\textquotedblright\ in the partially twisted eight-dimensional theory seem quite
natural from the perspective of F-theory. \ It would be interesting to
determine how the intersection of surface operators is described in the four-dimensional
topological theory.

In the context of more phenomenological applications, it is also important to
study supersymmetry breaking in the present context. \ While we have
restricted attention to supersymmetry preserving vacua which satisfy a
modified Hitchin-like system with a source term along a six-dimensional defect,
non-supersymmetric instanton solutions of the internal portion of the eight-dimensional
theory would correspond to metastable or stable supersymmetry breaking vacua
in the four-dimensional effective theory. \ It would be interesting to develop additional
features of such a scenario.

Finally, while for concreteness the toy models presented have all been based on
cases where $S$ is a del Pezzo surface, much of the analysis we have presented
is of a more general scope and directly applies to a wide class of F-theory
compactifications which may not possess a heterotic dual.  Anticipating potential
applications to GUTs, note that when $h^{2,0}(S)\neq 0$, additional moduli corresponding
to the deformation of $S$ in the compactification will in general be present.  From
the perspective of the GUT group, such fields are charged under the adjoint
representation.  In the presence of a suitable background flux, these moduli can
be frozen to a non-zero vev.  These fields can then play the role of more traditional
GUT Higgs fields used to break the GUT group down to the Standard Model gauge group.

\section*{Acknowledgements}

We thank B. Andreas, D.R. Morrison, M. Wijnholt and E. Witten for helpful discussions.  We would also
like to thank the Fifth Simons Workshop in Mathematics and Physics, where this work was
initiated.\ The work of the authors is
supported in part by NSF grants PHY-0244821 and DMS-0244464. \ The research of
JJH\ is also supported by an NSF\ Graduate Fellowship.

\appendix

\section*{Appendices}

\section{Review of Hirzebruch and del Pezzo
  Surfaces\label{APPDELPEZZO}}

We now briefly review some properties of Hirzebruch and del Pezzo surfaces.
For $n\geq0$, the middle homology of the Hirzebruch surfaces $\mathbb{F}_{n}$
is generated by the effective classes $f$ and $\sigma$ with intersection pairings:
\begin{equation}
f\cdot f=0, f\cdot \sigma=1, \sigma\cdot \sigma=-n\text{.}
\end{equation}
The canonical divisor of $\mathbb{F}_{n}$ is given by:
\begin{align}
K_{S} = -c_{1}(S) = -(n+2)f - 2\sigma\text{.}
\end{align}
The effective classes correspond to 2-cycles $af + b\sigma$ such that
$a$ and $b$ are both non-negative with at least one non-trivial.

A del Pezzo surface is defined by the condition that $-K_{S}>0$. \
This condition is satisfied for the surfaces
$\mathbb{F}_{0}=\mathbb{P}^{1}\times\mathbb{P}^{1}$ and
$dP_{0}=\mathbb{P}^{2}$. \ There are eight additional del Pezzo
surfaces given by blowing up $\mathbb{P}^2$ at up to eight points in
general position. For $n>0$, The middle homology of the del Pezzo $n$
($dP_{n}$) surface is generated by the hyperplane class $H$ and the
exceptional classes $E_{1},...,E_{n}$ with intersection pairing:
\begin{equation}
H\cdot H=1, H\cdot E_{i}=0, E_{i}\cdot E_{j}=-\delta_{ij}\text{.}
\end{equation}
The canonical divisor on $dP_{n}$ is given by:
\begin{equation}
K_{S}=-c_{1}(dP_{n})=-3H+\underset{i=1}{\overset{n}{\sum}}E_{i}\text{.}
\end{equation}

The class of any effective curve $C$ in $dP_n$ reduces to a sum over the
generators $C_{i}$ of the K\"ahler cone as:
\begin{equation}
C=\underset{i}{%
{\displaystyle\sum}
}n_{i}C_{i}%
\end{equation}
where each $n_{i}\geq0$. \ The generators of the K\"ahler cone for
each del Pezzo $n\geq1$ are well-known in the mathematics literature:
\[%
\begin{tabular}
[c]{|l|l|}\hline
Surface & {Generators}\\\hline
$dP_{1}$ & $E_{1},H-E_{1}$\\\hline
$dP_{2}$ & $E_{i},H-\underset{j=1}{\overset{2}{\sum}}E_{i_{j}}$\\\hline
$dP_{3}$ & $E_{i},H-\underset{j=1}{\overset{2}{\sum}}E_{i_{j}}$\\\hline
$dP_{4}$ & $E_{i},H-\underset{j=1}{\overset{2}{\sum}}E_{i_{j}}$\\\hline
$dP_{5}$ & $E_{i},H-\underset{j=1}{\overset{2}{\sum}}E_{i_{j}},2H-\underset
{j=1}{\overset{5}{\sum}}E_{i_{j}}$\\\hline
$dP_{6}$ & $E_{i},H-\underset{j=1}{\overset{2}{\sum}}E_{i_{j}},2H-\underset
{j=1}{\overset{5}{\sum}}E_{i_{j}}$\\\hline
$dP_{7}$ & $E_{i},H-\underset{j=1}{\overset{2}{\sum}}E_{i_{j}},2H-\underset
{j=1}{\overset{5}{\sum}}E_{i_{j}},3H-2E_{i}-\underset{j=1}{\overset{6}{\sum}%
}E_{i_{j}}$\\\hline
$dP_{8}$ & $%
\begin{array}
[c]{c}%
E_{i},H-\underset{j=1}{\overset{2}{\sum}}E_{i_{j}},2H-\underset{j=1}%
{\overset{5}{\sum}}E_{i_{j}},3H-2E_{i}-\underset{j=1}{\overset{6}{\sum}%
}E_{i_{j}},\\
4H-2\left(  E_{i}+E_{j}+E_{k}\right)  -\underset{j=1}{\overset{5}{\sum}%
}E_{i_{j}}%
\end{array}
$\\\hline
\end{tabular}
\]
where all indices are distinct. \ We note in passing that the number of
generators of the K\"ahler cone for each del Pezzo are
$1,2,3,6,10,16,27,56,240$.

A common formula which is used in the text to compute the chiral matter content
induced in the presence of a supersymmetric line bundle is given by
the Todd genus on a surface $S$:
\begin{equation}
\underset{S}{\int} Td(S) = \underset{S}{\int}\frac{c_{1}(S)^2+c_{2}(S)}{12}=1-h^{0,1}+h^{0,2} = \chi(S,\mathcal{O}_{S})
\end{equation}
which is the holomorphic Euler characteristic.  On the Hirzebruch and
del Pezzo surfaces, $h^{0,1} = h^{0,2} = 0$ so that the Todd genus is $1$.

\section{On-Shell Twisted Supersymmetries of the\\ Eight-Dimensional
  Theory\label{TWISTEDSUSYS}}

In this appendix we present our conventions for the action of the
supercharges $Q_\alpha$ and $\bar Q{}_{\dot\alpha}$, for
${\alpha,\dot\alpha=1,2}$, on the bosons and fermions of the
eight-dimensional, partially twisted Yang-Mills theory.
As standard, we denote the action of $Q_\alpha$ and $\bar
Q{}_{\dot\alpha}$ by
${\delta_\alpha (\,\cdot\,)\,=\,[Q_\alpha\,,\,\cdot\,\}}$ and
${\bar\delta{}_{\dot\alpha}(\,\cdot\,) \,=\, [\bar
  Q{}_{\dot\alpha}\,,\,\cdot\,\}}$.

The action by $\delta_\alpha$ and $\bar\delta{}_{\dot\alpha}$ on
the bosons and fermions in the twisted Yang-Mills theory on
${\BR^{3,1} \times S}$ follows immediately from the reduction of the on-shell
supersymmetry transformations in the maximally supersymmetric
ten-dimensional Yang-Mills theory.

First, we find that the supersymmetry transformations of the
eight-dimensional gauge field $A$ and the twisted scalar fields
$(\varphi, \bar\varphi)$ are given by
\begin{align}\label{BSUSY}
&\delta_{\alpha} A_m \,=\, 0\,, & &\bar\delta{}_{\dot\alpha} A_m \,=\,
\sqrt{2} \, \bar\psi{}_{\dot\alpha\,m}\,,\cr
&\delta_{\alpha} A_{\bar m} \,=\, \sqrt{2} \, \psi_{\alpha\,\bar m}\,, &
&\bar\delta{}_{\dot\alpha} A_{\bar m} \,=\, 0\,,\cr
&\delta_{\alpha} A_\mu \,=\, i \, (\sigma_\mu)_{\alpha\dot\alpha}
\, \bar\eta{}^{\dot\alpha}\,, & &\bar\delta{}_{\dot\alpha} A_\mu \,=\,
-i \, (\sigma_\mu)_{\alpha\dot\alpha} \, \eta^{\alpha}\,,\cr
&\delta_{\alpha} \varphi_{m n} \,=\, \sqrt{2} \, \chi_{\alpha \, m n}\,, &
&\bar\delta{}_{\dot\alpha} \varphi_{m n} \,=\, 0\,,\cr
&\delta_{\alpha} \bar\varphi{}_{\bar m \bar n} \,=\, 0\,, &
&\bar\delta{}_{\dot\alpha} \bar\varphi{}_{\bar m \bar n} \,=\, \sqrt{2} \,
\bar\chi{}_{\dot\alpha \, \bar m \bar n}\,.
\end{align}
Here $A_\mu$ for ${\mu=0,\ldots,3}$ are the components of the gauge
field on $\BR^{3,1}$, and $(A_m, A_{\bar m})$ for ${m=1,2}$ are the
components of the gauge field on $S$.  Also,
$(\sigma_\mu)_{\alpha\dot\alpha}$ for ${\mu=0,\ldots,3}$ are the
standard Pauli matrices which represent the Clifford algebra on
$\BR^{3,1}$, as defined for instance in Appendix B of \cite{WESSBAGGER}.
Up to rescalings of the fermions, the supersymmetry
transformations in \eqref{BSUSY} are the only possibility consistent
with the topological twist.  Finally, we have included various factors
of $\sqrt{2}$ in \eqref{BSUSY} so that our conventions for the
effective ${\CN=1}$ supersymmetry algebra on $\BR^{3,1}$ agree
with the conventions in \cite{WESSBAGGER}.

Similarly, the on-shell supersymmetry transformations of the fermions
are given by
\begin{align}\label{FSUSY}
&\delta_\alpha \eta_\beta \,=\, \big(\sigma^{\mu
\nu}\big){}_{\alpha \beta} \, F_{\mu \nu} \,-\, i \, \epsilon_{\alpha
\beta}\, \CD\,, &
&\bar\delta{}_{\dot\alpha} \eta_\beta \,=\, 0\,,\cr
&\delta_\alpha \bar\eta_{\dot\beta} \,=\, 0\,, &
&\bar\delta{}_{\dot\alpha} \bar\eta{}_{\dot\beta} \,=\,
(\bar\sigma^{\mu \nu}){}_{\dot\alpha \dot\beta} \, F_{\mu \nu} \,+\,
i\, \epsilon_{\dot\alpha \dot\beta}\,\CD\,,\cr
&\delta_\alpha \psi{}_{\beta\,\bar m} \,=\, \sqrt{2}\, \epsilon_{\alpha \beta}
\, (\bar\partial{}_A^\dagger \bar\varphi){}_{\bar m}\,, &
&\bar\delta{}_{\dot\alpha} \psi_{\beta\,\bar m} \,=\, i \sqrt{2} \,
(\sigma^\mu){}_{\beta \dot\alpha} \, F_{\mu \bar m}\,,\cr
&\delta_\alpha \bar\psi{}_{\dot\beta\,m} \,=\, i \sqrt{2} \,
(\sigma^\mu)_{\alpha\dot\beta} \, F_{\mu m}\,, &
&\bar\delta{}_{\dot\alpha} \bar\psi{}_{\dot\beta\,m} \,=\, \sqrt{2}
\, \epsilon_{\dot\alpha \dot\beta} \, (\partial{}_A^\dagger\varphi)_m\,,\cr
&\delta_\alpha \chi_{\beta\,m n} \,=\, -\sqrt{2} \, \epsilon_{\alpha
\beta} \, F_{m n}\,, &
&\bar\delta{}_{\dot\alpha} \chi_{\beta\,m n} \,=\, i \sqrt{2} \,
(\sigma^\mu)_{\beta\dot\alpha} \, D_\mu \varphi_{m n}\,,\cr
&\delta_\alpha \bar\chi{}_{\dot \beta \, \bar m \bar n} \,=\, i
\sqrt{2} \, (\sigma^\mu)_{\alpha\dot\beta} \, D_\mu \bar\varphi_{\bar m \bar
  n}\,, &
&\bar\delta{}_{\dot\alpha} \bar\chi_{\dot \beta \, \bar m \bar n}
\,=\, -\sqrt{2} \, \epsilon_{\dot\alpha \dot\beta} \, F_{\bar m \bar n}\,.
\end{align}
Let us explain the notation in \eqref{FSUSY}.  First, $\CD$
is shorthand for the section of $\ad(P)$ on $S$ given by
\begin{equation}\label{BIGD}
\CD \,=\, -\*_S\!\left(\omega \^ F_S \,+\, {i\over
  2}\,[\varphi\,,\bar\varphi]\right)\,.
\end{equation}
In the above, $\omega$ is the K\"ahler form on $S$, and $\*_S$ denotes the
duality operator on $S$ defined by the given K\"ahler metric.  Also,
we let ${F_S \equiv F|_S}$ denote the restriction of the curvature to
$S$, so that $\omega \^ F_S$ is a top-form on $S$ upon which $\*_S$
naturally acts.  The overall sign in \eqref{BIGD} is just a
convention.

Throughout the paper, we use $(\partial_A, \bar\partial{}_A)$ to
indicate the $(1,0)$ and $(0,1)$ components of the covariant
derivative defined by the gauge field on $S$, and we let
$(\partial{}_A^\dagger,\bar\partial{}_A^\dagger)$ be the adjoint
operators defined using $\*_S$.  Explicitly, ${\bar\partial{}_A^\dagger
= -\*_S\,\bar\partial{}_A^{}\*_S}$, and similarly for
$\partial_A^\dagger$.  In local coordinates $(s^m, \bar s{}^{\bar m})$
on $S$, ${(\partial_A^\dagger \varphi){}_m \,=\, g^{n \bar n}
  \,(\bar\partial{}_A){}_{\bar n}\,\varphi_{m n}}$, where $g_{n \bar
  n}$ is the K\"ahler metric.

Finally, we use $D_\mu$ to indicate the covariant derivative defined
by the gauge field along $\BR^{3,1}$, and we normalize the self-dual
and anti-self-dual projection operators
$(\sigma^{\mu\nu})_{\alpha\beta}$ and
$(\bar\sigma^{\mu\nu})_{\dot\alpha\dot\beta}$ appearing in
\eqref{FSUSY} as in \cite{WESSBAGGER}.  Also, $\epsilon_{\alpha\beta}$
and $\epsilon_{\dot\alpha\dot\beta}$ are the usual anti-symmetric tensors on
two indices.

\section{Partially Twisted Action of the Seven-Brane
  Theory\label{8DACTION}}

In this Appendix we determine the Lagrangian of the partially twisted
Yang-Mills theory on $\mathbb{R}^{3,1}\times S$ with gauge group
$G_{S}$. In the limit where the fields of the partially twisted theory
are independent of the coordinates on $\BR^{3,1}$, the reduction to
$S$ is given by the twisted version of four-dimensional, ${\CN=4}$
supersymmetric Yang-Mills theory studied in \cite{VafaWitten}.  One
purpose of the present Appendix is to explicitly demonstrate the
quasi-topological nature of the partially twisted eight-dimensional
theory.  Other twisted eight-dimensional theories on $\Spin(7)$
manifolds and Calabi-Yau fourfolds have been studied for instance in
\cite{SingerEIGHTD, AcharyaDB}.

The supersymmetric Lagrangian for the twisted Yang-Mills theory on
${\BR^{3,1} \times S}$ is most conveniently written once we pass to a
(mostly) off-shell formulation of the supersymmetry algebra.  The
existence of an off-shell formulation of the supersymmetry
transformations in \eqref{BSUSY} and \eqref{FSUSY} is not so
surprising, since the off-shell formulation is modeled on the standard
off-shell formulation of ${\CN=1}$ supersymmetry in four dimensions.

\bigskip\noindent{\it Auxiliary Fields}\smallskip

We first introduce auxiliary bosonic fields $(\CG\,, \CH\,, \CD)$.  Here
${\CG = \CG_{\bar m} \, d\bar s{}^{\bar m}}$ is a complex boson which
transforms on $S$ as a section of ${\bar\Omega^1_S \otimes \ad(P)}$,
and ${\CH = \CH_{m n} \, ds^m \^ ds^n}$ is another complex boson which
transforms on $S$ as a section of ${\Omega^2_S \otimes \ad(P)}$.  From
the perspective of the four-dimensional effective theory on
$\mathbb{R}^{3,1}$, $\CG$ and $\CH$ are the auxiliary components of
${\CN=1}$ chiral superfields
\begin{align}\label{CHRL}
{\bf A}_{\bar m} \,&=\, A_{\bar m} \,+\, \sqrt{2} \, \theta
\psi_{\bar m} \,+\, \theta\theta \, \CG_{\bar m} \,+\,
\cdots\,,\cr
{\mathbf \Phi}_{m n} \,&=\, \varphi_{m n} \,+\, \sqrt{2} \, \theta
\chi_{m n} \,+\, \theta\theta \, \CH_{m n} \,+\, \cdots\,,
\end{align}
where we use `$\cdots$' to indicate the usual higher-order terms in the
chiral superfields.  While the presentation in \eqref{CHRL} is
convenient for describing certain properties of the four-dimensional
effective theory, we note that $A_{\bar m}$ does not transform
covariantly under arbitrary gauge transformations on ${\BR^{3,1}
  \times S}$ but rather only under those which are constant along $S$.
As a result, we will have to be careful about eight-dimensional
gauge-invariance when we construct the action.

Along with $\CG$ and $\CH$, we also introduce the conjugate bosons
${\bar\CG = \bar\CG{}_m\,ds^m}$ and ${\bar\CH = \bar\CH{}_{\bar m \bar
    n} \, d\bar s{}^{\bar m} \^ d\bar  s{}^{\bar n}}$ which transform
as sections of ${\Omega^1_S \otimes \ad(P)}$ and
${\bar\Omega{}^2_S\otimes \ad(P)}$, respectively.  These bosons appear
as the auxiliary components in anti-chiral superfields conjugate to
those in \eqref{CHRL}.

We finally introduce a real scalar field $\CD$ which transforms on $S$ as a
section of $\ad(P)$.  In the four-dimensional effective theory, $\CD$ is the
auxiliary field of the ${\CN=1}$ vector multiplet, appearing in WZ
gauge as
\begin{equation}\label{VECT}
V \,=\, -\theta \sigma^\mu \bar\theta \, A_\mu \,+\, i\,
\theta\theta\bar\theta\bar\eta \,-\, i\,
\bar\theta\bar\theta\theta\eta \,+\, \frac{1}{2}\,
\theta\theta\bar\theta\bar\theta\,\CD\,.
\end{equation}
Eventually, the auxiliary scalar field $\CD$ in \eqref{VECT} will
be identified on-shell with the expression in \eqref{BIGD}.

In terms of the auxiliary bosons $(\CG\,, \CH\,, \CD)$, the
supersymmetry transformations \eqref{FSUSY} of the twisted fermions
are now given by
\begin{align}\label{FSUSYAUX}
&\delta_\alpha \eta_\beta \,=\, \big(\sigma^{\mu
\nu}\big){}_{\alpha \beta} \, F_{\mu \nu} \,-\, i \, \epsilon_{\alpha
\beta}\, \CD\,, &
&\bar\delta{}_{\dot\alpha} \eta_\beta \,=\, 0\,,\cr
&\delta_\alpha \bar\eta_{\dot\beta} \,=\, 0\,, &
&\bar\delta{}_{\dot\alpha} \bar\eta{}_{\dot\beta} \,=\,
(\bar\sigma^{\mu \nu}){}_{\dot\alpha \dot\beta} \, F_{\mu \nu} \,+\,
i\, \epsilon_{\dot\alpha \dot\beta}\,\CD\,,\cr
&\delta_\alpha \psi{}_{\beta\,\bar m} \,=\, -\sqrt{2}\, \epsilon_{\alpha\beta}
\,\CG_{\bar m}\,, &
&\bar\delta{}_{\dot\alpha} \psi_{\beta\,\bar m} \,=\, i \sqrt{2}
(\sigma^\mu){}_{\beta \dot\alpha} \, F_{\mu \bar m}\,,\cr
&\delta_\alpha \bar\psi{}_{\dot\beta\,m} \,=\, i \sqrt{2}
(\sigma^\mu)_{\alpha\dot\beta} \, F_{\mu m}\,, &
&\bar\delta{}_{\dot\alpha} \bar\psi{}_{\dot\beta\,m} \,=\, -\sqrt{2}\,
\epsilon_{\dot\alpha\dot\beta} \, \bar\CG_m\,,\cr
&\delta_\alpha \chi_{\beta\,m n} \,=\, -\sqrt{2}\,\epsilon_{\alpha\beta}
\, \CH_{m n}\,, &
&\bar\delta{}_{\dot\alpha} \chi_{\beta\,m n} \,=\, i \sqrt{2}
(\sigma^\mu)_{\beta\dot\alpha} \, D_\mu \varphi_{m n}\,,\cr
&\delta_\alpha \bar\chi{}_{\dot \beta \, \bar m \bar n} \,=\, i \sqrt{2}
(\sigma^\mu)_{\alpha\dot\beta} \, D_\mu \bar\varphi_{\bar m \bar
  n}\,, &
&\bar\delta{}_{\dot\alpha} \bar\chi_{\dot \beta \, \bar m \bar n}
\,=\,-\sqrt{2}\, \epsilon_{\dot\alpha\dot\beta} \, \bar\CH_{\bar m \bar n}\,.
\end{align}

Ideally, the supersymmetry transformations of the auxiliary bosons
$(\CG\,, \CH\,, \CD)$ are determined by the condition that
$\delta_\alpha$ and $\bar\delta{}_{\dot\alpha}$ satisfy the ${\CN=1}$
supersymmetry algebra
\begin{equation}\label{OFFALG}
\left\{\delta_\alpha\,,\delta_\beta\right\} \,=\,
\left\{\bar\delta{}_{\dot\alpha}\,, \bar\delta{}_{\dot\beta}\right\} \,=\,
0\,,\qquad\qquad
\left\{\delta_\alpha\,,\bar\delta{}_{\dot\alpha}\right\} \,=\, 2 i \,
(\sigma^\mu)_{\alpha\dot\alpha} \, D_\mu\,.
\end{equation}
However, as is well-known, the algebra in \eqref{OFFALG} cannot be
realized off-shell on the ${\CN=1}$ vector multiplet in WZ gauge.  We
know of two ways to work around this problem, such that we either preserve
manifest supersymmetry or manifest gauge-invariance at all stages of
the computation.

One strategy suggested in \cite{SeigelTEND,WackerGregoire,DijkgraafVafaDecon}
is to treat the eight-dimensional fields of the partially twisted
theory as a collection of four-dimensional fields
on $\mathbb{R}^{3,1}$ labelled by coordinates on $S$.  Organizing
these four-dimensional fields into ${\CN=1}$ superfields, the
superspace action for the partially twisted Yang-Mills theory
on ${\BR^{3,1} \times S}$ manifestly preserves supersymmetry.
However, the superspace description of the twisted Yang-Mills action
on ${\BR^{3,1} \times S}$ also obscures the higher-dimensional gauge
invariance of the theory.

Because we prefer to maintain manifest gauge-invariance over manifest
supersymmetry, in this Appendix we pursue an alternative strategy.
Rather than ask for an off-shell formulation of the full ${\CN=1}$
supersymmetry algebra in \eqref{OFFALG}, we consider only the simpler algebra
\begin{equation}\label{OFFALGII}
\left\{\bar\delta_{\dot\alpha}\,,\bar\delta_{\dot\beta}\right\} \,=\,
0\,.
\end{equation}
The simple algebra in \eqref{OFFALGII} does close off-shell when the
auxiliary bosons $(\CG,\CH,\CD)$ transform as
\begin{align}\label{AUXSUSY}
&\bar\delta{}_{\dot\alpha} \CG_{\bar m} \,=\, i \sqrt{2} \,
(\sigma^\mu)_{\alpha\dot\alpha} \, D_\mu \psi^\alpha_{\bar m} \,+\, 2
\big(\bar\partial{}_A \bar\eta{}_{\dot\alpha}\big){}_{\bar m}\,,\cr
&\bar\delta{}_{\dot\alpha} \bar\CG{}_m \,=\, 0\,,\cr
&\bar\delta{}_{\dot\alpha} \CH_{m n} \,=\,  i \sqrt{2}
(\sigma^\mu)_{\alpha\dot\alpha} \, D_\mu \chi^\alpha_{m n} \,+\, 2
\big[\varphi_{m n}\,,\bar\eta{}_{\dot\alpha}\big],\cr
&\bar\delta{}_{\dot\alpha} \bar\CH{}_{\bar m \bar n} \,=\, 0\,\cr
&\bar\delta{}_{\dot\alpha} \CD \,=\, -(\sigma^\mu)_{\alpha\dot\alpha}\,
D_\mu\eta^\alpha\,.
\end{align}
The supersymmetry transformations in \eqref{AUXSUSY} are merely
covariant versions of the usual transformations for the auxiliary
bosons in the ${\CN=1}$ chiral and vector multiplets.

For future reference, we also recall the conjugate supersymmetry
transformations under $\delta_\alpha$,
\begin{align}\label{AUXSUSYII}
&\delta_\alpha \CG_{\bar m} \,=\, 0\,,\cr
&\delta_\alpha \bar\CG{}_m \,=\, i \sqrt{2} \,
(\sigma^\mu)_{\alpha\dot\alpha} \, D_\mu \bar\psi{}^{\dot\alpha}_m + 2
\big(\partial{}_A \eta_\alpha\big){}_m\,,\cr
&\delta_\alpha \CH_{m n} \,=\, 0\,\cr
&\delta_\alpha \bar\CH{}_{\bar m \bar n} \,=\, i \sqrt{2} \,
(\sigma^\mu)_{\alpha\dot\alpha} \, D_\mu \bar\chi{}^{\dot\alpha}_{\bar
  m \bar n} \,+\, 2 \big[\bar\varphi{}_{\bar m \bar
  n}\,,\eta_\alpha\big]\,,\cr
&\delta_\alpha \CD \,=\, -(\sigma^\mu)_{\alpha\dot\alpha}\,
D_\mu \bar\eta{}^{\dot\alpha}\,.
\end{align}

\bigskip\noindent{\it Constructing the Supersymmetric Action}\smallskip

Despite the fact that \eqref{FSUSYAUX} and \eqref{AUXSUSY} only
provide an off-shell realization for the simple algebra involving
$\bar\delta_{\dot\alpha}$ in \eqref{OFFALGII}, the partially twisted
action on ${\BR^{3,1} \times S}$ must be invariant under both
$\delta_\alpha$ and $\bar\delta{}_{\dot\alpha}$.  To construct an
action invariant under both types of transformations, we use the fact
that $\delta_\alpha$ is conjugate to $\bar\delta{}_{\dot\alpha}$.
As a result, any action $I_S$ which is both real and annihilated by
$\delta_\alpha$ is automatically annihilated by $\bar\delta{}_{\dot\alpha}$.

At weak coupling, the partially twisted action serves to enforce the BPS
equations of motion described in subsection \ref{BPSEOM}.  In order to
make the quasi-topological nature of the partially twisted theory manifest, we
organize the action into three types of terms.  The first two types of terms
correspond to $\bar\delta{}^{2}$ or $\bar\delta{}_{\dot\alpha}$ exact
operators so that by construction such terms are annihilated by
$\bar\delta{}_{\dot\alpha}$.  In terms of the effective theory in four
dimensions, the final class of terms corresponds to integrating the
superpotential over chiral superspace.  These terms are
$\bar\delta{}_{\dot\alpha}$ closed but not exact.

We now present the $\bar\delta{}^{2}$ trivial terms of $I_S$.
Many of these terms contribute to terms in the action serve to enforce
the BPS equations motion.  To this end, we first introduce the
operator
\begin{equation}\label{OONE}
\CO^{(1)} \,=\, \frac{1}{4} \, \underset{\BR^{3,1} \times S}{\int} d^4x \>
\Tr\!\left(\CH \^ \bar\varphi\right)\,.
\end{equation}
We now compute
\begin{align}\label{IONE}
I_1 \,&=\, \bar\delta{}^2 \CO^{(1)}\,,\cr
&=\, \underset{\BR^{3,1} \times S}{\int} d^4 x \> \Tr\Big(
\CH \^ \bar\CH \,-\, D^\mu \bar\varphi \^ D_\mu \varphi \,+\, i\,
  [\varphi\,,\bar\varphi] \, \CD \,+\,\cr
&\qquad\qquad\,+i\, D_\mu \chi^\alpha \, (\sigma^\mu)_{\alpha \dot\alpha} \,
\bar\chi^{\dot\alpha} \,-\,  \sqrt{2} \, \eta^\alpha \, [\bar\varphi\,,
\chi_\alpha] \,-\, \sqrt{2} \, \bar\eta{}_{\dot\alpha} \, [\varphi\,,
\bar\chi{}^{\dot\alpha}]\Big)\,.
\end{align}
As expected, the integrands of \eqref{OONE} and \eqref{IONE} are differential
forms of top-degree on $S$.

The role of $I_1$ is to produce a Gaussian action for $\CH$ and
$\bar\CH$, along with certain kinetic terms for $\varphi$ and the
fermions.  In order to reproduce the relations $F^{(2,0)}_S = 0$ and
${F^{(0,2)}_S = 0}$, we next introduce source terms for $\CH$ and
$\bar\CH$ which enforce these conditions when $\CH$ and $\bar\CH$ are
integrated out.  One of these source terms is also
$\bar\delta{}^2$-trivial and descends from the operator
\begin{equation}
\CO^{(2)} \,=\, -{1 \over 4} \, \underset{\BR^{3,1} \times S}{\int} d^4 x \>
\Tr\Big(F_S^{(2,0)} \^ \bar\varphi\Big)\,,
\end{equation}
so that its contribution to the action is
\begin{align}\label{ITWO}
I_2 \,&=\, \bar\delta{}^2 \CO^{(2)}\,,\cr
&=\, -\underset{\BR^{3,1} \times S}{\int} d^4 x \> \Tr\Big(
F_S^{(2,0)} \^ \bar\CH \,+\, \bar \CG \^ \partial_A \bar\varphi \,+\, \cr
&\qquad\qquad\qquad\qquad+\,
\bar\chi{}_{\dot\alpha} \^ \partial_A \bar\psi{}^{\dot\alpha} \,-\,
\frac{1}{2}\,\bar\psi{}_{\dot\alpha} \, \big[\bar\varphi\,,
\bar\psi{}^{\dot\alpha}\big]\Big)\,.
\end{align}

The final $\bar\delta{}^2$-trivial term which contributes to $I_{S}$
serves to reproduce the standard action for the four-dimensional
$\mathcal{N}=1$ vector multiplet in $\mathbb{R}^{3,1}$.  Explicitly,
this term descends from the operator
\begin{equation}\label{OTHREE}
\CO^{(3)} \,=\, \frac{1}{8} \underset{\BR^{3,1} \times S}{\int} d^4 x \>
\omega\^\omega\,\Tr\big(\bar\eta{}_{\dot\alpha}\,\bar\eta{}^{\dot\alpha}\big)\,.
\end{equation}
We note that because $\bar\eta_{\dot\alpha}$ is a zero-form on $S$, we
must introduce two powers of the K\"{a}hler form $\omega$ on $S$ to
obtain an appropriate measure for the integral in \eqref{OTHREE}.  As
completely standard for the ${\CN=1}$ vector multiplet in four
dimensions, we obtain
\begin{align}\label{ITHREE}
I_3 \,&=\, \bar\delta{}^2 \CO^{(3)}\,,\cr
&= \underset{\BR^{3,1} \times S}{\int} d^4x \> \omega\^\omega \,\,
\Tr\Big( \ha \, \CD^2 \,-\, \frac{1}{4}\,F^{\mu \nu} \, F_{\mu
  \nu} \,-\, {i \over 8} \, \epsilon^{\mu \nu \rho \sigma} \, F_{\mu
  \nu} F_{\rho \sigma} \,+\,\cr
&\qquad\qquad\qquad\qquad +\, i\, (\sigma^\mu)_{\alpha\dot\alpha} \, D_\mu
\eta^\alpha \, \bar\eta{}^{\dot\alpha}\Big)\,.
\end{align}
In \eqref{ITHREE}, $\epsilon^{\mu\nu\rho\sigma}$ is the anti-symmetric
tensor associated to the Pontryagin density for the curvature on
$\BR^{3,1}$.  While this naively appears to fix a specific value of
the $\theta_{YM}$ angle in the four-dimensional effective theory, we
note that because the Pontryagin density is a topological term, it is
automatically supersymmetric.  Therefore, we may add an additional
topological term of this type to the quasi-topological action in order
to achieve a four-dimensional effective theory with an arbitrary
$\theta_{YM}$ angle.

Besides terms which are $\bar\delta{}^2$-trivial, the action $I_S$
also contains a set of terms of the form
\begin{equation}
I_4 \,=\, \epsilon^{\dot\alpha\dot\beta} \,
\bar\delta{}_{\dot\alpha} \CO^{(4)}_{\dot\beta}\,,
\end{equation}
where $\CO^{(4)}_{\dot\beta}$ for ${\dot\beta=1,2}$ is again a
gauge-invariant functional of the fields.  In order that
$\bar\delta_{\dot\alpha}$ annihilate $I_4$, the functional
$\CO^{(4)}_{\dot\beta}$ cannot be arbitrary but must satisfy
\begin{equation}\label{SUSYOF}
\bar\delta_{\dot\alpha} \CO^{(4)}_{\dot\beta} \,=\, -\bar\delta_{\dot\beta}
\CO^{(4)}_{\dot\alpha}\,.
\end{equation}

With a little bit of calculation, one can check that the following
choice for $\CO^{(4)}_{\dot\beta}$ satisfies \eqref{SUSYOF},
\begin{equation}\label{OFOUR}
\CO_{\dot\beta}^{(4)} \,=\, -\frac{i}{\sqrt{2}} \, \underset{\BR^{3,1}
  \times S}{\int} d^4x \>
\omega \^ \Tr\Big( \bar\psi{}_{\dot\beta} \^ \CG \,-\, i\,
F_{S\,\mu}^{(1,0)} \, (\sigma^\mu)_{\alpha\dot\beta} \^ \psi^\alpha
\,-\, \sqrt{2} \, F_S^{(1,1)} \, \bar\eta{}_{\dot\beta}\Big)\,,
\end{equation}
where ${F_{S\,\mu}^{(1,0)} \,=\, F_{\mu m} \, ds^m}$ is determined by
the mixed components of the curvature on ${\BR^{3,1} \times S}$.
Given the first term appearing in $\CO^{(4)}_{\dot\beta}$, the
coefficients of the remaining two terms are fixed uniquely by the
anti-symmetry condition in \eqref{SUSYOF}.

Of course, one might ask how we arrived at \eqref{OFOUR}.  The
significance of \eqref{OFOUR} is that the corresponding expression for
$I_4$ contains a term which is quadratic in the auxiliary fields
$\CG$ and $\bar\CG$,
\begin{align}
I_4 \,&=\, \underset{\BR^{3,1} \times S}{\int} d^4x \> 2 \,\omega \^
\Tr\Big( i\,\CG \^ \bar\CG \,-\, i\, F^{(1,0) \, \mu}_S \^
F^{(0,1)}_{S \, \mu} \,+\, F_S^{(1,1)} \, \CD \,\,+\,\cr
&\qquad\qquad\,+\, D_\mu \bar\psi{}^{\dot\alpha}
(\sigma^\mu)_{\alpha\dot\alpha} \^ \psi^\alpha \,-\, i \sqrt{2}\,
\bar\psi{}_{\dot\alpha} \^ \bar\partial_A \bar\eta^{\dot\alpha} \,-\,
i \sqrt{2}\, \partial_A \eta^\alpha \^ \psi_\alpha\Big).
\end{align}
From the perspective of the ${\CN=1}$ supersymmetric theory on
$\BR^{3,1}$, the terms in $I_4$ give rise to D-terms in the
low-energy effective action.  However, we do not know of any way to
write $I_4$ in the form ${I_4=\bar\delta{}^2 \CO'}$ for any
gauge-invariant expression $\CO'$.  This feature is presumably related
to the difficulties first encountered in \cite{SeigelTEND} in
providing a fully local gauge-invariant description in ${\CN=1}$ superspace
for the ten-dimensional super-Yang-Mills action.

The final term which appears in $I_S$ is simply the conjugate to
$I_2$ in \eqref{ITWO}, which we denote by
\begin{equation}\label{BIGW}
W_S \,=\, -\underset{\BR^{3,1}\times S}{\int}\!\!d^4x \,
\Tr\Big(F_S^{(0,2)}\^\CH \,+\, \CG \^ \bar\partial_A \varphi \,+\,
\chi^\alpha\^\bar\partial_A\psi_\alpha \,+\, \frac{1}{2}\,\psi^\alpha
\big[\varphi\,,\psi_\alpha\big]\Big)\,.
\end{equation}
As one can check, though $W_S$ is not $\bar\delta_{\dot\alpha}$-exact,
$W_S$ is nonetheless annihilated by $\bar\delta_{\dot\alpha}$.  Our
notation for $W_S$ is no accident, since $W_S$ can be written as an
integral over superspace of the superpotential for the twisted Yang-Mills theory on
${\BR^{3,1} \times S}$,
\begin{equation}
W_S \,=\, -\underset{\BR^{3,1} \times S}{\int}\!\!d^4 x \,
d^2\theta \> \Tr\Big( {\bf F}_S^{(0,2)} \^ {\mathbf \Phi} \Big)\,.
\end{equation}
Here ${\mathbf \Phi}$ is the chiral superfield with lowest bosonic
component $\varphi$ as introduced in \eqref{CHRL}, and
${\bf F}^{(0,2)}_S$ is the chiral superfield with lowest bosonic
component $F^{(0,2)}_S$.  Explicitly, in terms of the chiral
superfield ${\bf A}_{\bar m}$ in \eqref{CHRL},
\begin{align}
{\bf F}^{(0,2)}_S \,&=\, \bar\partial {\bf A} \,+\, {\bf A} \^ {\bf A}\,,\cr
&=\, F_S^{(0,2)} \,+\, \sqrt{2} \, \theta \,
\bar\partial{}_A \psi \,+\, \theta \theta \left(\bar\partial{}_A
\CG \,-\, \frac{1}{2} [\psi^\alpha\,,\psi_\alpha]\right) \,+\, \cdots\,.
\end{align}
Of all the terms that will appear in $I_S$, the terms in \eqref{BIGW}
are most important, since these terms determine the effective
superpotential on $\BR^{3,1}$ which describes F-theory compactified on
$X$.

The supersymmetric action $I_S$ for the partially twisted Yang-Mills
theory on ${\BR^{3,1} \times S}$ is now given by
\begin{align}\label{BIGIS}
I_S \,&=\, I_1 \,+\, I_2 \,+\, I_3 \,+\, I_4 \,+\, W_S\,,\cr
&=\, \underset{\BR^{3,1} \times S}{\int} d^4x \,\Tr\Biggr[
\omega\^\omega\,\Big(\ha \, \CD^2 - \frac{1}{4} F^{\mu \nu} F_{\mu \nu}\Big)
- 2 i \, \omega \^ F^{(1,0)}_{S\,\mu} \^ F^{(0,1)\, \mu}_S \,-\,
D^\mu\bar\varphi\^D_\mu\varphi \,+\,\cr
&\,+\, 2 i \, \omega\^\CG\^\bar\CG \,+\, \CH \^ \bar\CH \,-\,
F^{(2,0)}_A \^\bar\CH \,-\, F_S^{(0,2)}\^\CH \,-\,
\bar\CG\^\partial_A\bar\varphi \,-\, \CG\^\bar\partial_A\varphi
\,+\,\cr
&\,+\, 2 \Big(\omega\^F_S^{(1,1)} \,+\, {i\over 2}
[\varphi\,,\bar\varphi]\Big)\,\CD\Biggr]\,+\,\cdots\,.
\end{align}
This action deserves a number of comments.  As indicated by the
`$\cdots$', we omit from \eqref{BIGIS} terms involving fermions as
well as the Pontryagin density on $\BR^{3,1}$ which appears in
\eqref{ITHREE}.  Modulo the Pontryagin term, $I_S$ is purely
real.  Since $\delta_\alpha$ and $\bar\delta{}_{\dot\alpha}$ annihilate this term,
it now follows that the action $I_S$ is annihilated by $\delta_\alpha$ and
$\bar\delta{}_{\dot\alpha}$ as claimed.

By the same token, we are free to add to $I_S$ any other
topological terms involving characteristic classes of the surface $S$
and the $G_{S}$-bundle $P$.  Along with the effective $\theta_{YM}$-angle on
$\BR^{3,1}$, the coefficients of such terms are fixed by closed string
moduli which appear as parameters in the local F-theory backgrounds
under consideration.  Such topological terms will not play a role in the
present discussion.

Finally, if we integrate out the auxiliary bosons appearing in
\eqref{BIGIS}, the off-shell supersymmetry algebra in \eqref{FSUSYAUX}
reduces to the on-shell algebra in \eqref{FSUSY}.  By construction,
the on-shell action then enforces the F- and D-term supersymmetry
conditions at weak-coupling.  Finally, as is more or less apparent and
dictated by supersymmetry, the on-shell action derived from
\eqref{BIGIS} provides a twisted version of the maximally
supersymmetric Yang-Mills action on ${\BR^{3,1} \times S}$.

\section{Partially Twisted Action of the\\ Six-Dimensional Defect Theory\label{6DACTION}}

In this Appendix we present the action for a six-dimensional
hypermultiplet charged under a
gauge group $G_{S}\times G_{S^{\prime}}$ associated with the partially twisted
six-dimensional defect theory which originates from the intersection of seven-branes along
$\mathbb{R}^{3,1}\times\Sigma$. \ As opposed to most topological field theories, the on-shell
supersymmetry transformations of the fields do not descend directly from a reduction of super
Yang-Mills theory in ten dimensions. For this reason, it is convenient to treat the six-dimensional defect
theory in a superspace formalism which preserves four-dimensional $\mathcal{N}=1$ off-shell supersymmetry.
Indeed, in this Appendix we will effectively reverse the usual order of discussion by instead using the
off-shell supersymmetry transformations to determine the on-shell transformations of fields in the six- and
eight-dimensional theories.

In the spirit of \cite{SeigelTEND,WackerGregoire,DijkgraafVafaDecon}, we now treat the six-dimensional fields of the
defect theory as a collection of four-dimensional fields labelled by points of $\Sigma$.  In terms of the four-dimensional
effective theory, the six-dimensional hypermultiplet organizes into two collections of $\mathcal{N}=1$ chiral multiplets
$(\sigma,\lambda,\mathcal{K})$ and $(\sigma^{c},\lambda^{c},\mathcal{K}^{c})$ which transform in dual
representations of $G_{S}\times G_{S^{\prime}}$.  The off-shell supersymmetry transformations of the bosonic degrees of freedom are given by the
usual expressions for an $\mathcal{N}=1$ chiral multiplet.  As for the eight-dimensional theory, we follow the conventions of \cite{WESSBAGGER}:
\begin{align}\label{6DBOSONS}
&\delta_\alpha \sigma \,=\, \sqrt{2} \lambda_{\alpha}\,, &
&\bar\delta{}_{\dot\alpha} \sigma \,=\, 0\,,\cr
&\delta_\alpha \bar\sigma \,=\, 0\,, &
&\bar\delta{}_{\dot\alpha} \bar\sigma \,=\, \sqrt{2} \bar\lambda_{\dot\alpha}\,,\cr
&\delta_\alpha \sigma^{c} \,=\, \sqrt{2} \lambda_{\alpha}^{c}\,, &
&\bar\delta{}_{\dot\alpha} \sigma^{c} \,=\, 0\,,\cr
&\delta_\alpha \bar\sigma^{c} \,=\, 0\,, &
&\bar\delta{}_{\dot\alpha} {\bar\sigma}^{c} \,=\, \sqrt{2} {\bar\lambda^{c}}_{\dot\alpha}\,.
\end{align}
Similarly, the fermions transform as:
\begin{align}\label{6DFERMIONS}
&\delta_\alpha \lambda_{\beta} \,=\, -\sqrt{2}\epsilon_{\alpha \beta}\mathcal{K}\,, &
&\bar\delta{}_{\dot\alpha} \lambda_{\beta} \,=\, i\sqrt{2}({\sigma}^{\mu})_{\beta \dot\alpha}D_{\mu}\sigma \,,\cr
&\delta_\alpha {\bar\lambda}_{\dot\beta} \,=\, i\sqrt{2}({\sigma}^{\mu})_{\alpha \dot\beta}D_{\mu}{\bar\sigma}\,, &
&\bar\delta{}_{\dot\alpha} {\bar\lambda}_{\dot\beta} \,=\, -\sqrt{2}\epsilon_{\dot\alpha \dot\beta} \overline{\mathcal{K}} \,,\cr
&\delta_\alpha \lambda^{c}_{\beta} \,=\, -\sqrt{2}\epsilon_{\alpha \beta}\mathcal{K}^{c}\,, &
&\bar\delta{}_{\dot\alpha} \lambda^{c}_{\beta} \,=\, i\sqrt{2}({\sigma}^{\mu})_{\beta \dot\alpha}D_{\mu}\sigma^{c} \,,\cr
&\delta_\alpha \overline{\lambda^{c}}_{\dot\beta} \,=\, i\sqrt{2}({\sigma}^{\mu})_{\alpha \dot\beta}D_{\mu} \overline{\sigma^{c}}\,, &
&\bar\delta{}_{\dot\alpha} \overline{\lambda^{c}}_{\dot\beta} \,=\, -\sqrt{2}\epsilon_{\dot\alpha \dot\beta} \overline{\mathcal{K}^{c}}\,.
\end{align}
Finally, the auxiliary fields transform as:
\begin{align}\label{6DAUX}
&\delta_\alpha \mathcal{K} \,=\, 0\,, &
&\bar\delta{}_{\dot\alpha} \mathcal{K} \,=\, i\sqrt{2} (\sigma^{\mu})_{\alpha \dot\alpha}D_{\mu}\lambda^{\alpha}
+ 2\big[\sigma\,,\bar\eta_{\dot \alpha}] \,,\cr
&\delta_\alpha \overline{\mathcal{K}} \,=\, i\sqrt{2}
(\sigma^{\mu})_{\alpha \dot\alpha}D_{\mu}\overline{\lambda}^{\dot\alpha}
+ 2\big[\bar\sigma\,,\eta_{\alpha}]\,, &
&\bar\delta{}_{\dot\alpha} \overline{\mathcal{K}} \,=\, 0 \,,\cr
&\delta_\alpha \mathcal{K}^{c} \,=\, 0\,, &
&\bar\delta{}_{\dot\alpha} \mathcal{K}^{c} \,=\, i\sqrt{2} (\sigma^{\mu})_{\alpha \dot\alpha}D_{\mu}{\lambda^{c}}^{\alpha}
+ 2\big[\sigma^c\,,\bar\eta_{\dot \alpha}\big] \,,\cr
&\delta_\alpha \overline{\mathcal{K}^{c}} \,=\, i\sqrt{2} (\sigma^{\mu})_{\alpha \dot\alpha}D_{\mu}\overline{\lambda^{c}}^{\dot\alpha}
+ 2\big[\bar\sigma^c\,,\eta_{\alpha}\big]\,, &
&\bar\delta{}_{\dot\alpha} \overline{\mathcal{K}^{c}} \,=\, 0 \,.
\end{align}
Organizing these fields into $\mathcal{N}=1$ chiral superfields and
a background vector superfield which descends from the bulk eight-dimensional theory on $S$,
we have:%
\begin{align}
\bf\Lambda &  =\sigma+\sqrt{2}\theta\lambda+\theta\theta\mathcal{K}+\cdots\\
{\bf\Lambda}^{c} &  =\sigma^{c}+\sqrt{2}\theta\lambda^{c}+\theta\theta\mathcal{K}%
+\cdots\\
V &  =-\theta\sigma^{\mu}\overline{\theta}A_{\mu}+i\theta\theta\overline{\theta
}\overline{\eta}-i\overline{\theta\theta}\theta\eta+\frac{1}{2}\theta\theta\overline{\theta\theta}\mathcal{D}%
\end{align}
where for notational simplicity we have presented the background vector
multiplet in WZ gauge. \ There is a similar background vector superfield
$V^{\prime}$ which descends from the bulk theory on $\mathbb{R}^{3,1}\times S^{\prime}$.
Finally, to present gauge covariant interactions in the internal directions, we also introduce
the gauge covariant derivative superfield defined in equation \eqref{BIGDA} which we
reproduce here for the convenience of the reader:
\begin{align}
\bar\partial_{{\bf A} + {\bf A}'} \,&=\, \bar\partial \,+\, {\bf A}
\,+\, {\bf A}'\,,\cr
&=\, \bar\partial_{A + A'} \,+\, \sqrt{2} \theta \!\left(\psi \,+\, \psi'\right) \,+\, \theta\theta \!\left(\CG \,+\, \CG'\right)\,+\, \cdots\,.
\end{align}

The partially twisted action now follows from the known result for coupling a
six-dimensional hypermultiplet to a background gauge field in flat space
presented in \cite{WackerGregoire}. \ Indeed, because manifest off-shell
supersymmetry is preserved at all stages, for our purposes it is enough to
repackage this result in the partially twisted theory. In ${\CN=1}$
superspace the action is
\begin{align}
I_{\Sigma} &  =\underset{%
\mathbb{R}
^{3,1}\times\Sigma}{\int} d^{4}x \, d^{4}\theta \> \omega\^\Big[\left(
{\mathbf{\Lambda}},e^{-2V - 2V'}\mathbf{\Lambda}\right)
+\left({\mathbf \Lambda}^c,e^{2V + 2V'}{\mathbf
    \Lambda}^c\right)\Big] \label{I6D}\\
&+\underset{%
\mathbb{R}
^{3,1}\times\Sigma}{\int} d^{4}x d^2\theta \left\langle
\mathbf{\Lambda}^{c},\overline{\partial}_{\mathbf{A}+\mathbf{A}^{\prime}%
}\mathbf{\Lambda}\right\rangle \,+\, h.c.\label{W6D}%
\end{align}
In order to define the D-terms appearing in \eqref{I6D}, we
introduce a hermitian metric $(\,\cdot\,,\,\cdot\,)$ on the bundle
${K_\Sigma^{1/2} \otimes \SU \otimes \SU'}$ in which the chiral
superfield ${\mathbf \Lambda}$ is valued, and similarly for the bundle
${K_\Sigma^{1/2} \otimes \SU^* \otimes (\SU')^*}$ in which ${\mathbf
\Lambda^c}$ is valued.  If $h_{\bar a a}$ represents the hermitian
metric on ${K_\Sigma^{1/2} \otimes \SU \otimes \SU'}$ in a local basis,
then
\begin{equation}\label{METRICH}
\left({\mathbf \Lambda}, e^{-2V -2V'} {\mathbf \Lambda}\right) \,=\,
h_{\bar a a}\,\, \bar{\mathbf \Lambda}{}^{\bar a} e^{-2V -2V'} {\mathbf
  \Lambda}^a\,.
\end{equation}
We emphasize that the pairing $(\,\cdot\,,\,\cdot\,)$ includes an
implicit complex conjugation on one argument, as apparent above.
Because the quantity in \eqref{METRICH} transforms as a scalar on
$\Sigma$, we use the pullback of the K\"ahler form $\omega$ on $S$ to
define an integration measure over $\Sigma$ in \eqref{I6D}.

In contrast to \eqref{I6D}, the expression in \eqref{W6D} is defined
without reference to a metric on either the bundles $\SU$ and $\SU'$
or on $\Sigma$.  Thus $\langle\,\cdot\,,\,\cdot\,\rangle$
indicates the canonical dual pairing between ${\SU \otimes \SU'}$ and
${\SU^*\otimes (\SU')^*}$, so that $\left\langle
\mathbf{\Lambda}^{c},\overline{\partial}_{\mathbf{A}+\mathbf{A}^{\prime}
}\mathbf{\Lambda}\right\rangle$ transforms as a differential form of
type $(1,1)$ appropriate to integrate over $\Sigma$.  Indeed, this is
an expected consequence of twisting the field content of the
six-dimensional theory.

Whereas the bulk gauge fields ${\mathbf A}$ and ${\mathbf A}'$
naturally couple to the six-dimensional fields via the superpotential
in \eqref{W6D}, a natural superpotential coupling involving ${\mathbf
  \Phi}$ does not appear to exist, since ${\mathbf \Phi}$ transforms
as a $(2,0)$-form on $S$.

To proceed further, we next expand $I_{\Sigma}$ in terms of component
fields. This yields
\begin{align}\label{6DOFFSHELL}
I_{\Sigma} &  =\underset{%
\mathbb{R}
^{3,1}\times\Sigma}{\int}d^{4}x \,\,\omega \^
\Big[\left({\mathcal{K}},\mathcal{K}\right)+\left(  \mathcal{K}^{c}%
,{\mathcal{K}^{c}}\right) -\left(  D_{\mu}{\sigma
},D^{\mu}\sigma\right)-\left(  D_{\mu}\sigma^{c},D^{\mu}%
{\sigma^{c}}\right)\\
&\qquad\qquad\quad\qquad -\left({\sigma},(\CD + \CD')\cdot\sigma\right)
+\left(\sigma^{c},(\CD + \CD')\cdot{\sigma^{c}}\right)\Big]\,+\,\\
&+\,\underset{\BR^{3,1}\times\Sigma}{\int} d^4 x\,\Big[\left\langle
  \mathcal{K}^{c},\overline{\partial}_{A+A^{\prime}}%
\sigma\right\rangle+\left\langle \sigma^{c},\overline{\partial
}_{A+A^{\prime}}\mathcal{K}\right\rangle+\left\langle \sigma
^{c},(\mathcal{G}+\mathcal{G}^{\prime})\cdot\sigma\right\rangle\\
&\qquad\qquad
+\left\langle\bar\CK^c,\partial_{A + A'}\bar\sigma\right\rangle
+\left\langle\bar\sigma^c, \partial_{A + A'}\bar\CK\right\rangle
+\left\langle\bar\sigma^c, (\bar\CG +
  \bar\CG')\cdot\bar\sigma\right\rangle\Big]+\cdots\,,
\end{align}
where the \textquotedblleft$\cdots$\textquotedblright\ indicate
additional fermionic terms in $I_\Sigma$.  Also, in expressions such
as ${\CD \cdot \sigma}$, we indicate the action of elements in the Lie
algebra of the group $G_S$ on the representation $U$, and similarly
for $G_{S'}$ and $U'$.

We now determine both the on-shell supersymmetry transformations for
the six- and eight-dimensional fields as well as the resulting BPS
equations.  Integrating out the six-dimensional auxiliary
fields yields the conditions
\begin{align}
\overline{\mathcal{K}}  & = \*_\Sigma \,
\partial_{A+A^{\prime}} \bar\sigma^{c}\text{, \ \ \ \ \ \ \ \ \ \ \ \
  \ \ \ \ }\overline{\mathcal{K}^{c}
}= -\*_\Sigma \, \partial_{A+A^{\prime}}\bar\sigma\text{,}\\
\mathcal{K}  & = \*_\Sigma \, \bar\partial_{A+A^{\prime}}\sigma^{c}\text{,
\ \ \ \ \ \ \ \ \ \ \ \ \ \ \ \ }\mathcal{K}^{c}= -\*_\Sigma \,
\bar\partial_{A+A^{\prime}}{\sigma}\text{.}%
\end{align}
Here $\*_\Sigma$ denotes the duality operator acting on sections
of ${K^{1/2}_\Sigma \otimes \SU \otimes \SU'}$ and
${K^{1/2}_\Sigma\otimes \SU^* \otimes (\SU')^*}$ as determined by the
given metrics on $\Sigma$ and ${\SU \otimes \SU'}$.

Whereas the supersymmetry transformations for the six-dimensional bosons such
as $\sigma$ are the same both on and off-shell, the on-shell supersymmetry
transformations of the fermions follow by substituting in the values
of the auxiliary fields in the off-shell transformations of
(\ref{6DFERMIONS}) so that:
\begin{align}\label{6DFERMIONSONSHELL}
&\delta_\alpha \lambda_{\beta} \,=\, -\sqrt{2}\epsilon_{\alpha \beta}
\, \*_\Sigma \, \bar\partial_{A+A^{\prime}}\sigma^{c}\,, &
&\bar\delta{}_{\dot\alpha} \lambda_{\beta} \,=\,
i\sqrt{2}({\sigma}^{\mu})_{\beta \dot\alpha}D_{\mu}\sigma \,,\cr
&\delta_\alpha {\bar\lambda}_{\dot\beta} \,=\, i\sqrt{2}({\sigma}^{\mu})_{\alpha \dot\beta}D_{\mu}{\bar\sigma}\,, &
&\bar\delta{}_{\dot\alpha} {\bar\lambda}_{\dot\beta} \,=\,
-\sqrt{2}\epsilon_{\dot\alpha \dot\beta}\, \*_\Sigma \,
\partial_{A+A^{\prime}} \bar\sigma^{c} \,,\cr
&\delta_\alpha \lambda^{c}_{\beta} \,=\, \sqrt{2}\epsilon_{\alpha
  \beta}\, \*_\Sigma \,\bar\partial_{A+A^{\prime}}{\sigma} \,, &
&\bar\delta{}_{\dot\alpha} \lambda^{c}_{\beta} \,=\, i\sqrt{2}({\sigma}^{\mu})_{\beta \dot\alpha}D_{\mu}\sigma^{c} \,,\cr
&\delta_\alpha \overline{\lambda^{c}}_{\dot\beta} \,=\, i\sqrt{2}({\sigma}^{\mu})_{\alpha \dot\beta}D_{\mu} \overline{\sigma^{c}}\,, &
&\bar\delta{}_{\dot\alpha} \overline{\lambda^{c}}_{\dot\beta} \,=\,
\sqrt{2}\epsilon_{\dot\alpha \dot\beta}\,\*_\Sigma
\, \partial_{A+A^{\prime}}\bar\sigma \,.
\end{align}

Some of the on-shell supersymmetry transformations of the
eight-dimensional fields also change in the presence of the
six-dimensional defect theory.  Integrating out the
auxiliary fields $\CG$ and $\overline{\CG}$ from the total action ${I
  = I_S + I_\Sigma}$ implies
\begin{align}\label{MODCG}
2 i\,\omega\^\overline{\mathcal{G}} \,&=\,
\overline{\partial}_{A}\varphi \,-\, \delta_\Sigma\,
\llangle\sigma^{c},\sigma\rrangle_{\ad(P)}\,,\cr
-2 i\,\omega\^\CG \,&=\, \partial_A\bar\varphi \,-\, \delta_\Sigma \,
\llangle\bar\sigma^c, \bar\sigma\rrangle_{\ad(P)}\,.
\end{align}
Here we use $\llangle\,\cdot\,,\,\cdot\,\rrangle_{\ad(P)}$ to denote
the canonical `outer-product' determined by the action of $G$ on the
representation $U$,
\begin{equation}
\llangle\,\cdot\,,\,\cdot\,\rrangle_{\ad(P)}: \big[\SU \otimes
  \SU'\big] \otimes \big[\SU^* \otimes (\SU')^*\big]
\,\longrightarrow\, \ad(P)\big|_\Sigma\,.
\end{equation}
Explicitly, if $(T^I)^a_{a'}$ for ${I=1,\ldots,\dim(G_S)}$ represent the
generators of $G_S$ acting on $U$ in a given basis, then
${\llangle\sigma^c,\sigma\rrangle_{\ad(P)} \,=\,\sigma^c_a \,
  (T^I)^a_{a'} \, \sigma^{a'}}$.  Also, $\delta_\Sigma$ is a two-form
with delta-function support which represents the Poincar\'e dual of
the holomorphic curve $\Sigma$ inside $S$.

Based upon \eqref{MODCG}, the on-shell variations of the
eight-dimensional fermions $\psi_{\alpha \, \bar m}$ and
$\bar\psi{}_{\dot\alpha\, m}$ are now
\begin{align}
&\delta_\alpha \psi_\beta \,=\, \sqrt{2} \, \epsilon_{\alpha \beta}
\left( \bar\partial{}_A^\dagger \bar\varphi \,+\,
\*_S\,\delta_{\Sigma} \llangle\sigma^c,\sigma\rrangle_{\ad(P)}\right)\,, &
&\bar\delta{}_{\dot\alpha} \psi_{\beta} \,=\,
i\sqrt{2}(\sigma^\mu){}_{\beta \dot\alpha} \, F_{\mu \, S}^{(0,1)}\,,\cr
&\bar\delta{}_{\dot\alpha} \bar\psi{}_{\dot\beta} \,=\,
\sqrt{2}\epsilon_{\dot\alpha \dot\beta}
\left(\partial{}_A^\dagger\varphi \,+\, \*_S\,\delta_{\Sigma
}\llangle\bar\sigma^c,\bar\sigma\rrangle_{\ad(P)}\right)\,, &
&\delta_\alpha \bar\psi{}_{\dot\beta\,m} \,=\,
i\sqrt{2}(\sigma^\mu)_{\alpha\dot\beta} \, F_{\mu \, S}^{(1,0)}\,.
\end{align}

Similarly, by integrating out the eight-dimensional auxiliary field
$\mathcal{D}$ from the total action ${I=I_S+I_\Sigma}$,
we arrive at the modified relation
\begin{equation}\label{NEWDDEF}
\mathcal{D}=-\,\*_S\left( \omega \^\!\left[F_{S}^{(1,1)} \,-\, \ha
    \,\delta_\Sigma\,\mu(\bar\sigma,\sigma) \,+\, \ha \,
    \delta_\Sigma\,\mu(\bar\sigma^c,\sigma^c)\right]
+\frac{i}{2}\left[
\varphi,\overline{\varphi}\right]\right)\,.
\end{equation}
Here $\mu(\bar\sigma,\sigma)$ denotes the moment map associated to the
action of $G_S$ on the representation $U$, and similarly for
$\mu(\bar\sigma^c,\sigma^c)$.  In terms of the local generators
$(T^I)^a_{a'}$ which we introduced to describe the outer-product
$\llangle\,\cdot\,,\,\cdot\,\rrangle_{\ad(P)}$, the moment map is given as
usual by ${\mu(\bar\sigma,\sigma) = \bar\sigma{}^{\bar a} (T^I)_{\bar a
    \, a} \sigma^a}$.  The on-shell supersymmetry transformations for
the eight-dimensional zero-form fermions are then given by
\eqref{FSUSY} but now with $\mathcal{D}$ as in equation
\eqref{NEWDDEF},
\begin{align}\label{FSUSYAUXDEF}
&\delta_\alpha \eta_\beta \,=\, \big(\sigma^{\mu
\nu}\big){}_{\alpha \beta} \, F_{\mu \nu} \,-\, i \, \epsilon_{\alpha
\beta}\, \CD\,, &
&\bar\delta{}_{\dot\alpha} \eta_\beta \,=\, 0\,,\cr
&\delta_\alpha \bar\eta_{\dot\beta} \,=\, 0\,, &
&\bar\delta{}_{\dot\alpha} \bar\eta{}_{\dot\beta} \,=\,
(\bar\sigma^{\mu \nu}){}_{\dot\alpha \dot\beta} \, F_{\mu \nu} \,+\,
i\, \epsilon_{\dot\alpha \dot\beta}\,\CD\,.
\end{align}

Finally, by inspection of equation \eqref{6DOFFSHELL}, we note that
the auxiliary two-form fields $\CH$ and $\overline{\CH}$ do not couple
to the fields of the six-dimensional theory.  In particular, this
implies that the on-shell supersymmetry transformations of the
two-form fermions $\chi$ and $\overline{\chi}$ are unchanged in the
presence of the six-dimensional defect.

The BPS equations of motion now follow from the requirement that the on-shell variation of all fermions must vanish
in a supersymmetric vacuum.  At a practical level, these are simply the D- and F-flat conditions which arise when all auxiliary
fields have been set to zero.  The BPS equations of motion for the six-dimensional fields $\sigma$ and $\sigma^{c}$ are therefore:
\begin{equation}
\overline{\partial}_{A+A^{\prime}}\sigma=\overline{\partial}_{A+A^{\prime}%
}\sigma^{c}=0\text{.}%
\end{equation}
Because this derives from an F-term, it is protected from quantum corrections.  The other BPS equation of motion
which derives from an F-term corresponds to the equation of motion for $\varphi$ in the presence of the six-dimensional
defect theory:
\begin{equation}
\overline{\partial}_{A}\varphi \,=\,\delta_{\Sigma}\,\llangle
\sigma^{c},\sigma\rrangle _{\ad(P)},\qquad\qquad \partial_{A}\overline{\varphi
}\,=\,\llangle \overline{\sigma}^c,\overline{\sigma}\rrangle _{\ad(P)}\text{.}
\end{equation}
Note that the above expressions are independent of any
metric data.  The final BPS equation of motion on $S$ which derives from an F-term is
uncorrected by the presence of the six-dimensional defect because the auxiliary fields
$\CH$ and $\overline{\CH}$ do not couple to the defect:
\begin{equation}
F_{S}^{(2,0)}=0\,,\qquad\qquad F_{S}^{(0,2)}=0.
\end{equation}

Finally, the BPS equation of motion on $S$ which derives from a D-term
is given by
\begin{equation}
\omega\wedge F_{S}^{(1,1)}+\frac{i}{2}\left[  \varphi,\overline{\varphi
}\right] \,=\,\ha\,\omega\wedge\delta_{\Sigma}\Big[  \mu(  \overline{\sigma}%
,\sigma)\,-\,\mu(\bar\sigma^{c},\sigma^c)\Big]  \text{.}%
\end{equation}
Due to the explicit dependence on the K\"{a}hler
forms of $S$ and  $\Sigma$, away from the regime of large volume for both $S$
and $\Sigma$, we expect that this equation will generally
receive quantum corrections.  Also, though for simplicity we have
focused attention solely on the Yang-Mills theory which lives on the
surface $S$, the parallel BPS equations hold for the eight-dimensional
Yang-Mills theory on $S'$.

\section{A Vanishing Theorem\label{VanishAppendix}}

In this appendix, we establish the vanishing theorem used in Section
\ref{SPECTRUM} to constrain the zero mode content of the bulk fields
in the partially twisted Yang-Mills theory on ${\BR^{3,1} \times S}$. See 
for instance the proof of Theorem 6.1 in \cite{AHSvanish} for another 
appearance of this vanishing theorem.

As throughout, we take $S$ to be a smooth, compact K\"ahler
surface with K\"ahler form $\omega$ and canonical divisor $K_S$.  In
order to prove our vanishing theorem, we assume that ${-K_S \ge
  0}$, in the sense that the anti-canonical divisor $-K_S$ is
effective.  We also assume that ${h^{2,0}(S) = 0}$, as follows
automatically if $-K_S$ happens to be ample.  Examples of such
surfaces include both the del Pezzo surfaces $dP_n$ and the Hirzebruch
surfaces ${\mathbb F}_n$.

We now let $E$ be a complex vector bundle over $S$ which is endowed with a
hermitian metric and a compatible unitary connection $A$ solving the
hermitian Yang-Mills equations on $S$,
\begin{equation}\label{HYM}
F_A^{(2,0)}\,=\,F_A^{(0,2)}\,=\,\omega \cdot F_A \,=\, 0\,.
\end{equation}
Here ${\omega \cdot F_A}$ is shorthand for the contraction of $\omega$ with $
F_A$ as defined using the K\"ahler metric on $S$. In local coordinates $
(s^n,\bar s{}^{\bar n})$ on $S$,
\begin{equation}\label{DTERM}
\omega \cdot F_A \,=\, \omega^{n \bar n} \big[F_A\big]{}_{n \bar n}\,,
\end{equation}
where we sum over repeated indices ${n,\bar n = 1,2}$ in
\eqref{DTERM}.

Because $\omega$ is self-dual, the vanishing of ${\omega\cdot F_A}$ is
equivalent to the vanishing of ${\omega\^F_A}$.  Hence the equations
in \eqref{HYM} correspond to the bulk BPS equations \eqref{HOLP},
\eqref{HOLPHI}, \eqref{VAFWITTDTERM} on $S$ in a background with
${\varphi=0}$, where $E$ is any vector bundle associated to the
principal $G$-bundle $P$ on $S$.  In fact, if $S$ is a del Pezzo or
Hirzebruch surface, $\varphi$ necessarily vanishes for any solution of
the bulk BPS equations.  Otherwise, if $\varphi$ were nontrivial as an
element in $H^0_{\bar\partial}(S, K_S \otimes \ad(P))$, then at least
one Casimir of $\varphi$ would be non-vanishing.  Hence some positive power
of $K_S$ would admit a non-trivial holomorphic section.  But on del
Pezzo and Hirzebruch surfaces, no positive powers of $K_S$ admit
holomorphic sections, so ${\varphi = 0}$.

We now state the vanishing theorem.  First, we recall that because the
$(2,0)$ and $(0,2)$ components of $F_A$ vanish according to
\eqref{HYM}, $E$ admits a holomorphic structure. That is, if we let $d_A$ be
the covariant derivative defined by the unitary connection on $E$, then the $
(0,1)$ component ${\bar\partial{}_A \,=\, d_A^{(0,1)}}$ of the covariant
derivative satisfies ${\bar\partial{}^2_A = 0}$ and thereby defines a
holomorphic structure on $E$.

Under the assumptions on $S$ above, we now claim that $E$ satisfies
\begin{equation}
H^2_{\bar\partial}(S, E) \,=\, 0\,.\label{VANISH}
\end{equation}
Further, if $E$ is irreducible and non-trivial (meaning ${E \neq
  \CO_S}$), then also
\begin{equation}
H^0_{\bar\partial}(S, E) \,=\, 0\,.\label{VANISHII}
\end{equation}

\medskip\noindent{\it Proof of the Vanishing Theorem}\smallskip

To prove the vanishing theorem in \eqref{VANISH}, we first assume
without loss that $E$ is irreducible.  Otherwise, $E$ splits holomorphically
as a direct sum of irreducible bundles $E_{j}$ for $j$ running in some index
set $J$,
\begin{equation}
E\,=\,\bigoplus_{j\in J}\,E_{j}\,. \label{SUM}
\end{equation}
Each summand $E_{j}$ also carries a unitary connection satisfying the
hermitian Yang-Mills equations in \eqref{HYM}, and trivially
\begin{equation}
H_{\bar{\partial}}^{2}(S,E)\,=\,\bigoplus_{j\in
  J}\,H_{\bar{\partial}}^{2}(S,E_{j})\,. \label{SUMII}
\end{equation}
Thus to demonstrate \eqref{VANISH}, we restrict attention to the
irreducible summands of $E$.

The proof of the vanishing theorem now proceeds in two steps. In the first
step, we show that if the irreducible bundle $E$ is non-trivial, then
${H^{0}_{\bar\partial}(S, E) = 0}$.  Alternatively, if $E$ is
irreducible but trivial, so that ${E =\mathcal{O}_{S}}$, the vanishing
theorem follows immediately by our assumption that ${h^{2,0}(S) =
  0}$. In the second step, assuming ${E\neq\mathcal{O}_{S}}$, we apply
Serre duality to deduce the vanishing of $H^{2}_{\bar\partial}(S, E)$ from the
vanishing of $H^{0}_{\bar\partial}(S, E)$.

To show that ${H^{0}_{\bar\partial}(S, E) = 0}$ if $E$ is irreducible
and non-trivial, let us assume otherwise. Therefore $E$ admits a
non-trivial holomorphic section $s$.

We now recall a basic fact from Hodge theory.  First, we let
$d_{A}{}^{\dagger}$ and $\bar\partial{}_A^{\dagger}$ denote the
adjoints of $d_{A}$ and $\bar\partial{}_A$ defined using the K\"ahler
metric on $S$ and the hermitian metric on $E$.  Then as shown for
instance in Ch.~$4$ of \cite{Friedman}, we have the following Hodge
identity of linear operators acting on sections of $E$,
\begin{equation}\label{HODGE}
2 \, \bar\partial{}^{\dagger}_A \bar\partial{}^{}_A\,=\, d_{A}{}^{\dagger
}d_{A} \,-\, i\, \omega\cdot F_A \,.
\end{equation}
For completeness, we include a proof of \eqref{HODGE} towards the end of
this appendix.  Because ${\bar\partial_A s = \omega\cdot F_A = 0}$, we
deduce from \eqref{HODGE} that
\begin{equation}\label{HODGEII}
0 \,=\, 2 \underset{S}{\int} \big\|\bar\partial{}_A s\big\|^{2} \,=\,
\underset{S}{\int} \big\|d_{A} s\big\|^{2}\,.
\end{equation}
Therefore ${d_{A} s = 0}$, and $s$ is a covariantly constant section of $E$.

Because $s$ is covariantly constant, $s$ is a nowhere-vanishing holomorphic
section of $E$. Hence $s$ defines an inclusion of the trivial line bundle
$\mathcal{O}_{S}$ as a rank one holomorphic subbundle of $E$,
\begin{equation}
\label{TRVL}0\,\longrightarrow\,\mathcal{O}_{S}\, \overset{s}{\longrightarrow
}\,E\,.
\end{equation}
We let $\mathcal{O}_{S}^{\perp}$ denote the orthocomplement to $s$ in $E$,
defined using the hermitian metric on $E$. Because $d_{A}$ respects that metric,
$\mathcal{O}_{S}^{\perp}$ in turn carries an induced unitary
connection\footnote{Note that if $s^{\perp}$ is any section of $\mathcal{O}%
_{S}^{\perp}$, then ${ 0 \,=\, d(s,s^{\perp}) \,=\, (d_{A} s, s^{\perp}) \,+\,
(s, d_{A} s^{\perp}) \,=\, (s, d_{A} s^{\perp})}$, so $d_{A}$ restricts
immediately to a unitary connection on $\mathcal{O}_{S}^{\perp}$.} satisfying
the hermitian Yang-Mills equations \eqref{HYM}, and in particular
$\mathcal{O}_{S}^{\perp}$ is holomorphic. Consequently $E$ splits as a sum of
holomorphic bundles
\begin{equation}
\label{SPLTE}E \,=\, \mathcal{O}_{S} \oplus\mathcal{O}_{S}^{\perp}\,.
\end{equation}
Since ${E \neq\mathcal{O}_{S}}$, the bundle $\mathcal{O}_{S}^{\perp}$ must
have non-zero rank, but then \eqref{SPLTE} contradicts our assumption that $E$
be irreducible. Hence
\begin{equation}
\label{VANISHIIA}H^{0}_{\bar\partial}(S, E) \,=\, 0\,.
\end{equation}

We are left to consider \eqref{VANISHIIA} in light of Serre duality. Under
duality,
\begin{equation}
\label{SERRE}H^{2}_{\bar\partial}(S, E) \,\cong\, H^{0}_{\bar\partial}(S,
E^{*} \otimes \CO_S(K_S))^{*}\,.
\end{equation}
However, because ${-K_{S} \ge0}$ is effective, we also have an inclusion
\begin{equation}
\label{INCL}H^{0}_{\bar\partial}(S, E^{*}\otimes \CO_S(K_S))
\,\subseteq\, H^{0}_{\bar\partial}(S, E^{*})\,.
\end{equation}
The inclusion in \eqref{INCL} follows once we recall that sections of
$\CO_S(K_S)$ for ${K_S \le 0}$ can be interpreted as holomorphic
functions on $S$ which vanish along the effective divisor $-K_{S}$. As
a result, holomorphic sections of the product
$E^{*}\otimes\mathcal{O}_S(K_{S})$ are the same as holomorphic sections
of $E^{*}$ which similarly vanish along $-K_{S}$.

Now, if $E$ admits a non-trivial irreducible unitary connection satisfying the
hermitian Yang-Mills equations \eqref{HYM}, then so does the dual
$E^{\ast}$. Thus the vanishing result in \eqref{VANISHIIA} applies
equally well to $E^{\ast}$,
\begin{equation}
H_{\bar{\partial}}^{0}(S,E^{\ast})\,=\,0\,. \label{DUAL}
\end{equation}
Together, \eqref{SERRE}, \eqref{INCL}, and \eqref{DUAL} finally imply the
basic vanishing theorem
\begin{equation}
H_{\bar{\partial}}^{2}(S,E)\,=\,0\,. \label{VANISHA}
\end{equation}

\medskip\noindent{\it Corollaries}\smallskip

The vanishing theorem in \eqref{VANISHA} has a few immediate
corollaries which strongly constrain the low-energy spectrum of the
partially twisted Yang-Mills theory on ${\BR^{3,1} \times S}$.  First,
as we observed in relation to \eqref{DUAL}, if $E$ admits a hermitian
Yang-Mills connection satisfying \eqref{HYM}, then so does
$E^{\ast}$. Thus the vanishing theorem for $E$ immediately applies to
$E^{\ast}$ as well, so that
\begin{equation}
H_{\bar{\partial}}^{2}(S,E^{\ast})\,=\,0\,. \label{BARVAN}
\end{equation}
Equivalently by Serre duality,
\begin{equation}
H_{\bar\partial}^0(S, E \otimes \CO(K_S)) = 0\,.
\end{equation}

More generally, since $E$ is any vector bundle associated to the
principal $G$-bundle $P$ on $S$, the vanishing theorem trivially
applies to any complex vector bundle $\rho(E)$ constructed by taking
tensor products of $E$ and $E^*$, such as $\wedge^{2} E$ or
$\mathrm{End_{0}}(E)$.  Equivalently, the hermitian Yang-Mills
connection on $E$ induces a corresponding connection on $\rho(E)$.
Thus the vanishing theorem in \eqref{VANISHA} can be recast more broadly as
\begin{equation}
\label{VANISHR}H^{2}_{\bar\partial}(S, \rho(E)) \,=\, 0\,.
\end{equation}

Finally, the bulk Yukawa couplings \eqref{8DYUK} on $S$
always involve the appearance of $H^2_{\bar\partial}(S, E)^*$ for some
vector bundle $E$ associated to the principal $G$-bundle $P$ on $S$.  Hence
besides constraining the massless spectrum, the vanishing
theorem \eqref{VANISHA} implies that all bulk  Yukawa couplings vanish
when the K\"ahler surface $S$ satisfies ${-K_S \ge 0}$
and ${h^{2,0}(S) = 0}$.

\medskip\noindent{\it A Hodge Identity}\smallskip

For the sake of completeness, we include here a proof of the Hodge identity in
\eqref{HODGE}. This identity holds for sections of an arbitrary
holomorphic vector bundle $E$ over a compact K\"{a}hler manifold $S$.

To setup notation, let $\Lambda$ be the operator which acts on sections of
$\Omega^{*}_{S} \otimes E$ by contraction with the K\"ahler form
$\omega$. According to standard Hodge theory (see \S $0.7$ and \S
$1.2$ of \cite{Griffiths}), $\Lambda$ satisfies
\begin{equation}
\label{HODGII}\bar\partial{}^{\dagger}_A \,=\, i \, \big[\partial_A, \Lambda
\big]\,,\qquad\qquad\partial^{\dagger}_A \,=\, -i \, \big[\bar\partial_A,
\Lambda\big]\,.
\end{equation}
Here $\partial_A$ and $\bar\partial{}_A$ are the $(1,0)$ and
$(0,1)$ components of the covariant derivative $d_A$ acting on
sections of $\Omega^*_S \otimes E$,

According to \eqref{HODGII}, when $\bar\partial{}^{\dagger}$ and $\partial
_{A}^{\dagger}$ act on bundle-valued forms of type $(0,1)$ or $(1,0)$
(which are trivially annihilated by $\Lambda$), we can write
\begin{equation}
\label{DERII}\bar\partial{}^{\dagger}_A\,=\, -i \,
\Lambda\, \partial_{A}\,, \qquad\qquad \partial_{A}^{\dagger}\,=\, i
\, \Lambda\, \bar\partial_A\,.
\end{equation}
Thus, when acting on sections of $E$,
\begin{equation}
\label{LAPA}%
\begin{split}
d_{A}{}^{\dagger}d_{A} \,  &  =\, \left(  \partial_{A}^{\dagger}\,+\,
\bar\partial{}^{\dagger}_A\right)  \left(  \partial_{A} \,+\, \bar
\partial_A\right)  \,,\\
&  =\, \left(  i\,\Lambda\,\bar\partial_A\,-\, i\,\Lambda\,\partial_{A}\right)
\left(  \partial_{A} \,+\, \bar\partial_A\right)  \,,\\
&  = -i \, \Lambda\left(  \partial_{A} \, \bar\partial_A\,-\, \bar
\partial_A\,\partial_{A}\right)  \,.
\end{split}
\end{equation}
In passing to the final line of \eqref{LAPA}, we use that ${\partial_{A}^{2} =
\bar\partial{}^{2}_A = 0}$.

For the same reason, ${F_A\,=\,d_{A}^{2}\,=\,\partial_{A}\,\bar\partial_A
\,+\,\bar\partial_A\,\partial_{A}}$. So again when acting on sections of
$E$,
\begin{equation}
\omega\cdot F_A\,=\,\Lambda\;d_{A}^{2}\,=\,\Lambda\left(  \partial_A
\,\bar\partial_A\,+\,\bar\partial_A\,\partial_A\right)  \,. \label{CONF}
\end{equation}
Comparing \eqref{LAPA} to \eqref{CONF} and using the identity in
\eqref{DERII}, we conclude
\begin{equation}
\begin{split}
d_{A}{}^{\dagger}d_{A}\,  &  =\,i\,\omega\cdot F_A\,-\,2i\,\Lambda
\,\partial_{A}\,\bar\partial_A\,,\\
&  =i\,\omega\cdot F_A\,+\,2\,\bar{\partial}{}^{\dagger}_A\bar\partial_A\,.
\end{split}
\label{HODGIII}
\end{equation}

\medskip\noindent{\it Examples for Line Bundles on $S$}\smallskip

We now describe some examples of line bundles on $S$ which admit an
anti-self-dual connection satisfying the Hermitian Yang-Mills
equations in \eqref{HYM}. \ Although we do not present an exhaustive
list, there are a large number of candidate line bundles which allow
us to tune the chiral matter spectrum of the resulting theory. \ As a
warmup, we first characterize all candidate line bundles
for the Hirzebruch surfaces $\mathbb{F}_{n}$ for $n\geq0$.
A candidate K\"{a}hler class $\omega=af+b\sigma$ satisfies:%
\begin{equation}
\omega\cdot f\text{, }\omega\cdot \sigma>0
\end{equation}
or:%
\begin{equation}
b>0\text{, }a>bn \label{effbound}\,.
\end{equation}
Given a line bundle $L$ on ${\mathbb F}_{n}$ such that $c_{1}(L)=Af+B\sigma$,
the condition that $c_1(L)$ be anti-self-dual implies
\begin{equation}
\omega\cdot c_{1}(L)=Ab+B(a-bn)=0\,,
\end{equation}
which admits a non-trivial solution for $a$ and $b$ satisfying the
inequalities of (\ref{effbound}) if and only if $AB<0$.

We now consider line bundles $L$ on a del Pezzo surface $dP_n$ for
which $c_1(L)$ is given by
\begin{equation}\label{DPLS}
c_{1}(L)=\underset{i=1}{\overset{n}{%
{\displaystyle\sum}
}}a_{i}E_{i}\,.%
\end{equation}
We claim that if $a_{i}a_{j}<0$ for some $i\neq j$, then there exists
a parametric family of K\"{a}hler classes $\omega$ such that the
condition $\omega\cdot c_{1}(L)=0$ holds.

Clearly, the condition on the integers $a_{i}$ implies that there exist
positive integers $b_{i}>0$ such that
\begin{equation}
\underset{i=1}{\overset{n}{%
{\displaystyle\sum}
}}b_{i}\,a_{i}=0\text{.}%
\end{equation}
We next observe that the following choice for $\omega$ defines a
K\"ahler class on $dP_n$ if the parameter $A$ appearing below is
sufficiently large,
\begin{equation}
\omega=AH-\underset{i=1}{\overset{n}{%
{\displaystyle\sum}
}}b_{i} \, E_{i}\,.
\end{equation}
Indeed, by inspection of the generators of the K\"ahler cone given in
Appendix \ref{APPDELPEZZO}, we note that when $A$ is sufficiently large,
\begin{equation}
\omega\cdot G_{i}>0
\end{equation}
where $G_{i}$ is any effective divisor on $dP_n$. \ Finally, by
construction, ${\omega \cdot c_1(L) = 0}$.

We note that in the limit that $A$ is large, the volume of $S$ is also
large.  Although the D-term condition ${\omega \cdot c_1(L) = 0}$
might receive quantum corrections away from the large-volume limit,
we expect that the line bundles $L$ determined by \eqref{DPLS} are
still associated to supersymmetric solutions on $S$ when the volume of
$S$ is large enough.

\section{Explicit Deformations of an $E_{7}$ Singularity\label{Explicit}}

In this Appendix we collect the explicit expressions for the unfolding of
the $E_{7}$ singularity used in subsection \ref{GENUNFOLDING}. The leading
order behavior of each $\beta_{i}$ as a polynomial in the $t_{i}$ is:
\begin{align}
\beta_{2}\alpha^{2}  & =\alpha^{2}\left(  16(s_{1})^{2}+O(t_{i})\right)  \\
\beta_{4}\alpha^{2}  & =\alpha^{2}\left(  \frac{16}{3}(s_{2})^{2}-8s_{1}s_{3}%
+O(t_{i})\right)  \\
\beta_{6}\alpha^{3}  & =\alpha^{3}\left(  -\frac{128}{27}(s_{2})^{3}+\frac{32}{3}%
s_{1}s_{2}s_{3}-\frac{32}{3}\left(  s_{1}\right)  ^{2}s_{4}+O(t_{i}%
^{2})\right)  \\
\beta_{8}\alpha^{3}  & =\alpha^{3}\left(
\begin{array}
[c]{l}%
-\frac{8}{3}s_{2}(s_{3})^{2}+\frac{32}{9}(s_{2})^{2}s_{4}+\frac{8}{3}%
s_{1}s_{3}s_{4}-\frac{16}{3}s_{1}s_{2}s_{5}\\
+16(s_{1})^{2}s_{6}+O(t_{i}^{2})
\end{array}
\right)  \\
\beta_{10}\alpha^{4}  & =\alpha^{4}\left(
\begin{array}
[c]{l}%
\frac{32}{9}(s_{2})^{2}(s_{3})^{2}-\frac{8}{3}s_{1}(s_{3})^{3}-\frac{128}%
{27}(s_{2})^{3}s_{4}-\frac{16}{3}(s_{1})^{2}(s_{4})^{2}\\
+\frac{64}{9}s_{1}(s_{2})^{2}s_{5}+\frac{32}{3}(s_{1})^{2}s_{3}s_{5}+\frac
{32}{3}\left(  s_{1}\right)  ^{2}s_{2}s_{6}\\
-64(s_{1})^{3}s_{7}+O(t_{i}^{3})
\end{array}
\right)  \\
\beta_{12}\alpha^{4}  & =\alpha^{4}\left(
\begin{array}
[c]{l}%
\frac{1}{3}(s_{3})^{4}-\frac{8}{9}s_{2}(s_{3})^{2}s_{4}+\frac{16}{27}%
(s_{2})^{2}(s_{4})^{2}+\frac{16}{9}s_{1}s_{3}(s_{4})^{2}\\
-\frac{8}{3}s_{1}(s_{3})^{2}s_{5}-\frac{16}{9}s_{1}s_{2}s_{4}s_{5}+\frac
{16}{3}(s_{1})^{2}(s_{5})^{2}+8s_{1}s_{2}s_{3}s_{6}\\
-\frac{32}{3}(s_{1})^{2}s_{4}s_{6}-16s_{1}(s_{2})^{2}s_{7}+O(t_{i}^{3})
\end{array}
\right)  \\
\beta_{14}\alpha^{5}  & =\alpha^{5}\left(
\begin{array}
[c]{l}%
-\frac{8}{9}s_{2}(s_{3})^{4}+\frac{64}{27}(s_{2})^{2}(s_{3})^{2}s_{4}+\frac
{8}{9}s_{1}(s_{3})^{3}s_{4}\\
-\frac{128}{81}(s_{2})^{3}(s_{4})^{2}-\frac{128}{81}(s_{2})^{3}(s_{4}%
)^{2}-\frac{32}{9}s_{1}s_{2}s_{3}(s_{4})^{2}\\
+\frac{64}{27}(s_{1})^{2}(s_{4})^{3}+\frac{16}{9}s_{1}s_{2}(s_{3})^{2}%
s_{5}+\frac{128}{27}s_{1}(s_{2})^{2}s_{4}s_{5}\\
-\frac{32}{9}(s_{1})^{2}s_{3}s_{4}s_{5}+\frac{64}{9}(s_{1})^{2}s_{2}%
(s_{5})^{2}-\frac{32}{3}s_{1}(s_{2})^{2}s_{3}s_{6}\\
+\frac{16}{3}(s_{1})^{2}(s_{3})^{2}s_{6}-\frac{32}{9}(s_{1})^{2}s_{2}%
s_{4}s_{6}-\frac{64}{3}(s_{1})^{3}s_{5}s_{6}\\
+\frac{64}{3}s_{1}(s_{2})^{3}s_{7}-64(s_{1})^{2}s_{2}s_{3}s_{7}+\frac{256}%
{3}(s_{1})^{3}s_{4}s_{7}+O(t_{i}^{4})
\end{array}
\right)
\end{align}

\newpage

\begin{align}
\beta_{18}\alpha^{6}  & =\alpha^{6}\left(
\begin{array}
[c]{l}%
\frac{2}{27}(s_{3})^{6}-\frac{8}{27}s_{2}(s_{3})^{4}s_{4}+\frac{32}{81}%
(s_{2})^{2}(s_{3})^{2}(s_{4})^{2}\\
+\frac{16}{27}s_{1}(s_{3})^{3}(s_{4})^{2}-\frac{128}{729}(s_{2})^{3}%
(s_{4})^{3}-\frac{64}{81}s_{1}s_{2}s_{3}(s_{4})^{3}\\
+\frac{64}{81}(s_{1})^{2}(s_{4})^{4}-\frac{8}{9}s_{1}(s_{3})^{4}s_{5}%
+\frac{16}{27}s_{1}s_{2}(s_{3})^{2}s_{4}s_{5}\\
+\frac{64}{81}s_{1}(s_{2})^{2}(s_{4})^{2}s_{5}-\frac{64}{27}(s_{1})^{2}%
s_{3}(s_{4})^{2}s_{5}+\frac{32}{9}(s_{1})^{2}(s_{3})^{2}(s_{5})^{2}\\
+\frac{64}{27}(s_{1})^{2}s_{2}s_{4}(s_{5})^{2}-\frac{128}{27}(s_{1})^{3}%
(s_{5})^{3}+\frac{8}{3}s_{1}s_{2}(s_{3})^{3}s_{6}\\
-\frac{32}{9}s_{1}(s_{2})^{2}s_{3}s_{4}s_{6}-\frac{32}{9}(s_{1})^{2}%
(s_{3})^{2}s_{4}s_{6}-\frac{64}{27}(s_{1})^{2}s_{2}(s_{4})^{2}s_{6}\\
+\frac{32}{3}(s_{1})^{2}s_{2}s_{3}s_{5}s_{6}+\frac{128}{9}(s_{1})^{3}%
s_{4}s_{5}s_{6}+16(s_{1})^{2}(s_{2})^{2}(s_{6})^{2}\\
-\frac{16}{3}s_{1}(s_{2})^{2}(s_{3})^{2}s_{7}+\frac{64}{9}s_{1}(s_{2}%
)^{3}s_{4}s_{7}+\frac{128}{3}(s_{1})^{2}s_{2}s_{3}s_{4}s_{7}\\
-\frac{256}{9}(s_{1})^{3}(s_{4})^{2}s_{7}-\frac{128}{3}(s_{1})^{2}(s_{2}%
)^{2}s_{5}s_{7}+O(t_{i}^{5})
\end{array}
\right)  \text{.}%
\end{align}
By inspection, when a single $t_{i}$ develops a pole, we find that each
coefficient is regular in $\alpha$ and does not vanish when $\alpha$ equals zero.

\section{Higher Exotic Singularities\label{EXOTICA}}

The results of Section \ref{MULTINT} provide a gauge theory interpretation of
models where the singularity type $G_{S}$ of rank $r$ present at generic
points of the surface enhances to a gauge group of rank $r+2$ at a discrete
collection of points in $S$. \ Our expectation is that this analysis remains
valid provided the singularity at such a point is of $ADE$ type.
\ In this Appendix we discuss the more exotic possibility where
the singularity type degenerates further. \ An example of this phenomenon is
the minimal Weierstrass model:
\begin{equation}
y^{2}=x^{3}+\alpha xz^{3}+\beta z^{5}+\gamma z^{6}+ \gamma^{\prime}xz^{4}\text{.}%
\end{equation}
so that $G_{\Sigma}=E_{8}$ and $G_{S}=E_{7}$. \ In a consistent compact model,
the coefficients $\gamma$ and $\gamma^{\prime}$ control the separation between the
singularity located at $z=0$ and other singularities at infinity in the $z$ coordinate.

At the intersection point of the curves $(\alpha = 0)$ and $(\beta = 0)$, the Weierstrass
model degenerates further. \ This has been studied in detail in
\cite{MorrisonFlorea} where this singular behavior was matched to a
codimension three degeneration of an instanton bundle in the corresponding
heterotic dual. \ We first summarize the findings of \cite{MorrisonFlorea} and
also add a few points to its physical interpretation. \ As explained in
\cite{MorrisonFlorea}, in the limit where the K\"{a}hler class of the elliptic
fibration is of zero size, the resulting singularity cannot be resolved.
\ Because a consistent F-theory compactification has vanishing K\"{a}hler
class for the elliptic curve, we physically interpret this to mean that in a
four-dimensional theory we can expect exotic physical phenomena from such enhancements in
singularity type. \ However, as explained in \cite{MorrisonFlorea}, by
performing a partial blowup of the $E_{7}$ singularity, the exotic singularity
disappears. \ In F-theory this corresponds to the fact that if we compactify
the theory on a circle to three dimensions and turn on a Wilson line around
the circle, the exotic singularity disappears. \ In other words, no
\textit{extra} blowup is necessary to resolve the singularity. This suggests
that whatever the degrees of freedom responsible for the exotic singularity
may correspond to, these degrees of freedom are charged under the $E_{7}$
gauge group so that when an $E_{7}$ Wilson line develops a vev, the resulting
degrees of freedom develop a mass.

Perhaps one of the most interesting aspects of the blowup considered in
\cite{MorrisonFlorea} is that the minimal resolution of the singularity breaks
the $E_{7}$ group to a rank seven group which is given by $SU(3)\times SU(2)\times
U(1)\times G$ for some rank three group $G$. This is remarkable because this
breaking pattern is consistent with preserving the Standard Model gauge group!
\ We find this quite intriguing and worthy of further investigation. \ Here we
explain the relevant geometric facts about the blowup and explain how the
Standard Model gauge group remains intact after the blowup.

To better understand the singular degeneration locus, we first study a blowup
of the geometry near the singular locus $x=y=z=0$. \ Following
\cite{MorrisonFlorea}, we introduce an additional projective coordinate
$\lambda$ of weight $-1$ and perform a weighted blowup by assigning weights
$1,2,3$ to $z,x$ and $y$. \ Performing the change of coordinates:%
\begin{equation}
x\mapsto\lambda^{2}x,\text{ }y\mapsto\lambda^{3}y,\text{ }z\mapsto\lambda
z\text{,}%
\end{equation}
we note that this assignment of weights preserves the scaling of the
holomorphic four form. \ The geometry is now given by:%
\begin{equation}
\lambda y^{2}=\lambda x^{3}+\alpha xz^{3}+\beta z^{5}+\lambda^{2}\gamma z^{6}+\lambda^{2}\gamma^{\prime}xz^{4} \label{blownup}%
\end{equation}
which reduces to the original Weierstrass model when $\lambda=1$. \ Note,
however, that in the patch $\lambda=0$, the hypersurface becomes:%
\begin{equation}
0=\alpha xz^{3}+\beta z^{5}%
\end{equation}
which is trivially satisfied when $\alpha = \beta = 0$. \ Precisely at the point of
degeneration, the parameters $x,y$ and $z$ are unconstrained and parameterize
an entire weighted projective space $\mathbb{P}_{[1,2,3]}^{2}$. \ Note that
the presence of the higher order terms $\gamma$ and $\gamma^{\prime}$ does not alter
this conclusion so that the resulting physics remains insensitive to global
details of the compactification. \ Away from points in $S$ where $\alpha$ and $\beta$
both vanish, the blowup corresponds to a curve inside of $\mathbb{P}%
_{[1,2,3]}^{2}$ which collapses to a point in the blowdown.

Although the blowup does not change the rank of the gauge group, the end
result now contains additional simple group factors and a $U(1)$ factor at the
location of the blown up node of the Dynkin diagram. \ We now show that the
geometry still contains a $A_{2}\times A_{1}$ type singularity so that the
non-abelian gauge group factors of the Standard Model remain intact.
\ Restricting to the patch $x=1$ of equation (\ref{blownup}) and dropping
the irrelevant contribution from $\gamma$ and $\gamma^{\prime}$, note that
rescaling the remaining coordinates by $\zeta$ yields:%
\begin{equation}
\zeta^{5}\lambda y^{2}=\zeta^{-1}\lambda+\zeta^{3}\alpha z^{3}+\zeta^{5}\beta z^{5}%
\end{equation}
so that the above equation remains invariant when $\zeta^{2}=1$. \ Indeed, note that
the condition $x=1$ is preserved under the rescaling $x\mapsto\zeta^{2}x$. \ This
establishes the presence of an $A_{1}$ type singularity. \ Next restrict to
the patch $y=1$. \ In this case, rescaling by $\eta$ yields:%
\begin{equation}
\eta^{-1}\lambda=\eta^{5}\lambda x^{3}+\eta^{5}\alpha xz^{3}+\eta^{5}\beta z^{5}%
\end{equation}
so that the above equation remains invariant when $\eta^{6}=1$. \ On the other hand, the condition $y=1$
 is preserved under the rescaling $y\mapsto\eta^{3}y$ provided $\eta^{3}$.  This
establishes the presence of a $%
\mathbb{Z}
_{3}\times%
\mathbb{Z}
_{2}=A_{2}\times A_{1}$ orbifold singularity in the geometry.\newpage
\bibliographystyle{ssg}
\bibliography{fgutsi}

\end{document}